\documentclass[iop,apj,twocolappendix]{emulateapj}
\usepackage{amsmath,amssymb,amstext}

\usepackage[breaklinks,colorlinks,citecolor=blue,linkcolor=magenta]{hyperref} 

\usepackage[all]{hypcap} 
\usepackage{multirow}
\usepackage{booktabs}
\usepackage{enumitem}
\usepackage{rotating}

\usepackage{aas_macros}
\usepackage{natbib}
\usepackage[percent]{overpic}

\shorttitle{Star Formation History of NGC~4449}

\shortauthors{Sacchi et al.}

\begin{document}

\title{
Star Formation Histories of the LEGUS dwarf galaxies. II. Spatially resolved star formation history of the Magellanic irregular NGC~4449 \footnotemark[$\star$]} \footnotetext[$\star$]{Based on observations obtained with the NASA/ESA \textit{Hubble Space Telescope} at the Space Telescope Science Institute, which is operated by the Association of Universities for Research in Astronomy under NASA Contract NAS 5-26555.}%
\author{E. Sacchi$^{1,2,3}$, M. Cignoni$^{2,4,5}$, A. Aloisi$^{3}$, M. Tosi$^{2}$,
D. Calzetti$^{6}$,
J. C. Lee$^{3,7}$,
A. Adamo$^{8}$,
F. Annibali$^{2}$,
D. A. Dale$^{9}$,
B. G. Elmegreen$^{10}$,
D. A. Gouliermis$^{11,12}$,
K. Grasha$^{6}$,
E. K. Grebel$^{13}$,
D. A. Hunter$^{14}$,
E. Sabbi$^{3}$,
L. J. Smith$^{15}$,
D. A. Thilker$^{16}$,
L. Ubeda$^{3}$ and
B.C. Whitmore$^{3}$
}
\affil{$^{1}$Dipartimento di Fisica e Astronomia, Universit\`a degli Studi di Bologna, Via Gobetti 93/2, I-40129 Bologna, Italy\\
$^{2}$INAF--Osservatorio di Astrofisica e Scienza dello Spazio di Bologna, Via Gobetti 93/3, I-40129 Bologna, Italy\\
$^{3}$Space Telescope Science Institute, 3700 San Martin Drive, Baltimore, MD 21218, USA; esacchi@stsci.edu\\
$^{4}$Dipartimento di Fisica, Universit\`a di Pisa, Largo Bruno Pontecorvo, 3, 56127 Pisa, Italy\\
$^{5}$INFN, Sezione di Pisa, Largo Pontecorvo 3, 56127 Pisa, Italy\\
$^{6}$Department of Astronomy, University of Massachusetts -- Amherst, Amherst, MA 01003, USA\\
$^{7}$Visiting Astronomer, Spitzer Science Center, Caltech, Pasadena, CA, USA\\
$^{8}$Department of Astronomy, The Oskar Klein Centre, Stockholm University, Stockholm, Sweden\\
$^{9}$Department of Physics and Astronomy, University of Wyoming, Laramie, WY\\
$^{10}$IBM Research Division, T.J. Watson Research Center, Yorktown Hts., NY\\
$^{11}$Zentrum f\"ur Astronomie der Universit\"at Heidelberg, Institut f\"ur Theoretische Astrophysik, Albert-Ueberle-Str.\,2, 69120 Heidelberg, Germany\\
$^{12}$Max Planck Institute for Astronomy,  K\"{o}nigstuhl\,17, 69117 Heidelberg, Germany\\
$^{13}$Astronomisches Rechen-Institut, Zentrum f\"ur Astronomie der Universit\"at Heidelberg, M\"onchhofstr.\ 12-–14, 69120 Heidelberg, Germany\\
$^{14}$Lowell Observatory, Flagstaff, AZ\\
$^{15}$European Space Agency/Space Telescope Science Institute, Baltimore, MD\\
$^{16}$Department of Physics and Astronomy, The Johns Hopkins University, Baltimore, MD
}

\begin{abstract}
We present a detailed study of the Magellanic irregular galaxy NGC~4449 based on both archival and new photometric data from the Legacy Extragalactic UV Survey, obtained with the \textit{Hubble Space Telescope} Advanced Camera for Surveys and Wide Field Camera 3. Thanks to its proximity ($D=3.82\pm 0.27$ Mpc) we reach stars 3 magnitudes fainter than the tip of the red giant branch in the F814W filter. 
The recovered star formation history spans the whole Hubble time, but due to the age-metallicity degeneracy of the red giant branch stars, it is robust only over the lookback time reached by our photometry, i.e. $\sim 3$~Gyr. The most recent peak of star formation is around 10 Myr ago. The average surface density star formation rate over the whole galaxy lifetime is $0.01$~M$_{\odot}$~yr$^{-1}$~kpc$^{-2}$. From our study it emerges that NGC~4449 has experienced a fairly continuous star formation regime in the last 1 Gyr with peaks and dips whose star formation rates differ only by a factor of a few. The very complex and disturbed morphology of NGC~4449 makes it an interesting galaxy for studies of the relationship between interactions and starbursts, and our detailed and spatially resolved analysis of its star formation history does indeed provide some hints on the connection between these two phenomena in this peculiar dwarf galaxy.
\end{abstract}


\maketitle

\section{Introduction}
\setcounter{footnote}{16}
\begin{figure*}
\centering
 \begin{overpic}[width=0.99\textwidth]{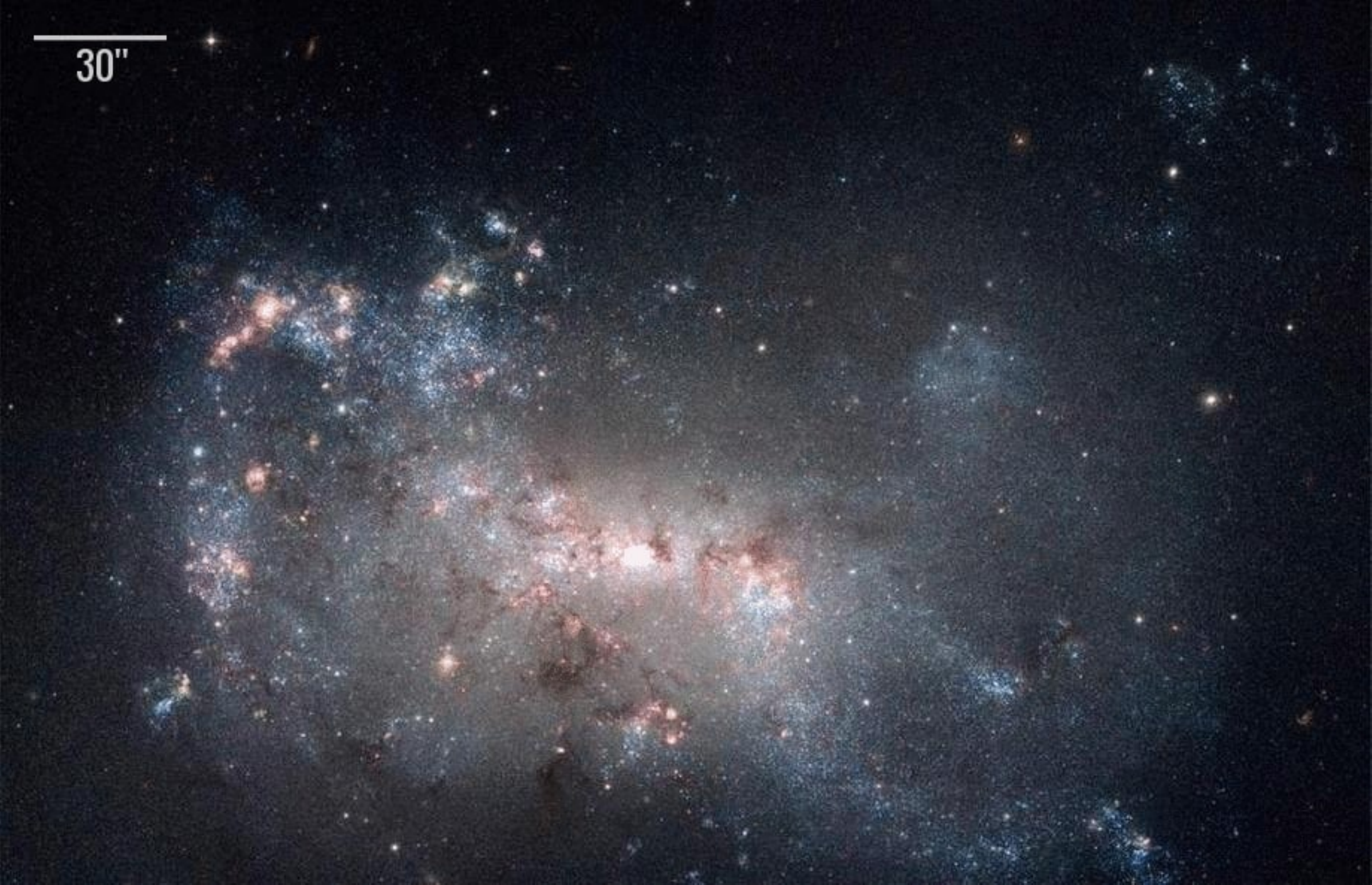}
     \put(91.5,0.5){\includegraphics[scale=0.4]{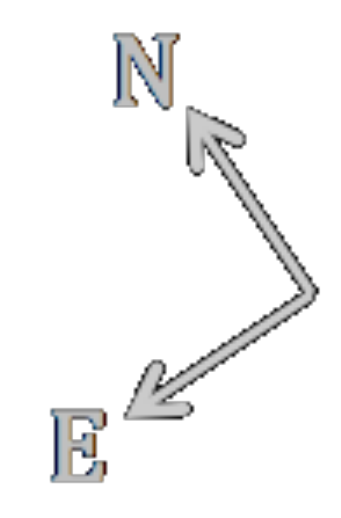}}
  \end{overpic}
\caption{Four color composite image of NGC~4449 from \textit{HST}/ACS observations: blue corresponds to F435W (B), green to F555W (broad V), red to F814W (I), magenta to F658N (H$\alpha$). In the upper left corner is indicated the angular scale (30 arcsec) which corresponds to $\sim 550$ pc at the adopted distance. Original figure from \cite{Annibali2008}.}
\label{3col}
\end{figure*}
\begin{figure}
\centering
\includegraphics[trim=0 -1.7cm 0 -1.8cm, clip, width=\linewidth]{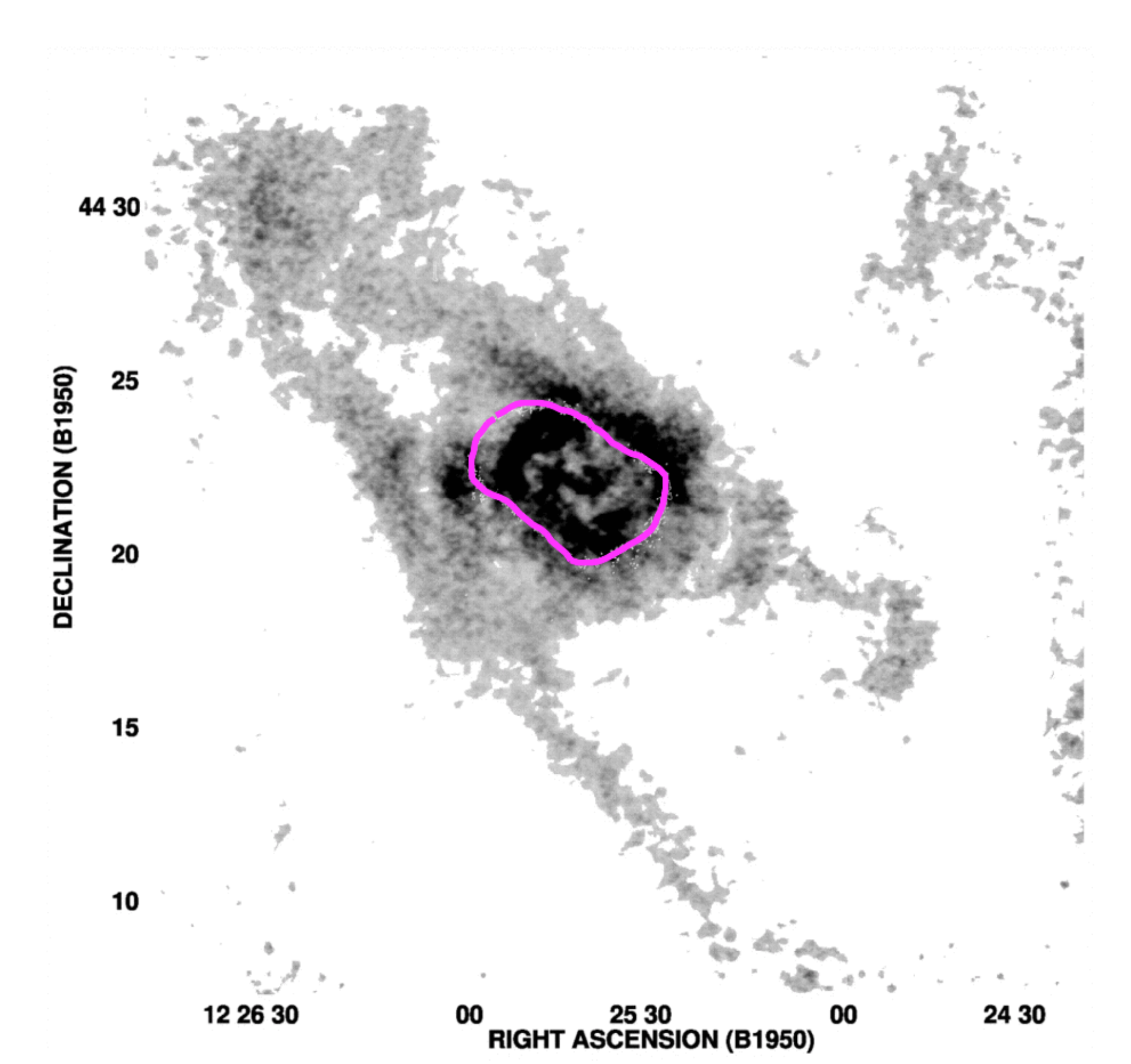}
\caption{Integrated H~\textsc{i} map of NGC~4449. The magenta line indicates the contour of the optical galaxy. Original figure from \cite{Hunter1999}; notice the different orientation angle (North up, East left) with respect to Figure \ref{3col}.}
\label{gas}
\end{figure}

The local Universe is characterized by a great variety of galaxies and galactic environments. Among them, dwarf galaxies are the most common but probably least understood objects when we consider their formation and evolution. In fact, in the comparison both among dwarfs and between dwarf and spiral galaxies, we can find a wide range of chemical, dynamical and star formation (SF) properties, and the evolutionary processes taking place in dwarfs are still highly debated. In particular, it is unclear which mechanisms can trigger strong bursts of star formation in such small systems, and what the impact of the starburst activity on the evolution of the galaxy itself is. Many previous studies have investigated possible mechanisms leading to the formation of a starburst, such as tidal interactions, mergers, or gas accretion \citep{Lelli2014c}, highlighting the close link between environment, gas distribution and star formation. On the other hand, simulations show that stellar feedback (gas out/in-flows) also plays an important role in determining the stellar kinematics and radial gradients in low-mass ($\mathrm{M_{\ast} \sim 10^{7-9.6}\ M_{\odot}}$) galaxies \citep{ElBadry2016}.

For nearby galaxies, we have the great advantage of resolving their stellar populations, so that we are able to perform star-by-star analyses and use the color-magnitude diagram (CMD) to recover their star formation histories (SFHs). Insight into the origin and impact of a starburst can be provided by deriving the detailed star formation history from young to old epochs based on deep photometric data. 

Here we present the SFH of NGC~4449, one of the key target galaxies of the LEGUS (Legacy ExtraGalactic Ultraviolet Survey) \textit{Hubble Space Telescope} (\textit{HST}) Treasury Program \citep{Calzetti2015}, a survey whose aim is to investigate and connect the different scales of star formation in the local Universe, from stellar clusters to galaxies, and to explore the relation of star formation with the environment. 
NGC~4449 is an extremely interesting and well-studied galaxy, which is known to be interacting and currently forming stars at a high rate ($\mathrm{SFR_{\, UV} \sim 0.94\ M_{\odot}\ yr^{-1}}$, $\mathrm{M_{\ast} \sim 1.1\times 10^9\ M_{\odot}}$ from \citealt{Calzetti2015}). It is classified as a Magellanic irregular galaxy; it has characteristics similar to those of the Large Magellanic Cloud (LMC), such as dimensions, $\mathrm{R_{\, opt} \sim 3.3}$~kpc \footnote{$\mathrm{R_{\, opt}}$ is defined as 3.2 times the exponential scale length \citep{Lelli2014b}.}, and metallicity, between $12+\log(O/H) = 8.26 \pm 0.09$ and $12+\log(O/H) = 8.37 \pm 0.05$ from spectroscopy of the ionized gas in its H~\textsc{ii} regions \citep{Berg2012,Annibali2017}, although its star formation activity is at least two times higher. Moreover, NGC~4449 seems to be interacting with neighboring galaxies, as shown by its stellar and gas distribution (see Figures \ref{3col} and \ref{gas}). A recent detection highlights a stellar tidal stream that is falling onto the galaxy and whose origin is thought to be the disruption of a smaller dwarf spheroidal (dSph) galaxy \citep{Rich2012,Martinez-Delgado2012}. Furthermore, \cite{Annibali2012} reported the discovery of a very massive ($\mathrm{M\sim 10^6\ M_{\odot}}$) elliptical old star cluster, apparently associated with two tails of blue stars, which may be the nucleus of a former gas-rich satellite galaxy undergoing tidal disruption by NGC~4449.

\begin{figure}
\centering
\includegraphics[width=\linewidth]{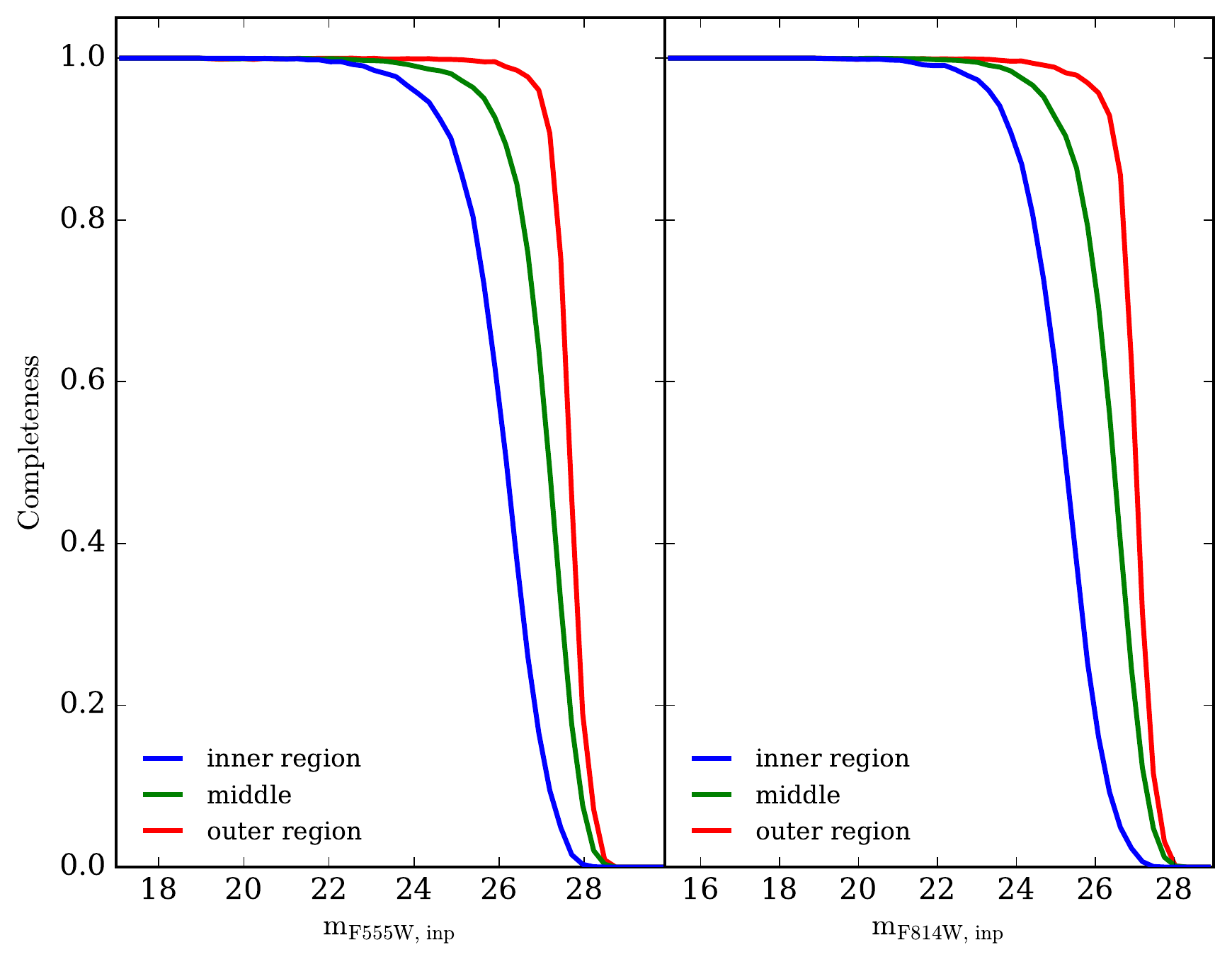}
\caption{Completeness in F555W (left panel) and F814W (right panel) from our artificial star tests in three different regions of the galaxy (inner in blue, middle in green, outer in red, see Section \ref{sec_sfh}) highlighting the different crowding conditions from inside out.}
\label{compl}
\end{figure}
\begin{figure}
\centering
\includegraphics[width=\linewidth]{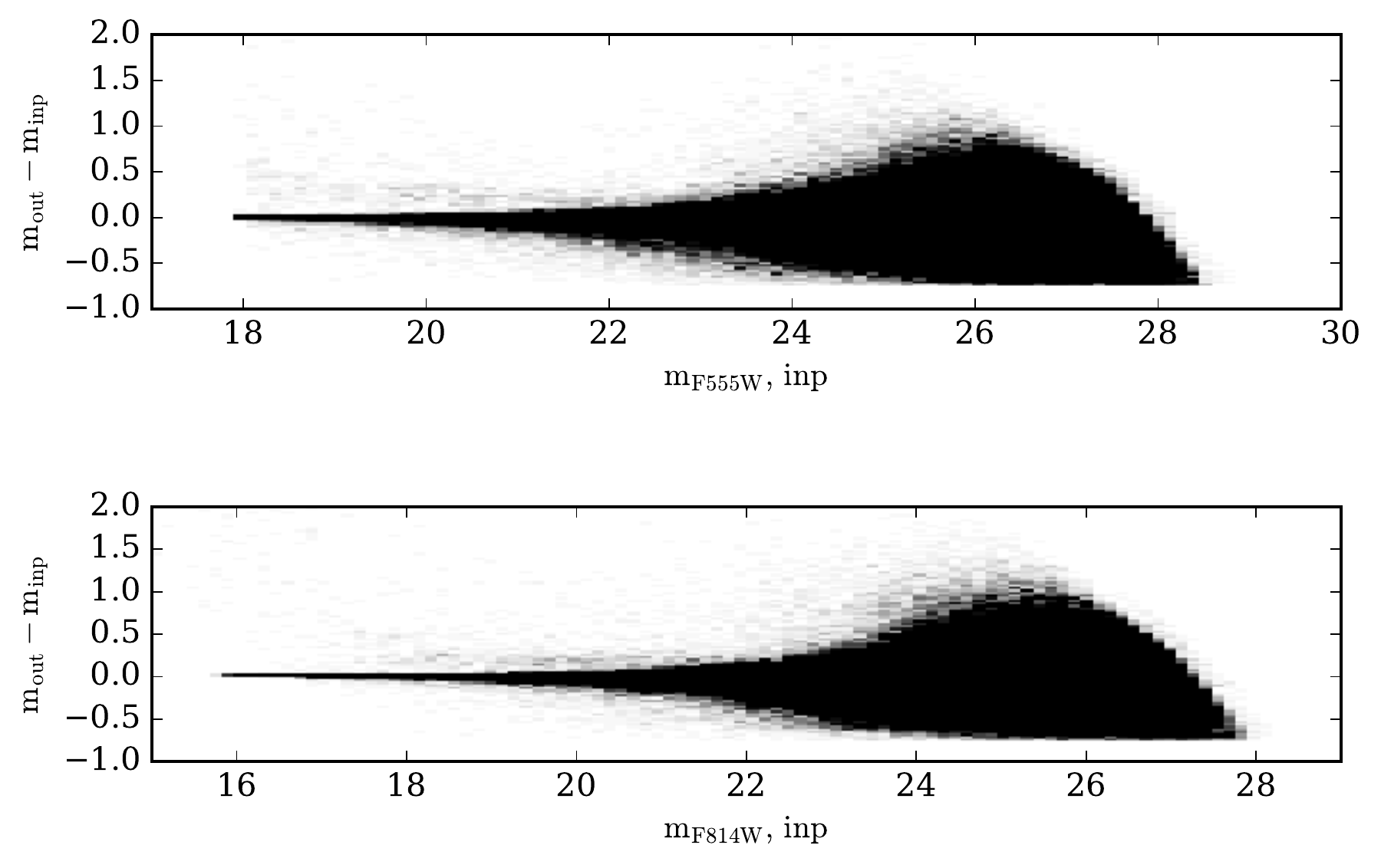}
\caption{Photometric errors in F555W (top panel) and F814W (bottom panel) from our artificial star tests.}
\label{deltamag}
\end{figure}

The H~\textsc{i} morphology is highly disturbed, with filaments and giant ring and shell-shaped complexes extending well beyond the optical galaxy (Fig. \ref{gas}). Moreover, the inner and outer parts of the neutral gas form two separate systems that are counter-rotating \citep{Hunter1999} which generally signals the recent accretion of gas.

LEGUS studies NGC~4449 under many aspects, e.g., its star forming regions, its star clusters, its resolved stellar populations, exploiting a multi-band (UV, U, B, V, I) approach unavailable until now. To this purpose, LEGUS has acquired deep UV, U and B photometry with the Wide Field Camera 3 (WFC3) on board \textit{HST}, and reduced archival data with the Advanced Camera for Surveys (ACS), with the same procedures and requirements for the entire 50 galaxy sample.

The SFH of NGC~4449 was already derived by \cite{McQuinn2010} as part of a study of 18 dwarf galaxies. It was based on the same archival V and I ACS images and refers to the whole galaxy. Here we provide a new SFH based on the same data, but inferred with the method for comparing synthetic and observational CMDsthat will be homogeneously applied to all the LEGUS dwarf galaxies, and with two different sets of the most updated stellar evolution models. The latter is important to assess systematic uncertainties \citep[see, e.g.,][]{Skillman2017}. The UV (from F336W and F555W filters) SFH for NGC~4449 is presented by \cite{Cignoni2018} together with those of other LEGUS galaxies, while here we focus on the optical one. Our analysis includes an accurate method to estimate photometric errors and completeness of the catalog, and is based on two different up-to-date sets of stellar evolution models. Moreover, we perform a spatially resolved SFH analysis of the galaxy, which can be compared to the results by \citet{McQuinn2012} who measured the concentration of SF in 15 nearby starburst galaxies, including NGC~4449.

\section{Observations and Data} \label{sec_obs}
Figure \ref{3col} shows the \textit{HST}/ACS Wide Field Channel (WFC) image of NGC~4449 presented by \cite{Annibali2008}. The images were acquired in November 2005 (GO program 10585) in the F435W, F555W, F814W and F658N filters, following a dither pattern. For each of these filters, images with eight exposures of 900~s, 600~s, 500~s and 90~s, respectively, were acquired. These data were included in the LEGUS 5-band image processing together with the new UV- and U-band ones (broadband filters F275W and F336W) acquired with the UVIS channel of the WFC3 (GO Treasury program 13364, PI D. Calzetti). Images in all five bands were aligned and drizzled onto the same grid, and photometry was performed using the DOLPHOT 2.0 package \citep{Dolphin2016}. No relevant differences were found with the photometry performed by \cite{Annibali2008}. The sensitivity was enough to reach at least one magnitude below the tip of the red giant branch (TRGB) with a photometric error smaller than 0.1 magnitudes, and detect stars as faint as $\sim 3$ magnitudes below the tip.

More details on the LEGUS stellar photometry are described by \citet{Calzetti2015} and  extensively treated in \citet{Sabbi2018} while the data are publicly available on MAST \footnote{\url{https://archive.stsci.edu/prepds/legus/dataproducts-public.html}}.

We selected our data with the following DOLPHOT parameters, in all filters: photometric error $\sigma \le 0.2$ (meaning a signal-to-noise ratio $\geq 5$), sharpness $\le 0.2$ (excluding objects with size significantly deviating from the point spread function), crowding $\le 2.25$ (excluding only objects in very crowded regions such as candidate clusters), object type $\le 2$ (excluding spurious objects such as background galaxies). The final catalog after the quality cuts includes $\sim 571\,000$ stars in both F555W and F814W and is complete down to $\sim 5$ M$_{\odot}$.

\begin{figure}
\centering
\includegraphics[width=\linewidth]{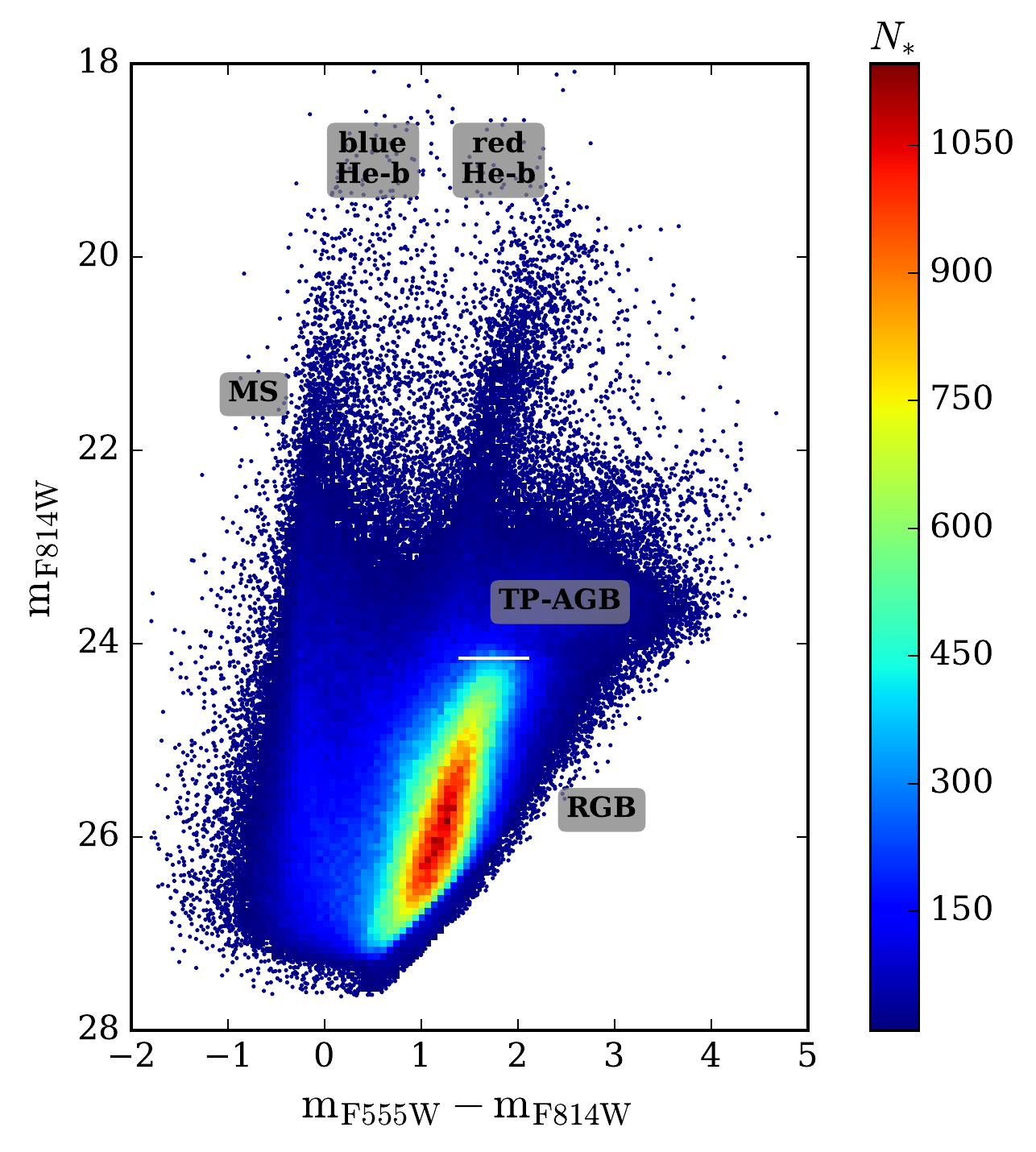}
\caption{Optical color-magnitude diagram of the whole field of NGC~4449 covered by the ACS imaging (after the quality cuts, see Section \ref{sec_obs}). The high density regions have been binned and color coded by number density (see the color bar) for a better visualization of the evolutionary features in the diagram. The main stellar evolutionary phases are indicated (see Section \ref{sec_pop}). The horizontal white line represents the magnitude of the red giant branch tip.}
\label{cmd_all}
\end{figure}

\section{Artificial Star Tests} \label{sec_ast}
To estimate the photometric errors and incompleteness of our data, we use a refined version of the artificial star test 
described in \cite{Cignoni2016} and here summarized. First, we put a spatially uniform distribution of artificial stars on the images (one star at a time) and re-process them with DOLPHOT every time we add a new source; this gives us a first estimate of the completeness of the catalog as a function of the magnitude. We then use this function to reconstruct the actual density profile of the galaxy and to add a new set of fake stars on the images. We applied this method to each filter separately and then combined the outputs to obtain information of every input star in all filters.

\begin{figure*}
\centering
\includegraphics[width=\linewidth]{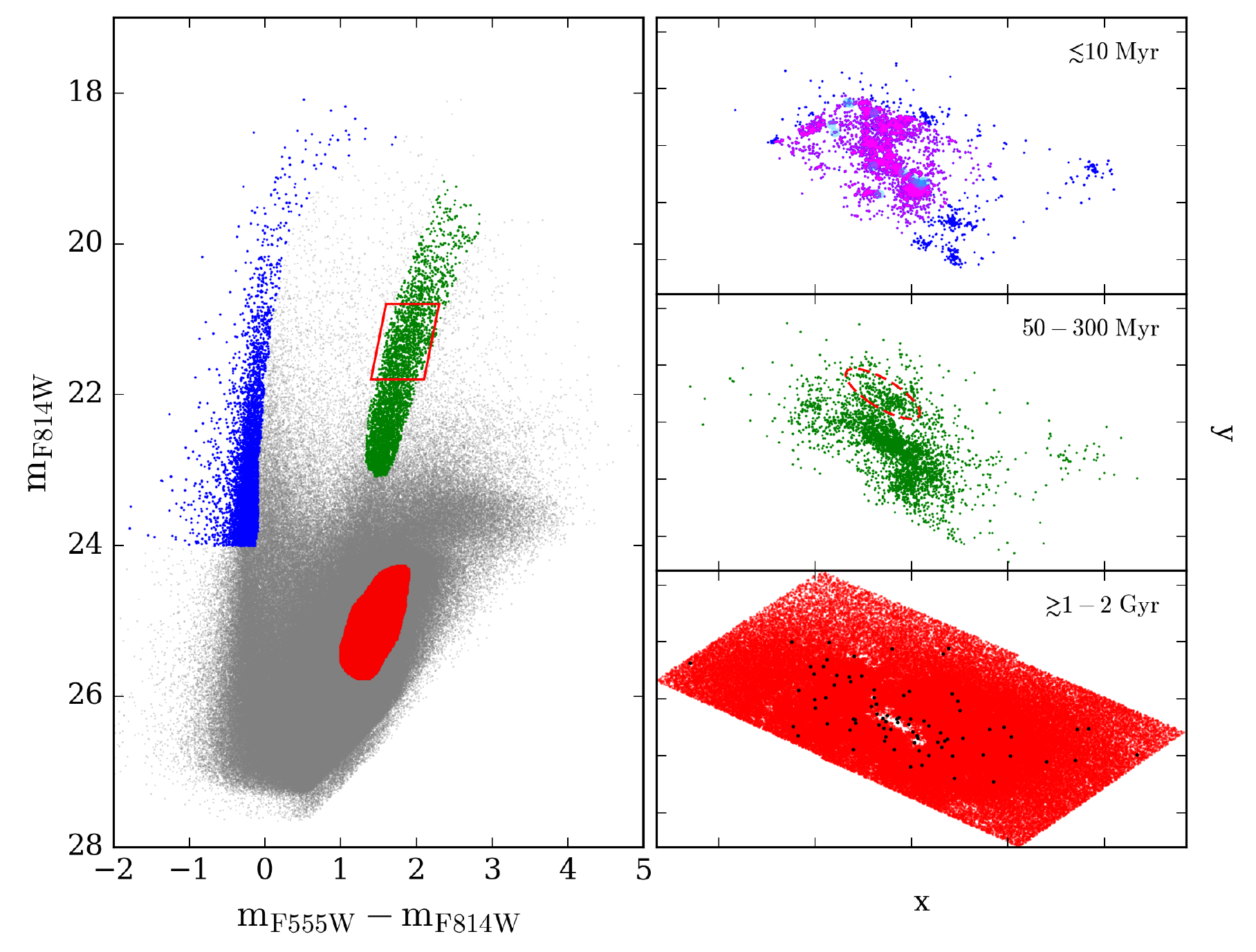}
\caption{Left panel. Selection of three different stellar populations in the CMD: in blue, stars with ages $\lesssim 10$ Myr; in green, stars with intermediate ages between $\sim 50$ and  $\sim 300$ Myr; in red, stars older than $\sim 1-2$ Gyr. Right panel. Spatial distribution of the age-selected stars: the magenta points in the top map represent stars that have also a measured flux in the F336W filter, while the light blue circles are clusters with thermal radio emission from \cite{Reines2008}; the red dashed ellipse in the middle map includes stars forming an elongated structure and with roughly the same F814W luminosity (inside the red parallelogram in the CMD); the black dots in the bottom map are the stellar clusters identified by \cite{Annibali2011}.}
\label{pop_distr}
\end{figure*}
With this iterative procedure, we obtain a more precise description of the completeness as a function of position and magnitude, and we are also able to put more artificial stars in regions that need to be explored in more detail because of their higher crowding. This way, in particular in galaxies with high density gradients from one region to another, we obtain a description of the completeness function not biased by the average density but describing the local incompleteness: the most crowded parts will have a lower and more realistic completeness than the one inferred from the uniform distribution, since they have been explored with a higher number of artificial stars. Notice that this procedure does not create artificial crowding, since we always add only one artificial star at a time. This allows to take into account the very different morphologies we can encounter with a very general procedure, which does not involve galaxy-by-galaxy treatment such as a division of the field following iso-density regions. An equivalent approach was adopted by, e.g., \cite{Johnson2016}, who choose input positions for the artificial stars distributed according to the surface brightness of the real stars.

In Figure \ref{compl} we plot the completeness derived from our artificial star tests in three different regions of the galaxy (inner, middle and outer, see Section \ref{sec_sfh}) to show how the higher crowding of the inner regions severely reduces their completeness. We point out, however, that the complex morphology of NGC 4449 makes crowding (and hence completeness) not a simple function of the galactocentric distance, but rather a patchy pattern following the major star forming regions. Figure \ref{deltamag} shows the $m_{output}-m_{input}$ versus $m_{input}$ distribution of the artificial stars in the two filters, that we use to estimate the photometric errors.

\section{Distribution of the stellar populations} \label{sec_pop}
The optical CMD of the whole catalog we obtained after the quality cuts is shown in Fig. \ref{cmd_all}. We can see a large variety of stellar populations and recognize the main features of CMDs typical of this kind of galaxies:
\begin{itemize}
\item[-] the blue plume ($-1 \lesssim \mathrm{m_{F555W}-m_{F814W}} \lesssim 0.5$) containing stars on the main sequence (MS) and at the hot edge of the helium burning phase;
\item[-] the red plume ($1.5 \lesssim \mathrm{m_{F555W}-m_{F814W}} \lesssim 2.5$, $\mathrm{m_{F814W}} \lesssim 24$) made of red helium burning and asymptotic giant branch (AGB) stars;
\item[-] the blue loops, populated by helium burning stars between the two plumes;
\item[-] the horizontal red tail ($\mathrm{m_{F555W}-m_{F814W}} \gtrsim 1.8$, $\mathrm{m_{F814W}} \lesssim 24$) made of thermally pulsing asymptotic giant branch (TP-AGB) stars;
\item[-] the red giant branch (RGB, $\mathrm{m_{F555W}-m_{F814W}} \gtrsim 0.5$, $\mathrm{m_{F814W}} \gtrsim 24$) with the oldest population reachable at this distance.
\end{itemize}

In order to better understand the evolutionary status of NGC~4449, we isolated some of these phases and inspected how they are spatially distributed across the body of the galaxy. We chose three age intervals, corresponding to: very young ($\lesssim 10$ Myr) blue MS stars, intermediate-age ($\sim 50 - 300$ Myr) helium burning and AGB stars, and old ($> 1-2$ Gyr) RGB stars. Fig. \ref{pop_distr} shows these selected stars, both in the CMD and in the spatial map. As a general behavior, we find that the younger the stars, the more concentrated and clumped in groups their spatial distribution, as naturally expected for star formation that proceeds hierarchically \citep{Gouliermis2017}, a behavior that is observed in star clusters as well \citep{Grasha2017a,Grasha2017b}.

In addition, there are several very interesting features revealed by the maps. First of all, the young stars seem to form a sort of S-shaped structure crossing the galactic center, that follows the distribution of the H$\alpha$ gas. These are the loci of very recent SF, as confirmed by the detection of stars in our catalog in the F336W filter (magenta points in the figure). Their distribution is clearly lopsided, suggesting an external event to cause the asymmetry and to trigger the formation of H~\textsc{ii} regions on the north side of the galaxy. Moreover, \cite{Reines2008} identified 13 clusters with thermal radio emission (shown with light blue circles in the figure); they estimate that these sources have ages $\lesssim 5$~Myr and stellar masses $\gtrsim 10^4$~M$_{\odot}$, and they represent the birthplace of the next generation of stars. The S-shaped structure is still visible in the intermediate-age population, that is however more diffuse. A clump of stars forming a shallow overdensity (highlighted by the red dashed ellipse in the figure) is also visible; if we select these stars on the map and look at where they are on the CMD, we find that their luminosities are all very similar to each other ($\mathrm{m_{F814W}} \sim 21.3$), meaning that they are stars of roughly the same age \footnote{During the helium burning phase of intermediate mass stars ($> 2$ M$_{\odot}$), there is a direct correlation between mass and luminosity of a star since the core is non-degenerate; this makes the luminosity of the blue loops a very fine age indicator.}; from a comparison with isochrones, the age of the stars in the parallelogram is between 80 and 150 Myr. This suggests that the structure originated from a specific episode of SF, maybe triggered by the inflow of new gas from some interaction, that created this small coeval population. Moreover, this region hosts a Wolf$-$Rayet (WR) massive cluster which is believed to be a young evolutionary stage of a super star cluster as 30 Doradus in the LMC \citep{Sokal2015}. The old stars are spread all over the galaxy, with holes in the distribution only due to incompleteness. 


\section{Star Formation History} \label{sec_sfh}
The SFH of NGC~4449 was studied through the synthetic CMD method, originally devised by \cite{Tosi1991}. We used the code SFERA (Star Formation Evolution Recovery Algorithm), developed by \cite{Cignoni2015}, a hybrid-genetic algorithm (genetic plus simulated annealing) based on a Poissonian approach \citep{Dolphin2002}, already applied to derive the SFH over both short and long lookback times \citep{Cignoni2015,Cignoni2016,Sacchi2016}, and described in detail in the Appendix. In particular, the application of SFERA to NGC~4449 is similar to what was done for the concentric sub-regions of DDO~68 \citep{Sacchi2016}, where the lookback time was also given by the RGB stars.

The theoretical CMDs used for the comparison with the observed one were built from two different sets of stellar evolution isochrones, PARSEC-COLIBRI \citep{Marigo2013,Rosenfield2016} and MIST \citep{Choi2016}. This allowed us to explore the uncertainties arising from the models, and to test the strength and reliability of different evolutionary phases. For both, we used the following input parameters:
\begin{itemize}
\item[-] Kroupa initial mass function \citep{Kroupa2001} from 0.1 to 300 M$_{\odot}$;
\item[-] 30\% binary fraction;
\item[-] metallicity in the range $Z = [0.0003 - 0.008]$ or [Fe/H] from $-1.7$ to $-0.3$ in steps of $0.1$.
\end{itemize}
We then convolved the models with the data properties:
\begin{itemize}
\item[-] distance modulus $(m-M)_0 =27.91\pm 0.15$ \citep{Annibali2008} which is allowed to vary in a range around the chosen literature value, in steps of 0.05 mag (this is not a parameter included in the minimization, see the Appendix for details);
\item[-] photometric errors, incompleteness and blending from the Artificial Star Tests;
\item[-] total reddening (foreground\footnote{$\mathrm{E(B-V)}=0.017$ from \cite{Schlafly2011}.} plus internal) varying from 0.05 to 0.20 mag in steps of 0.05;
\item[-] differential reddening (modeled as a Gaussian spread) varying from 0.1 to 0.2 mag; the differential reddening of the youngest MS population is modeled separately from the rest of the stars \citep{Dolphin2003}.
\end{itemize}

A crucial point in the analysis of this galaxy is the existence of a rich population of TP-AGB stars, which presents several challenges, both observational and theoretical, such as modeling circumstellar dust obscuration \citep{Boyer2009}, mass loss, and evolutionary timescales. Both isochrone sets we used have recently included a detailed treatment of this stellar phase, but they still lack an empirical calibration with a real galaxy (this is typically done with the LMC). 
Moreover, the models from the two groups are quite different, so we decided to mask the TP-AGB area in the CMD, in order to exclude it from the minimization procedure applied to search for the best SFH.

\subsection{Whole galaxy}
\begin{figure*}
\centering
\includegraphics[width=0.9\columnwidth]{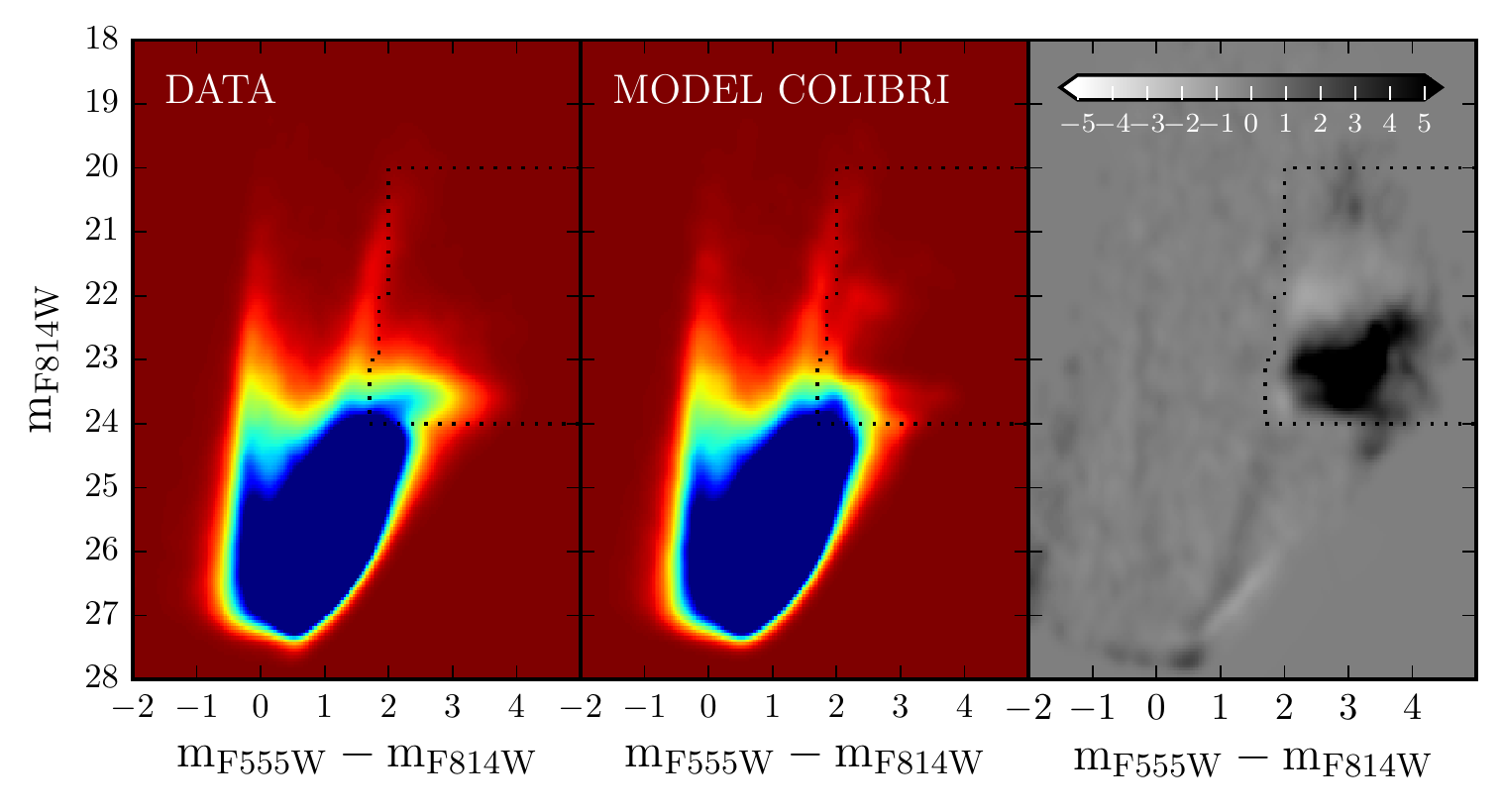}
\includegraphics[width=0.9\columnwidth]{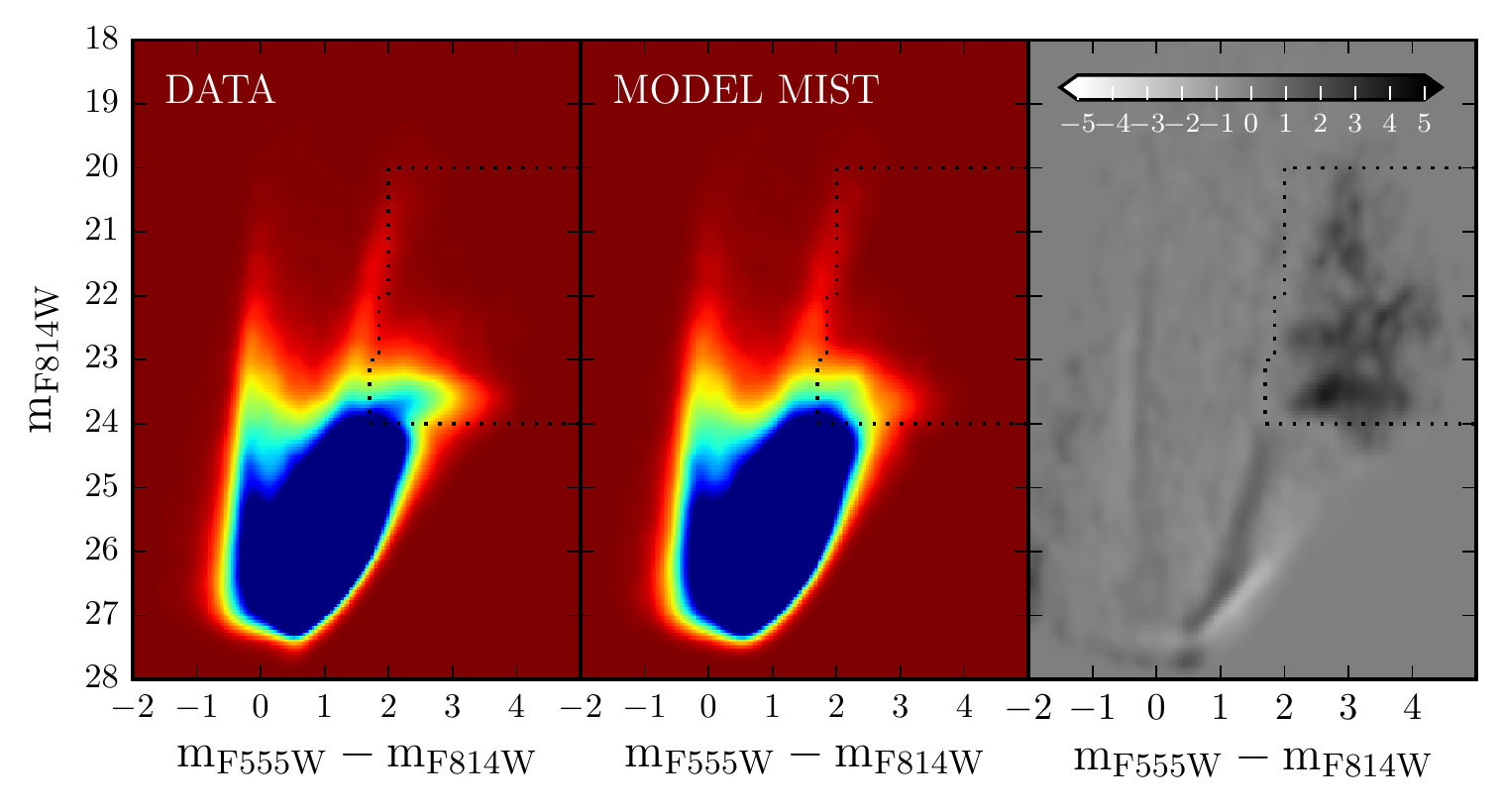}
\centering
\includegraphics[width=0.9\columnwidth]{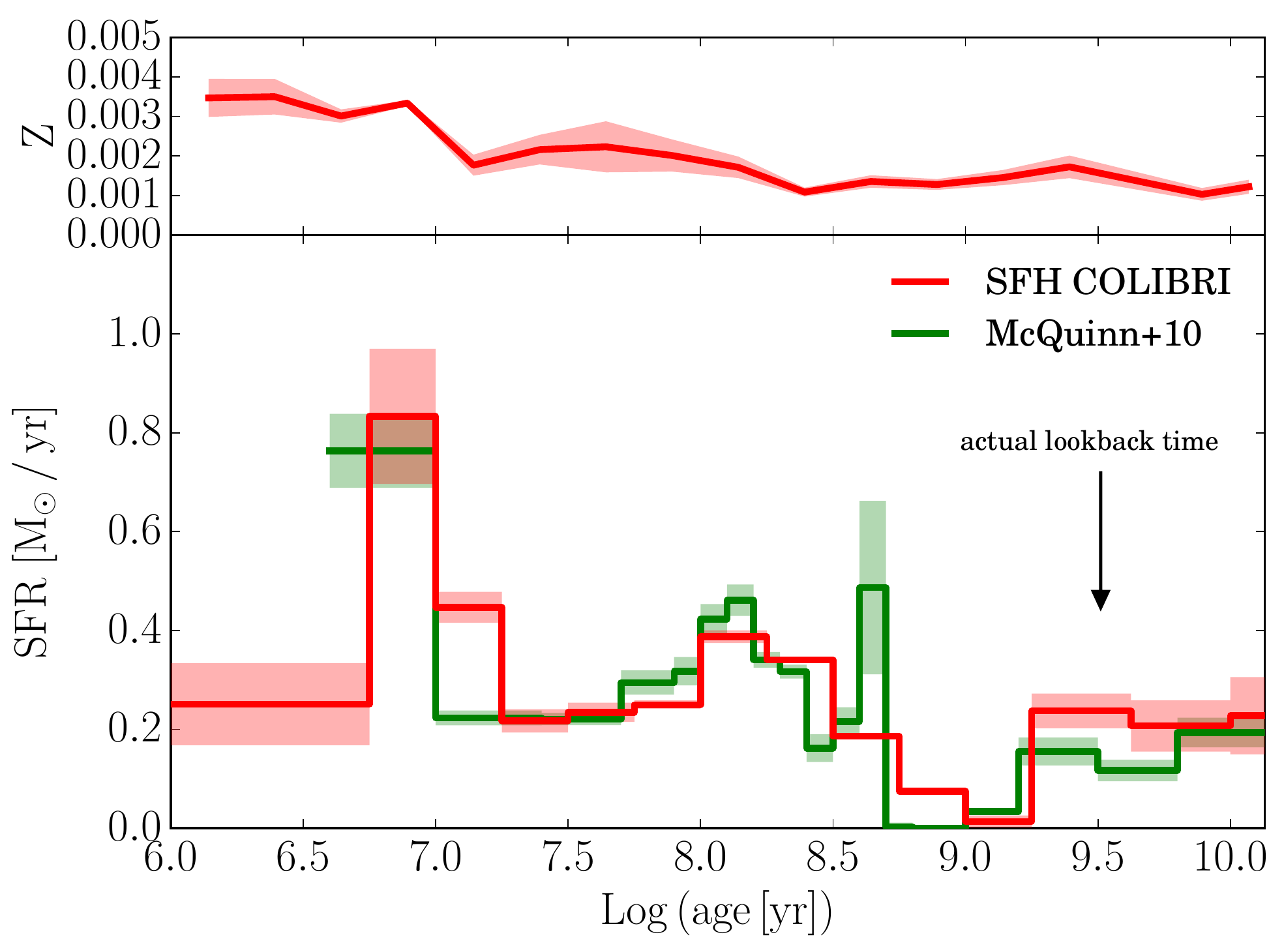}
\includegraphics[width=0.9\columnwidth]{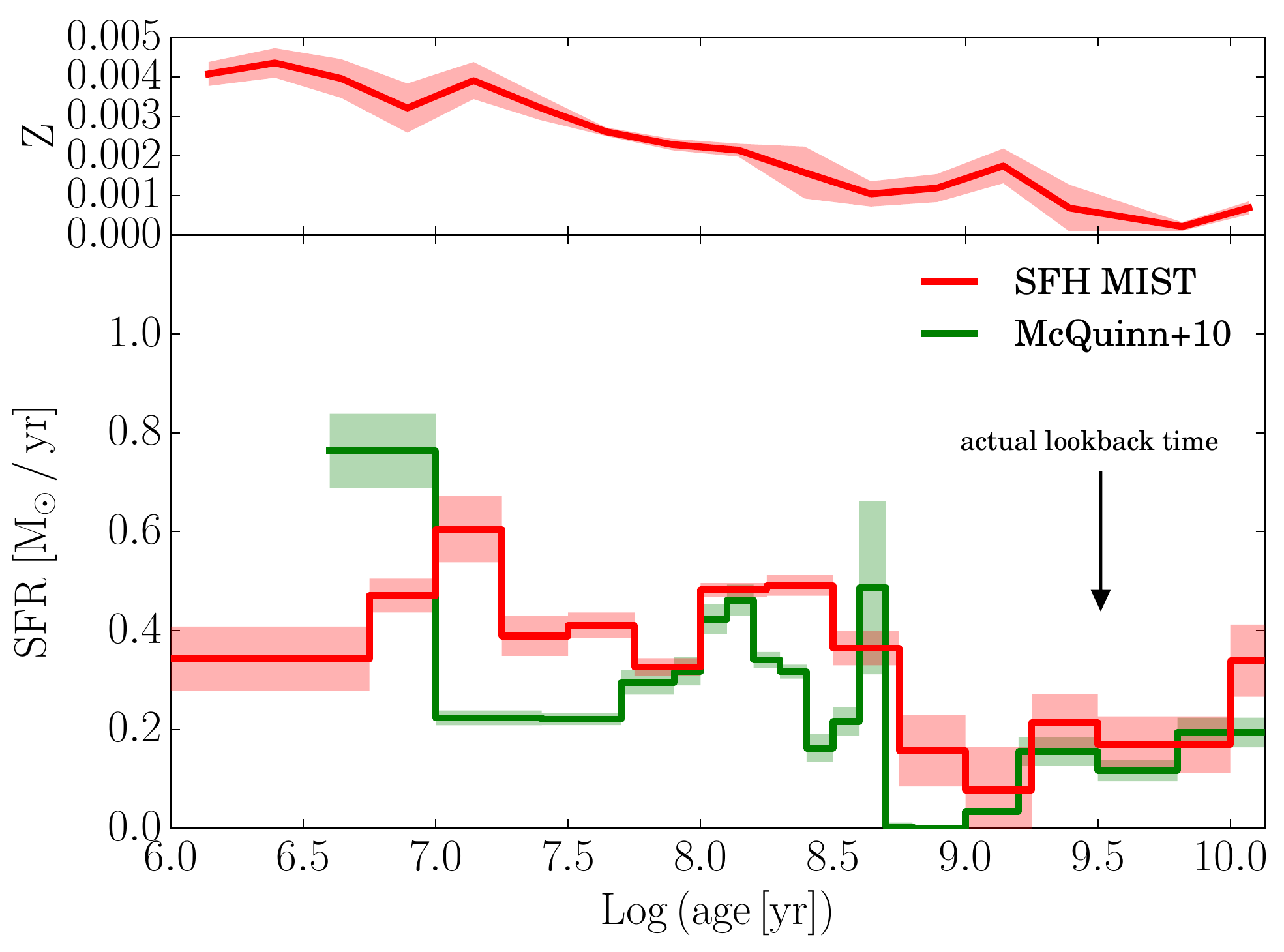}
\caption{Top panels. Hess diagrams of the whole optical field of NGC~4449: the observational one is on the left, the one reconstructed on the basis of different sets of models in the middle, and the residuals between the two (in units of Poisson uncertainties $(data-model)/\sqrt{model}$) on the right; the dotted line shows the region containing TP-AGB stars and excluded from the minimization. Bottom panels. Recovered star formation history and metallicity; in green, the SFH from \cite{McQuinn2010} based on older stellar models from the Padova group. The left plots refer to the COLIBRI models, the right ones to the MIST models.}
\label{sfh-total}
\end{figure*}

Figure \ref{sfh-total} shows the results for the total CMD of NGC~4449. In the top panels the observed and recovered CMDs are shown as Hess diagrams, i.e., density plots, as this is the way we perform the minimization of the residuals between them; the dotted black box indicates the TP-AGB region masked for the comparison. To analyze these Hess diagrams we also show the residuals between the observational and synthetic CMDs in units of Poisson uncertainties $(data-model)/\sqrt{model}$ (right panels).
In the bottom panels, we show the star formation history and the evolution of the metallicity as a function of the age. For the whole field we show the results for both the COLIBRI and MIST isochrones, to identify their differences and the features that may result from the chosen set of evolutionary tracks.

As suggested by the presence of stellar populations of all ages, the SF of NGC~4449 has been fairly continuous over the lifetime of the galaxy. In particular, we find peaks and dips, but the SFR at the top of the peaks is only a factor of a few higher than at the bottom of the dips. The duration of the peaks is similar to that of the dips. This behavior is not what we would call a ``bursting'' regime, where bursts should be short episodes, with SFR significantly higher than average, separated by long quiescent phases. What we find for NGC~4449 is instead quite similar to what is found in the Small Magellanic Cloud \citep[e.g.][]{Cignoni2012,Cignoni2013} and in other Local Group irregulars \citep[e.g.][]{Tolstoy2009,Gallart2015}: a ``gasping'' \citep{Marconi1995} more than a ``bursting'' regime. We find the main peak of SF at ages between 5 and 20 Myr ago, in the same interval found by \cite{McQuinn2010}, with a SFR almost 4 times higher than the average; some activity is also found in the most recent bin (last $5-6$ Myr) in agreement with the presence of H$\alpha$ emission in the galaxy. The SFH shows another enhancement around 100 Myr, in excellent agreement with the SFH by \cite{McQuinn2010}. Notice that their SFRs have been re-scaled to take into account the different IMFs used to construct the synthetic CMDs (Salpeter in their case, Kroupa in ours). The SFH by \cite{McQuinn2010} better agrees with our COLIBRI solution, as expected since they were computed from the same stellar evolutionary models, even though from an older version in their case. 

There are a few differences between the models, and between models and data: both synthetic CMDs reproduce the overall shape of the observed one, with some differences in the widths of the MS and RGB, in particular for the MIST models. This is likely caused by the relatively simplistic treatment of the reddening parameters, in particular the differential reddening, which is quite difficult to determine exactly. 
Even with the whole CMD fitted, all models fail to reproduce the TP-AGB phase, either in terms of CMD morphology and of star counts. This is due to the well-known difficulty in modeling such a rapid phase and is the reason why we decided to mask the entire phase to infer the proper SFH. 
The remaining features of the CMD are well fitted within the errors, and the SFH never shows signs of long interruptions within our time resolution. Although the trends are similar, the solutions from the two sets of stellar models show significant differences: the MIST SFH tends to be more constant where the COLIBRI one shows a higher variation around the average. This is a behavior we often find when comparing the two libraries, thus we consider this difference as an estimate of the systematic uncertainty in our SFH.

The metallicity shows a growing trend with time, and in the youngest bins it matches the spectroscopic value inferred from H~\textsc{ii} region observations \citep{Berg2012,Annibali2017}.

\begin{figure}
\centering
\includegraphics[width=\columnwidth]{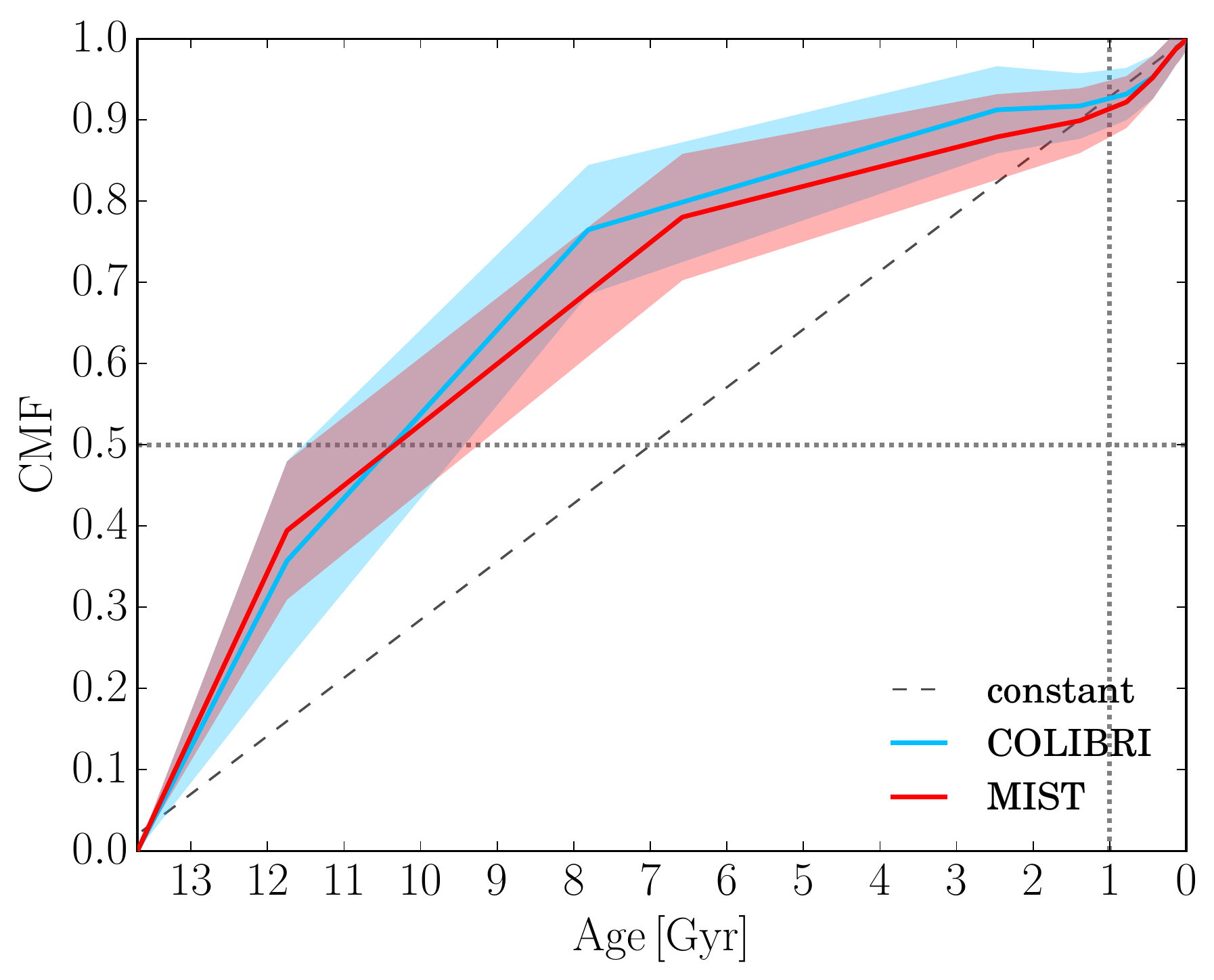}
\caption{Cumulative stellar mass fraction for the total field of NGC~4449 recovered with the COLIBRI (light blue) and MIST (red) tracks; the dotted vertical line indicates an age of 1 Gyr whilst the dotted horizontal line shows 50\% of the total mass. The dashed line corresponds to a constant SFH.}
\label{cdf}
\end{figure}
\begin{figure}
\centering
\includegraphics[width=\columnwidth]{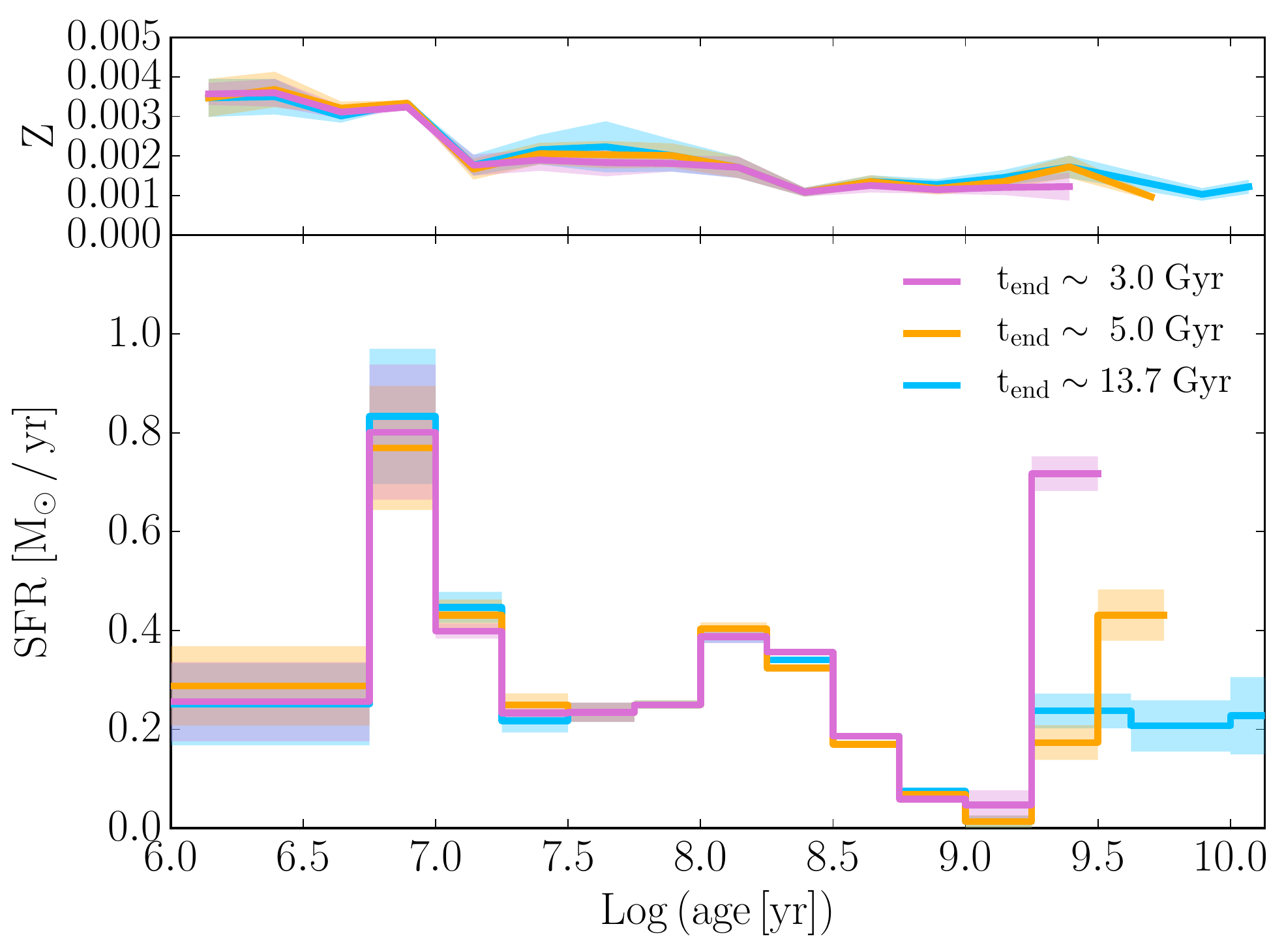}
\caption{Comparison among the SFH of the total field of NGC~4449 recovered within three different lookback times: 3 Gyr (pink line), 5 Gyr (orange line) and the whole Hubble time (light blue, same as in the left bottom panel of Fig. \ref{sfh-total}).}
\label{sfh-resolution}
\end{figure}
To better evaluate the differences between the models, and for an easier comparison with other literature results, we show in Figure \ref{cdf} the cumulative stellar mass fraction, which is less sensitive to the correlated errors between adjacent bins. The two functions are remarkably similar and clearly show that the bulk of SF is older than 1 Gyr, when roughly 90\% of the mass was already formed. 

As shown by the observational CMD, our photometry reaches the RGB population, but is not deep enough to identify other older features such as the red clump or the horizontal branch, i.e. the helium burning phases of low mass stars ($\lesssim 2$ M$_{\odot}$); this means that the actual lookback time safely sampled by our CMD is formally 1 Gyr (the minimum age of an RGB star) and in practice around 3 Gyr (since the quite low metallicity allows us to reduce the age-metallicity degeneracy affecting RGB stars and thus have an older age constraint). To quantify the uncertainties related to assuming different lookback times in the derivation of the SFH, we have re-run SFERA assuming the starting epoch of the SF activity to be either $\sim 5$ Gyr or $\sim 3$ Gyr ago instead of $\sim 13$ Gyr ago. In both cases the CMD is reproduced quite well, with no relevant differences with respect to the one in Figure \ref{sfh-total}, and the corresponding SFHs are shown in Figure \ref{sfh-resolution}. The shorter the lookback time, the higher the enhancement of the earliest episode in order to recover the right number of observed stars, while the SFR in all the other bins remains the same within the errors. We thus keep modeling the whole Hubble time in the following, but warn the reader that the actual SFR in the earliest $8-10$ Gyr may vary within the extremes shown in Fig. \ref{sfh-resolution}.

\begin{figure}
\centering
\includegraphics[width=\linewidth]{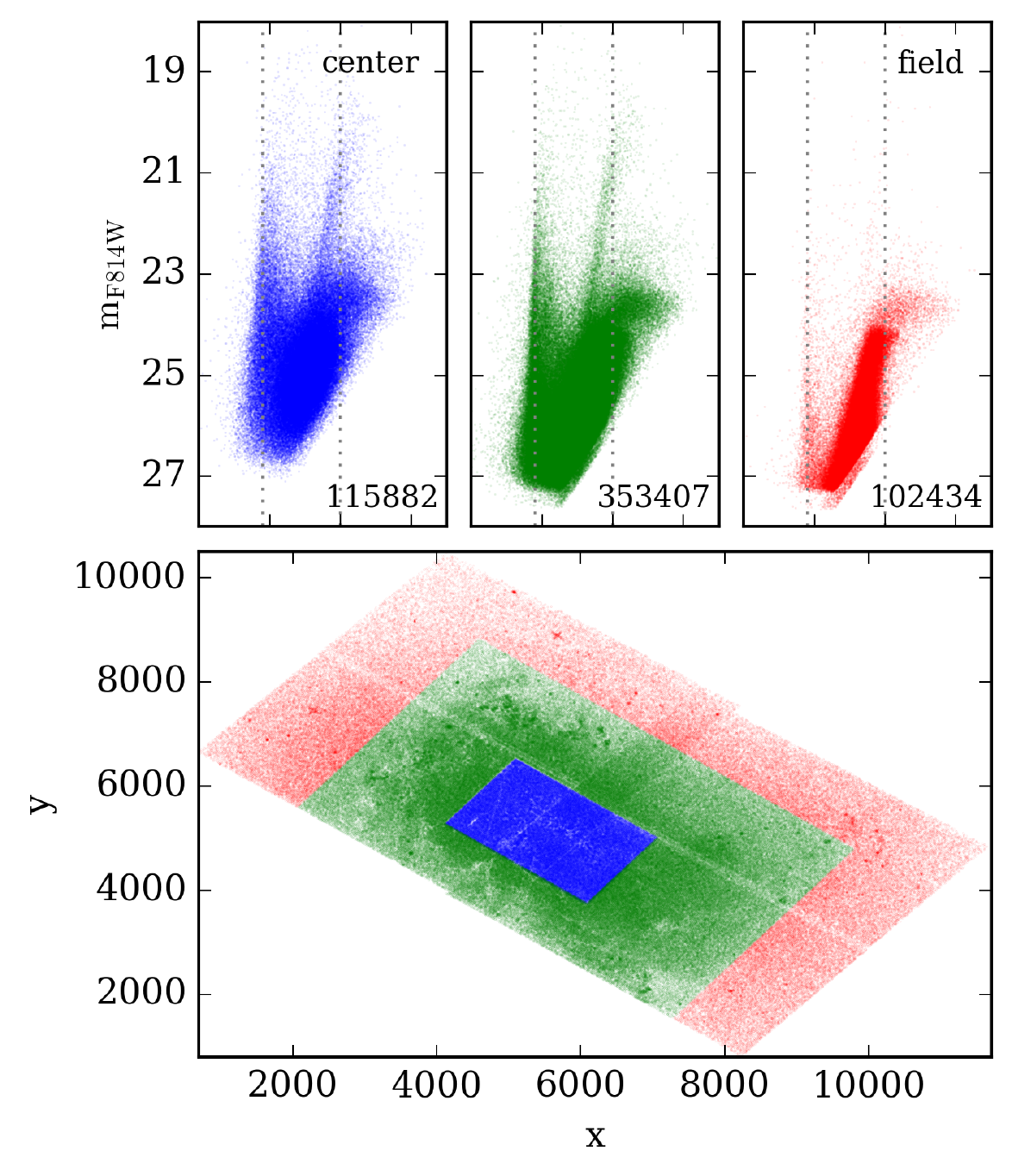}
\caption{Map and corresponding CMDs of the 3 regions we identified in the galaxy and used to recover the SFH; the dotted vertical lines are a rough reference of the blue edge of the MS ($\mathrm{m_{F555W}-m_{F814W}}=-0.2$) and the red edge of the RGB ($\mathrm{m_{F555W}-m_{F814W}}=2$). In the bottom right corner of every CMD the number of stars in the corresponding field is indicated.}
\label{out}
\end{figure}

Despite this observational limitation, it is quite reasonable to expect this galaxy to be older than 3 Gyr, possibly as old as a Hubble time, as already demonstrated by spectroscopic studies \citep{Strader2012,Karczewski2013,Annibali2018}. Thus, we are confident that future observations will allow the photometry to go deeper and detect older population tracers.

\begin{figure}
\centering
\includegraphics[trim=0 0 0 0, clip, width=\columnwidth]{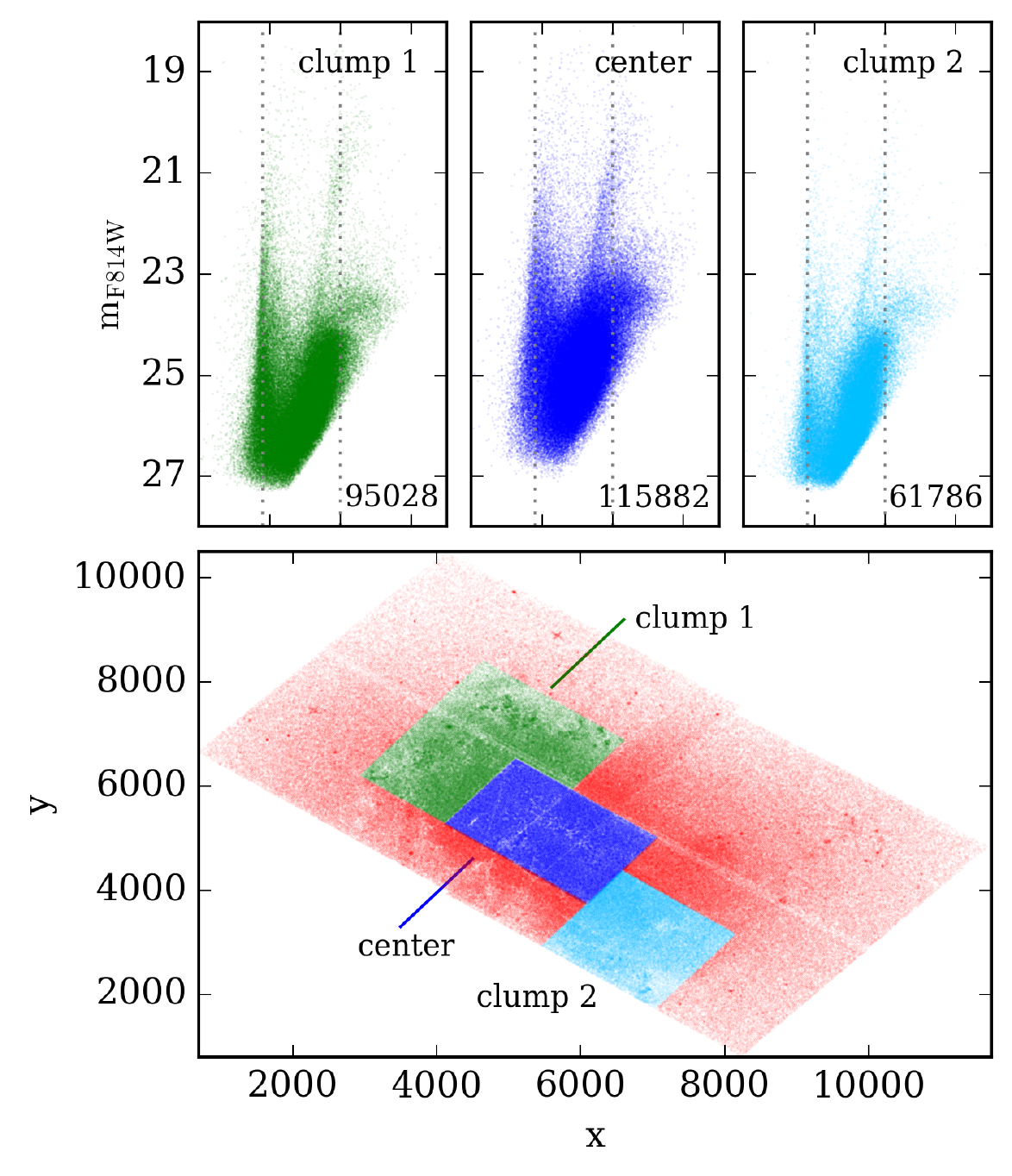}
\caption{Map and corresponding CMDs of the 3 central subregions we identified in the galaxy and used to recover the SFH; the lines and labels follow that of Figure \ref{out}.}
\label{fields}
\end{figure}
\begin{figure}
\centering
\includegraphics[trim=0 -0.5cm 0 -0.5cm, clip, width=\columnwidth]{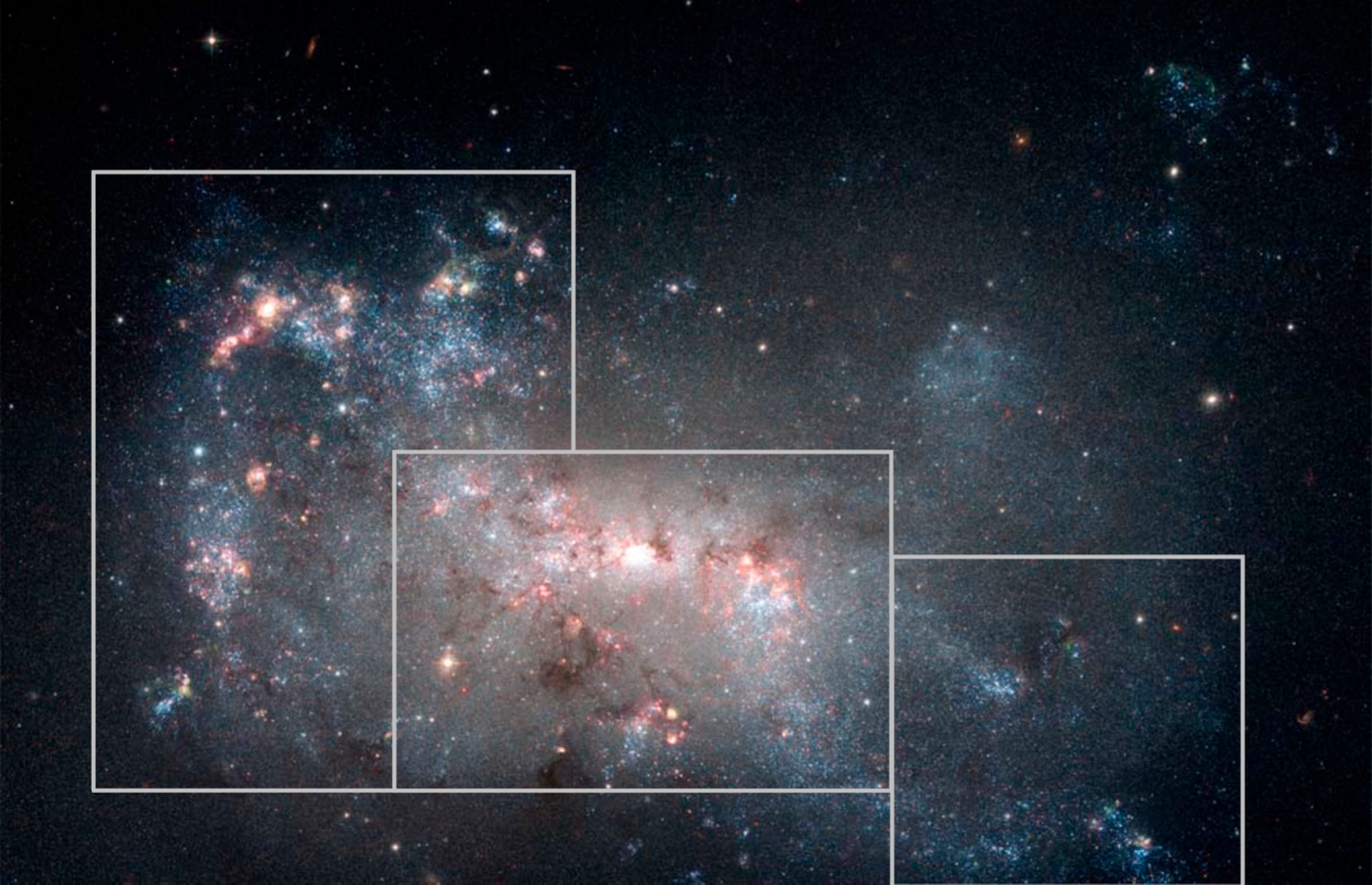}
\caption{Footprints of the four regions identified in Figures \ref{out} and \ref{fields} on the image of Fig. \ref{3col}.}
\label{fields_image}
\end{figure}

Given the richness and complexity of this galaxy, we divided the whole field in 3 separate regions, for which we performed the SFH analysis independently. Figure \ref{out} shows the division and the corresponding CMDs. Given the boxy shape of the galaxy, an elliptical selection of the regions would have been less representative of the SF region distribution, so we adopted this rectangular selection. The outer field was selected to avoid all the infalling features we see around the galaxy (see Figure \ref{3col}), while the central one includes the strongest H$\alpha$ emission. The CMDs of these three regions clearly reflect the changing stellar populations, with the central region containing lots of very bright young stars, almost absent in the outer part, and the different completeness conditions of the subregions (the CMD becomes deeper going from inside out, see also Figure \ref{compl}).

However, the presence of different substructures in the intermediate region made the SFH recovery process very difficult, in particular in the differential reddening treatment. For this reason, we do not show here the corresponding SFH, and we decided to split this field in two different subregions, one including the norther SF region, with the elongated structure discussed above, the other including the southern clumps of stars, 
as shown in Figures \ref{fields} and \ref{fields_image}. 

\subsection{Central regions}
From the point of view of the SF, the three central fields are the most interesting ones, showing all the main stellar populations. Moreover, the Center is also included in the UVIS field, so we can directly compare our results with the UV ones presented by \cite{Cignoni2018}. The UV data confirm the “gasping” scenario. In fact, no strong SF burst was detected in the last $180$ Myr, with a SFR increasing, at most, by a factor of $\sim 2$ over the 100 Myr-averaged SFR. The age resolution improves dramatically for ages less than 50 Myr, but strong variations over intervals of a few Myr are not observed either. This SF modality is typically seen in the last 100 Myr of many high-resolution studies \citep{McQuinn2010,Weisz2008} and is probably connected with the formation of star clusters and associations. Compared with other independent SFR tracers, the UV 100 Myr-averaged-SFR and 10 Myr-averaged-SFR are in very good agreement with FUV and H$\alpha$ rates, respectively.

\begin{figure*}
\centering
\includegraphics[width=0.9\columnwidth]{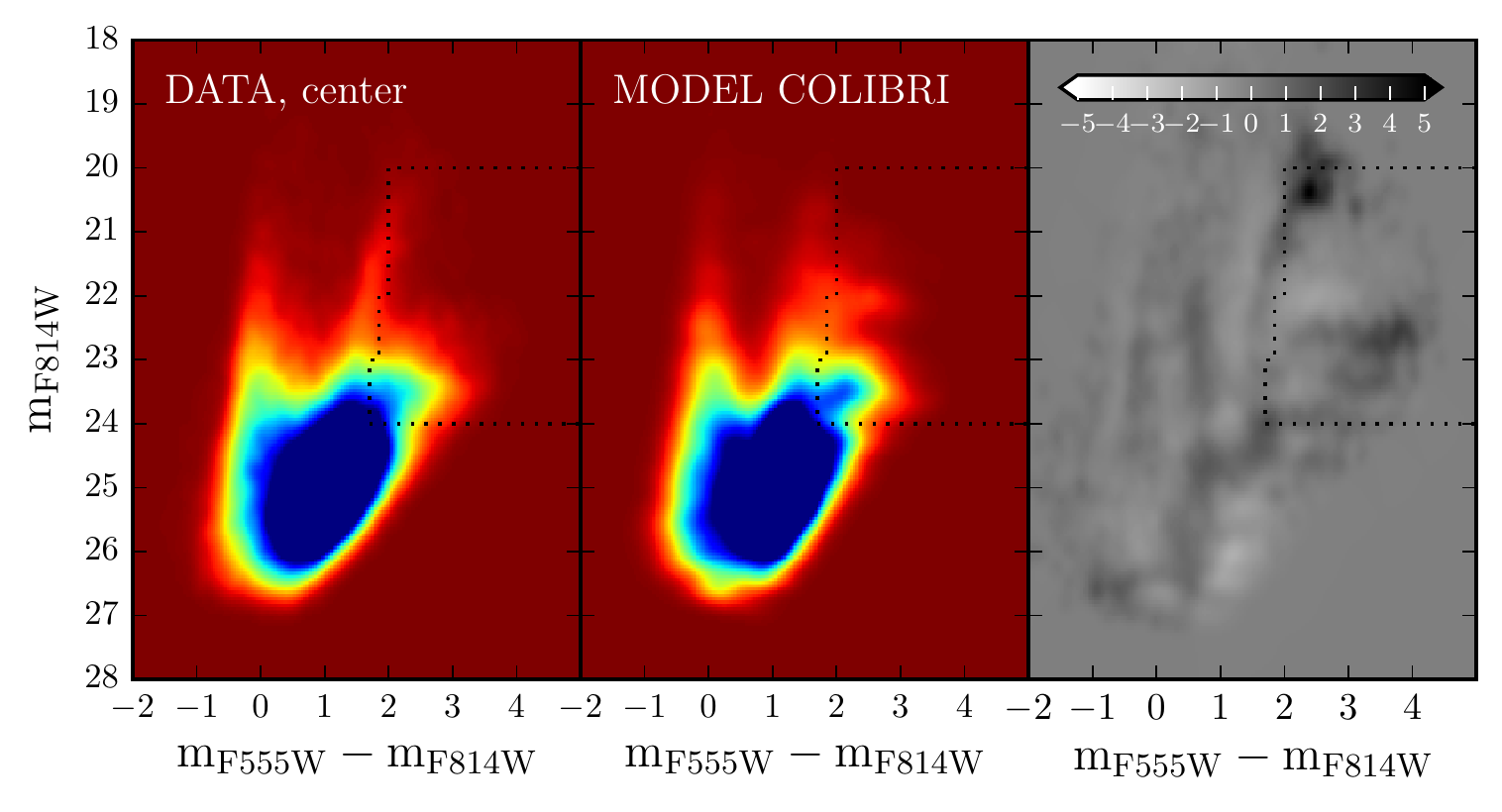}
\includegraphics[width=0.9\columnwidth]{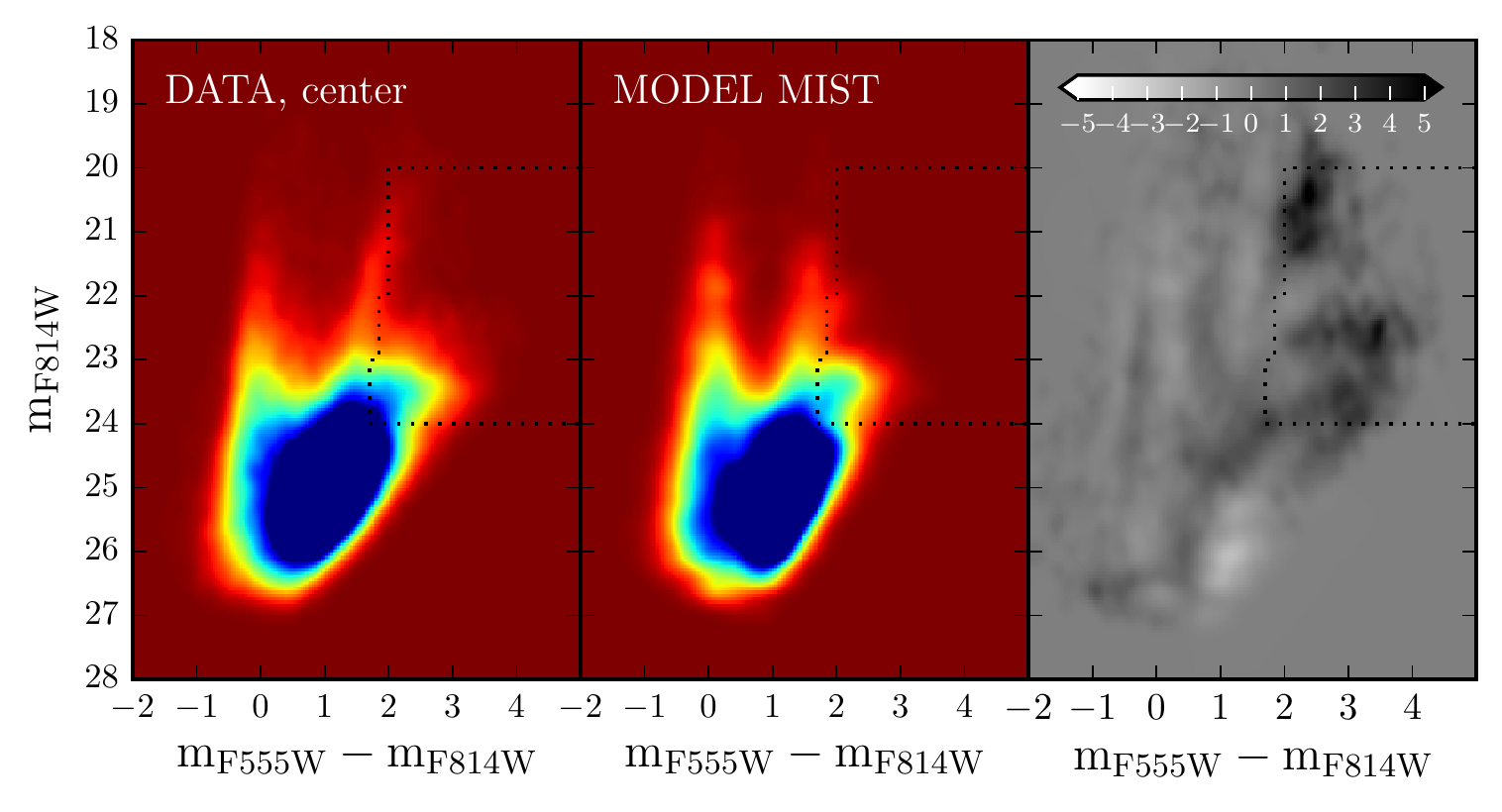}
\centering
\includegraphics[width=0.9\columnwidth]{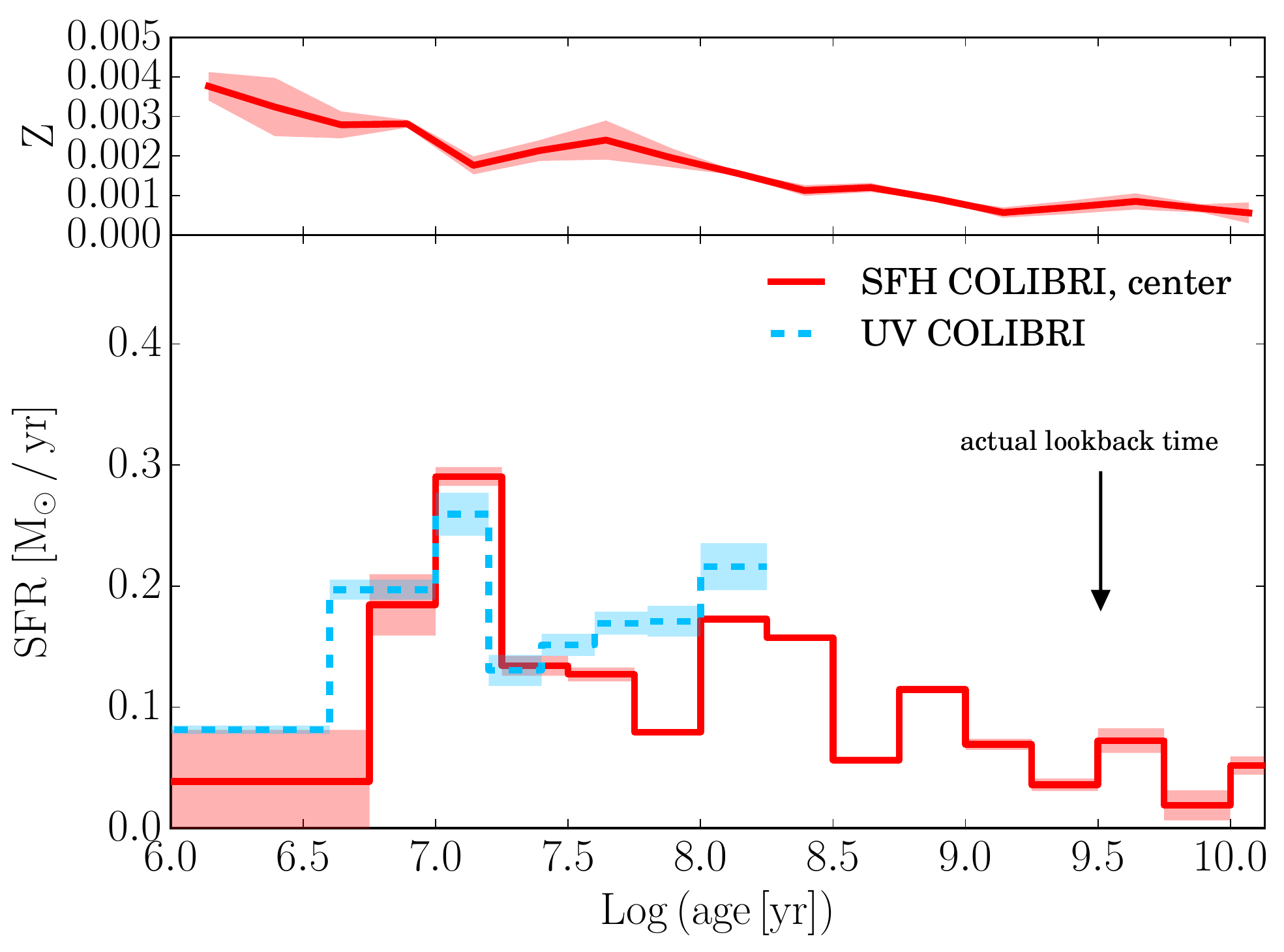}
\includegraphics[width=0.9\columnwidth]{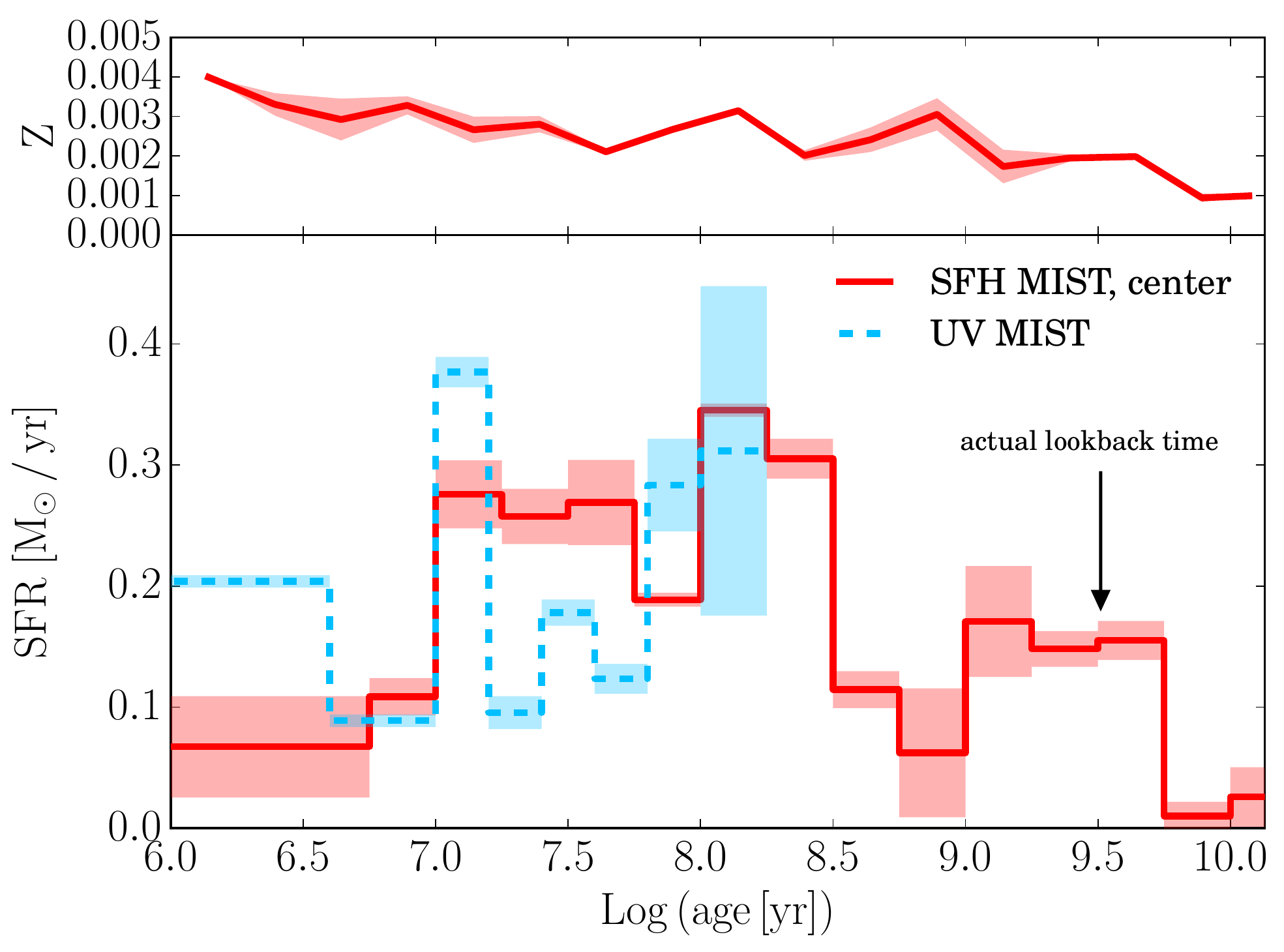}
\caption{Top panels. Hess diagrams of the Center field of NGC~4449: the observational one is on the left, the one reconstructed on the basis of different sets of models in the middle, and the residuals between the two (in units of Poisson uncertainties $(data-model)/\sqrt{model}$) on the right; the dotted line shows the region containing TP-AGB stars and excluded from the minimization. Bottom panels. Recovered star formation history and metallicity; in light blue, the UV SFHs from \cite{Cignoni2018}. Notice that the SFR scale is different from that in Fig. \ref{sfh-total}. The left plots refer to the COLIBRI models, the right ones to the MIST models.}
\label{sfh-center}
\end{figure*}

\begin{figure*}
\centering
\includegraphics[width=0.9\columnwidth]{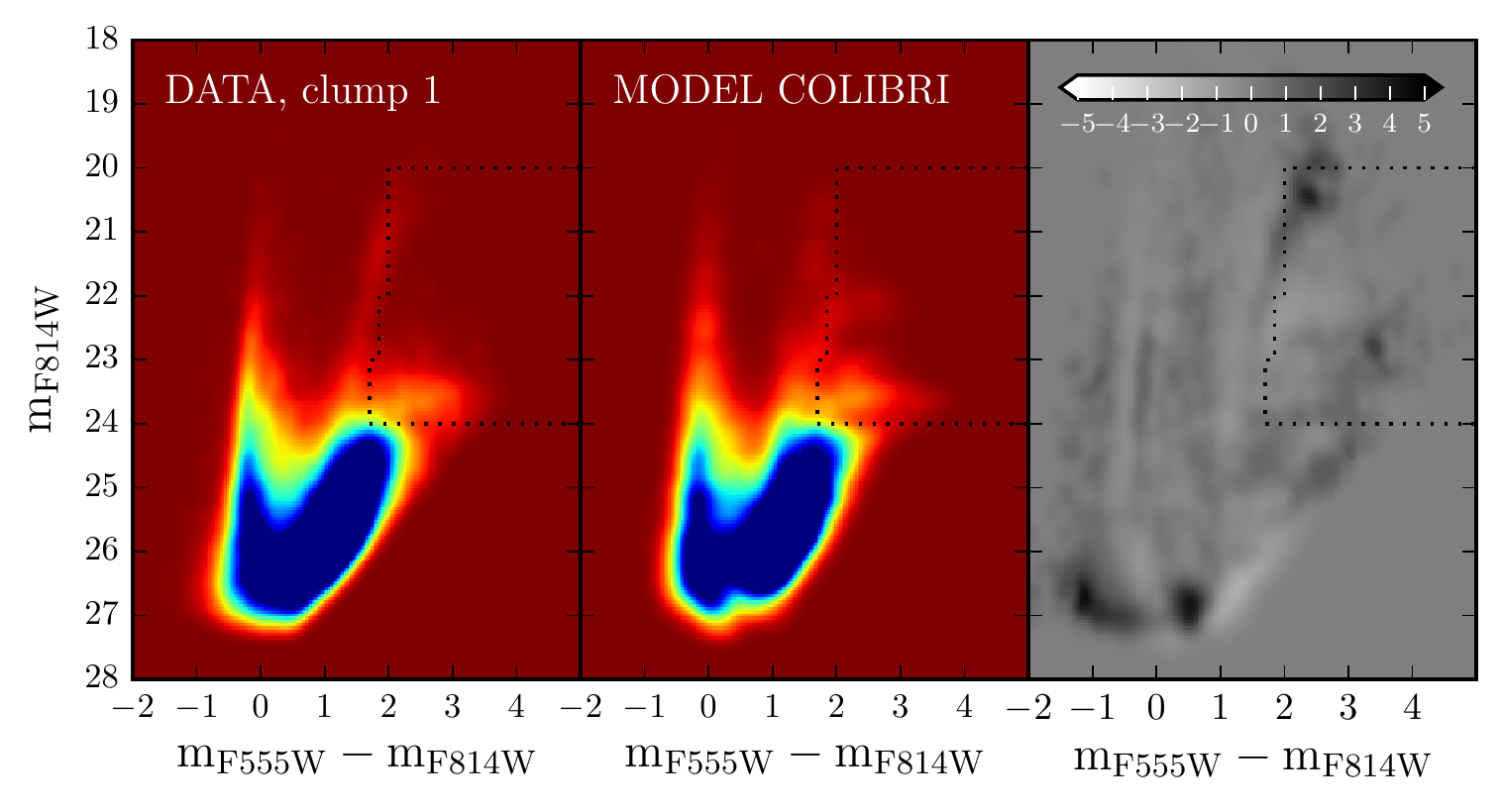}
\includegraphics[width=0.9\columnwidth]{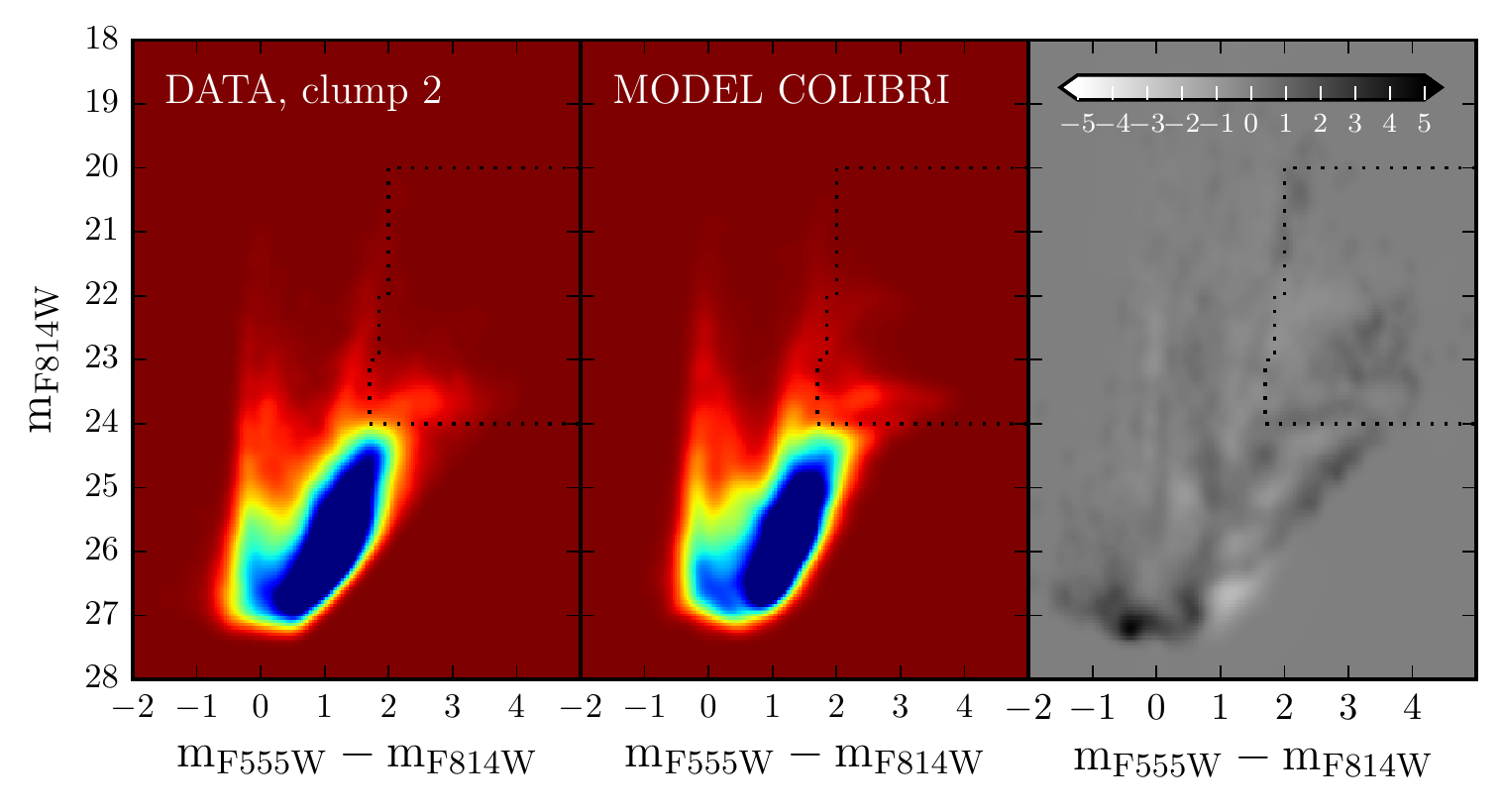}
\centering
\includegraphics[width=0.9\columnwidth]{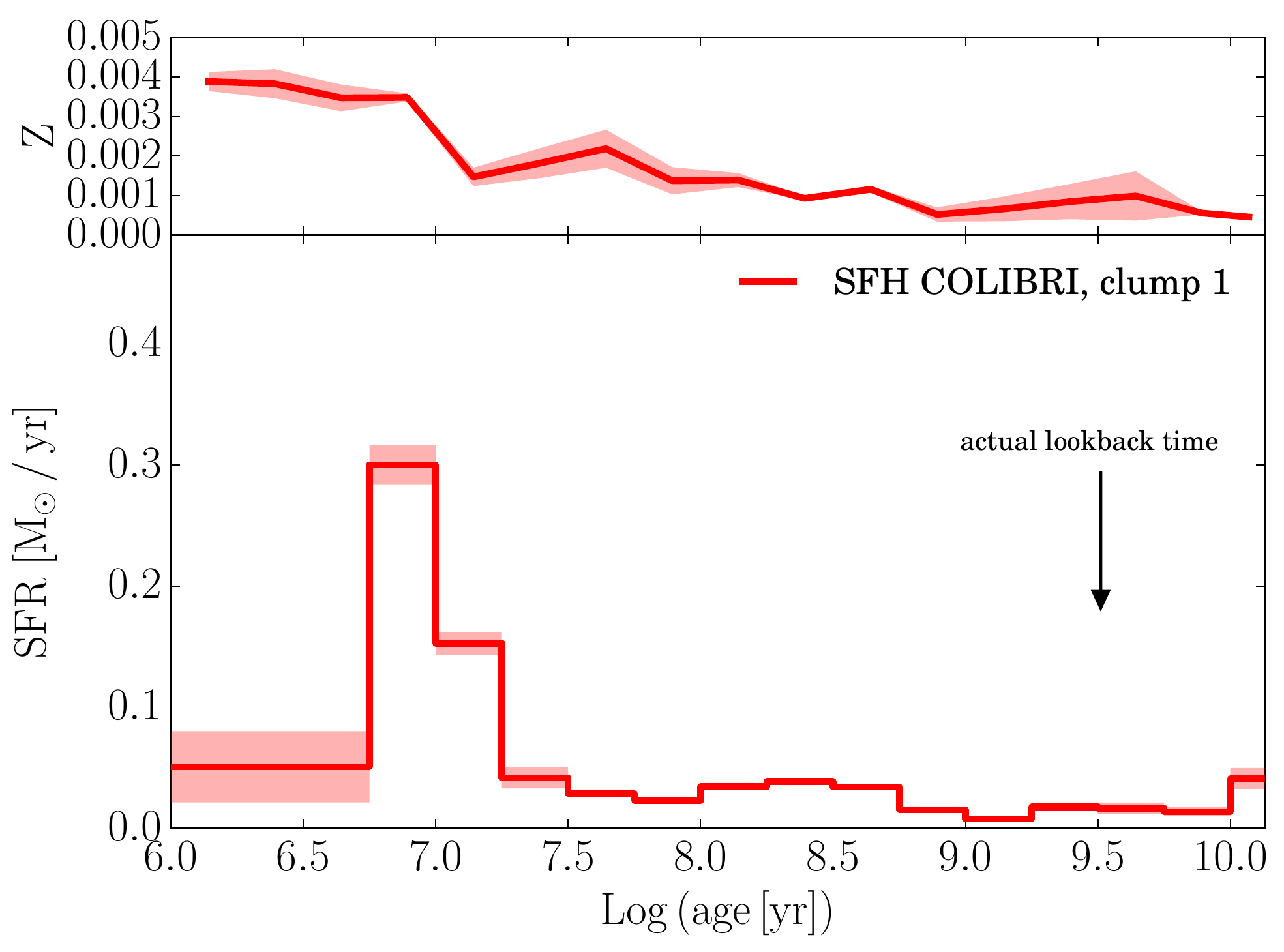}
\includegraphics[width=0.9\columnwidth]{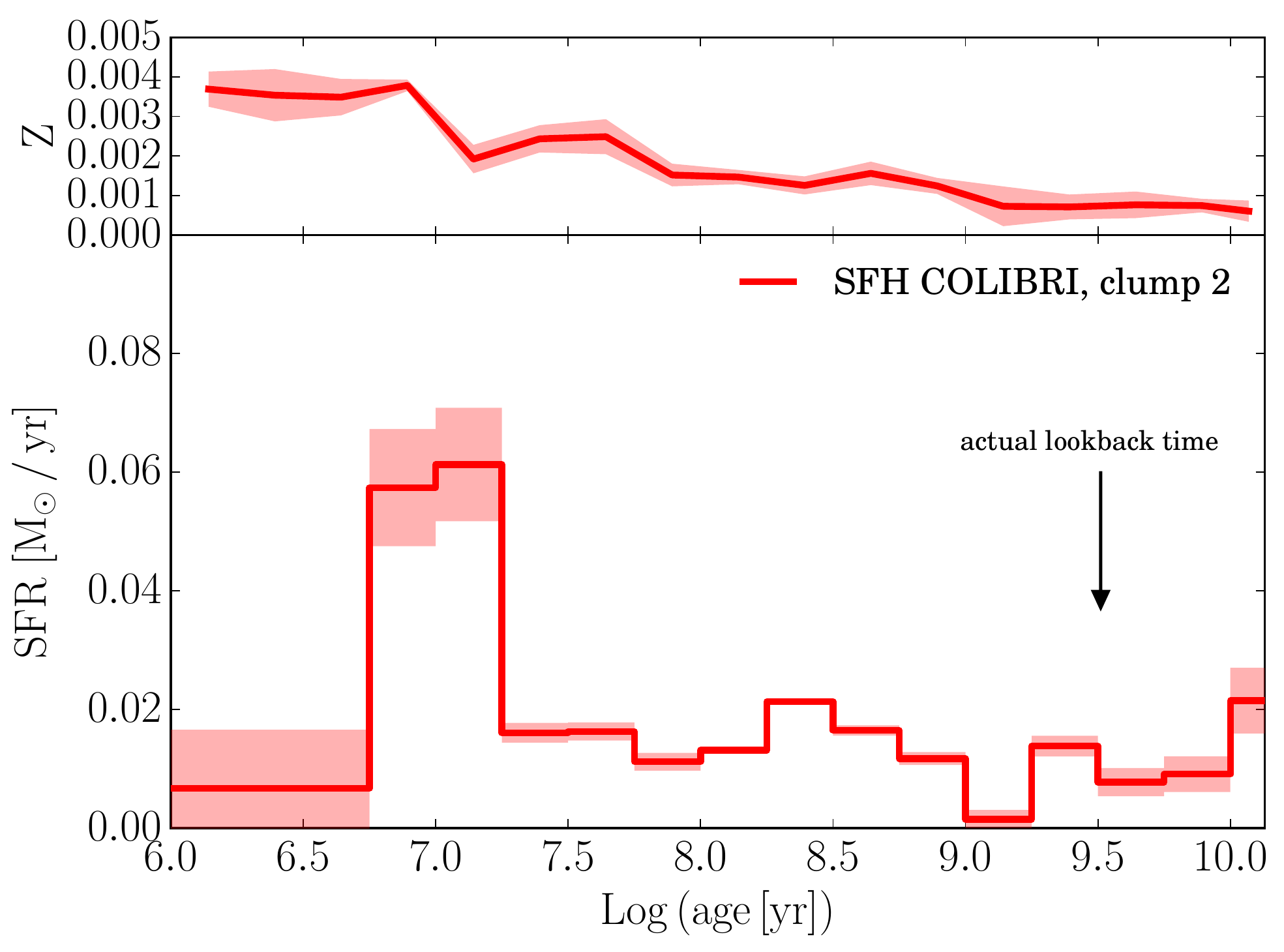}
\caption{Left panels. Hess diagrams, residuals, age-metallicity relation and star formation history of Clump 1. Right panels. Same for Clump 2. Notice that the SFR scales are different from each other and from those of the other figures.}
\label{sfh-clumps}
\end{figure*}

\begin{figure}
\centering
\includegraphics[width=0.9\columnwidth]{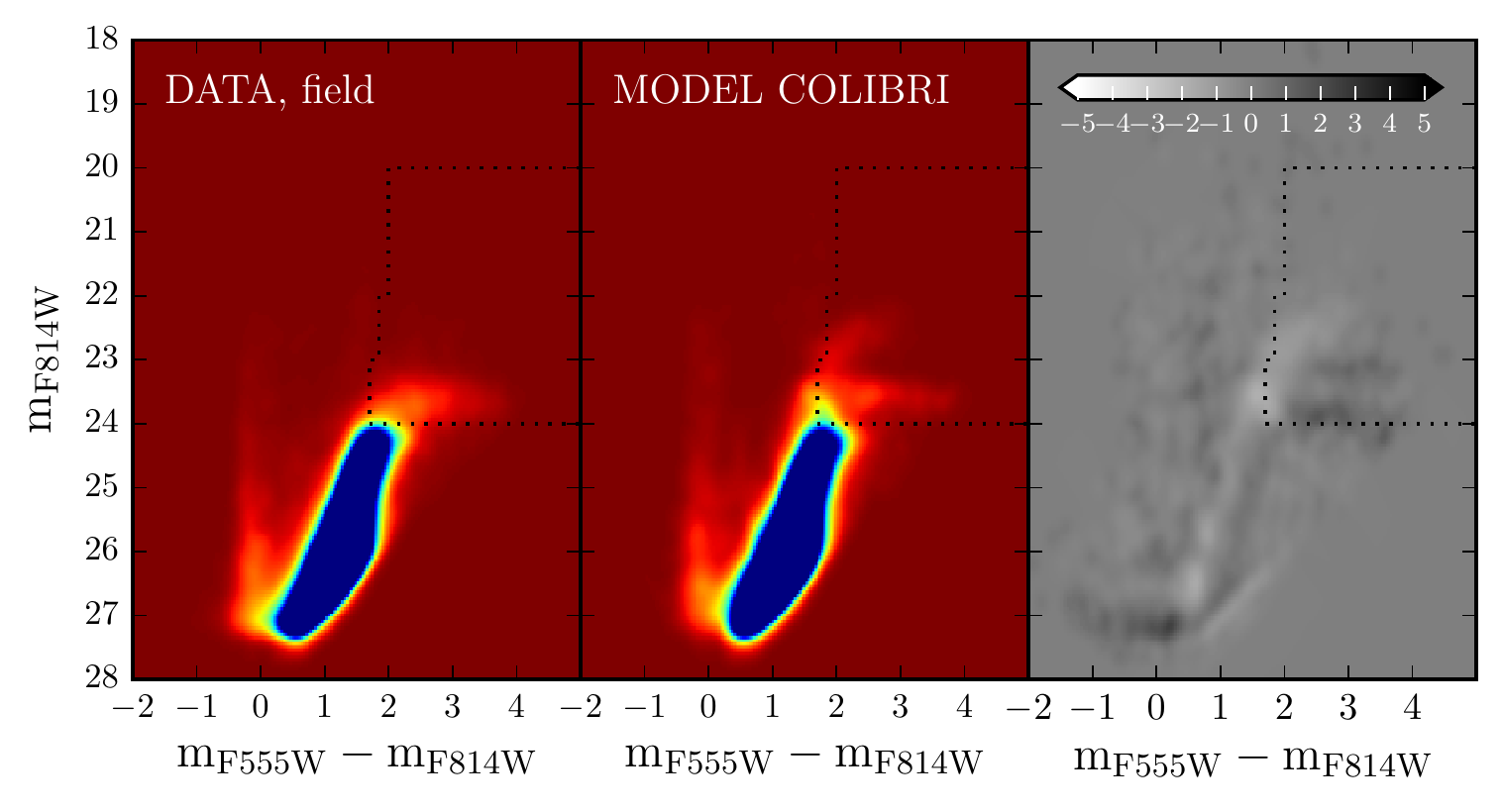}
\includegraphics[width=0.9\columnwidth]{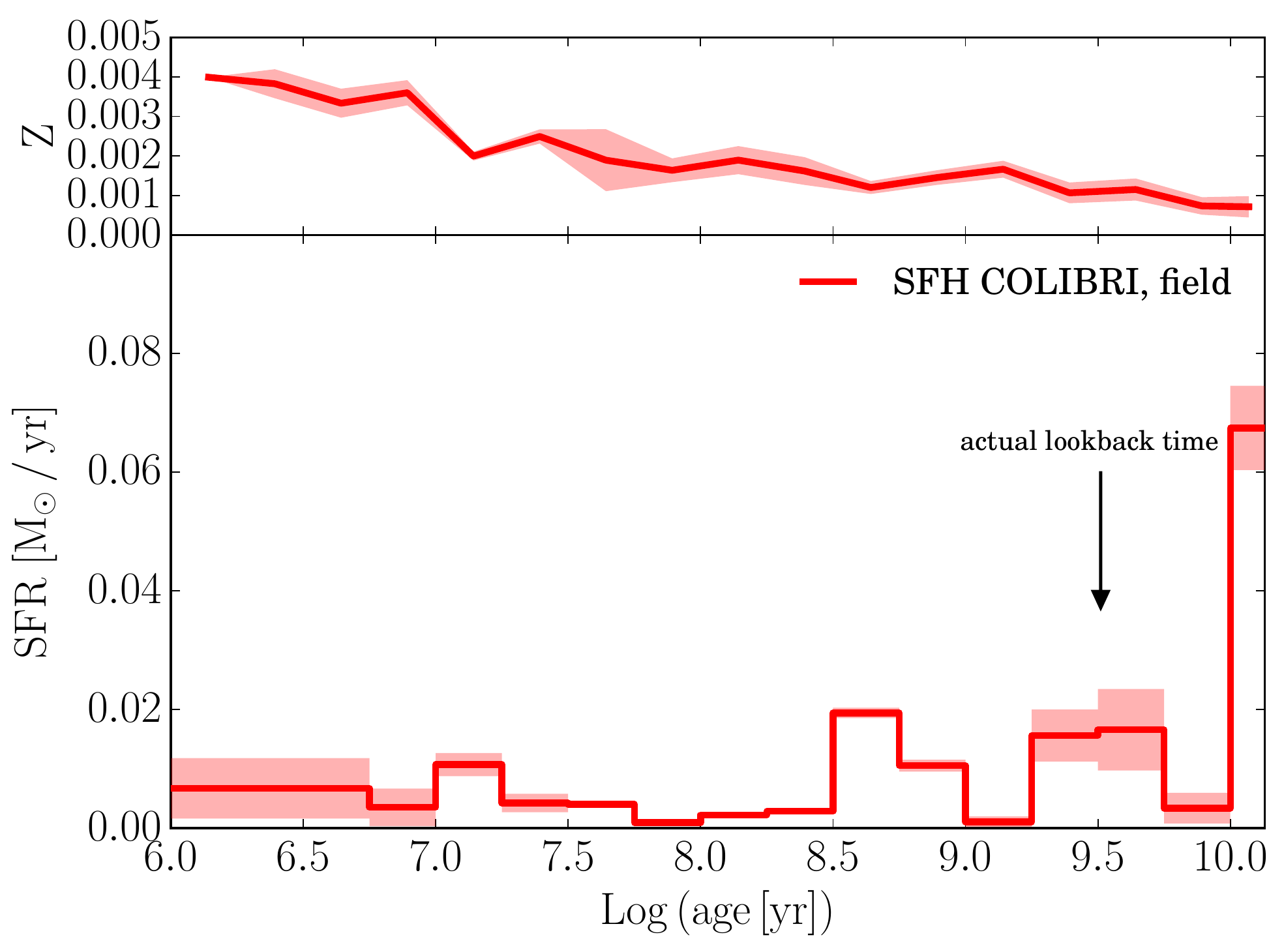}
\caption{Hess diagrams, residuals, age-metallicity relation and star formation history of the outer Field shown in Figure \ref{out}.}
\label{sfh-field-all}
\end{figure}

Figures \ref{sfh-center} and \ref{sfh-clumps} show the results for the center and the two clumps. 
\subparagraph{Center} For the Center, we again show the comparison of the two sets of tracks we used. Both models successfully reproduce the shape of the observed CMD, with the same difficulty in modeling the differential reddening found for the total field. Here the effect is more evident, which is not surprising given that we are analyzing the innermost region of the galaxy, which more likely contains dust and young massive stars. Despite the similar shape of the recovered CMDs, the solutions provided by the two sets of models are quite different both in the optical and in the UV (plotted in light blue in the figures) cases. They both show two peaks at 10 and 100 Myr, but their amplitude and duration are different; in the UV, the MIST solution shows another rise in the most recent time bin, which is not present in the COLIBRI solution. When comparing these SFHs, it is important to consider that is hard to disentangle differences due to systematic errors (e.g. the validity of the same extinction law in the different bands, the different mass fraction of binary stars in different mass ranges, or the statistical weight of different populations of stars). Indeed, UV and optical CMDs are prone to different uncertainties: the optical one is sensitive to the MS and entire He-burning phase (for stars younger than few 100 Myr), whereas in UV CMDs we are only sensitive to the MS and the \textit{blue} He-burning stars. Moreover, in terms of mass, UV CMDs measure MS stars down to $5-6$ M$_{\odot}$, while in the optical we reach $\lesssim 2$ M$_{\odot}$, so we are matching a much broader mass range, where UV just matches massive stars. Also, the time resolution  is different in the last 100 Myr (see also the Appendix for details on the optical time resolution and uncertainties, and \citealt{Cignoni2018}, for the complete UV analysis). Despite all these differences, the trends agree within $2 \sigma$ for the COLIBRI solutions, and within $3 \sigma$ for the MIST ones.

It is worth to notice that in the most recent bin the UV SFRs are higher than the corresponding optical ones. We believe that this behavior is due to the better 
temporal resolution of the UV CMDs, where the characterization of the youngest stellar phases is much more precise.
In future analyses of both optical and UV CMDs of other LEGUS galaxies we will be able to to check whether or not the UV SFR is systematically higher that the optical one at the most recent epochs, as in NGC~4449.

We find that the SFH of the Center is fairly similar to that of the whole galaxy, with the main peaks (one around 14 Myr ago, the other around 100 Myr ago) and dips in SFR occurring at the same epochs. This is not surprising since the Center contains more than $20\%$ of all the stars measured in NGC~4449. Since the two different sets of stellar evolution models consistently provide the same kind of results in the various fields, for sake of simplicity in the following we show only the results based on the COLIBRI models.

\subparagraph{Clump 1} This region includes the small elongated structure and the WR cluster mentioned in Section \ref{sec_pop}. As apparent from Figure \ref{fields}, Clump 1 is the area with the tightest and brightest blue and red plumes, suggesting that its SF activity has been more peaked at certain epochs. The CMD in Figure \ref{sfh-clumps} clearly reveals the presence of a well populated MS, and a small over-density in the red plume due to the coeval stars in the elongated structure. The different shape of the faintest edge of the CMD is due to the fact that we do not fit the area below $\mathrm{m_{F814W}} \sim 26.5$ since it is more than 50\% incomplete. 
The main peak of SF is between 5 and 20 Myr ago, and a very low activity is found in older time bins. 
Interestingly, the metallicity appears to decrease with time between 3 and 1 Gyr ago; since this is one of the most active and interacting regions of the galaxy, this could be due to gas infall or a merging with another lower metallicity body.

\subparagraph{Clump 2} The CMDs in Fig. \ref{sfh-clumps} show a very good agreement in the fitted area. 
In the SFH we can find again a peak around 10 Myr, but with lower relative importance over the activity at earlier epochs than in Clump 1. This is consistent with the almost zero H$\alpha$ emission found in this region.

\begin{figure*}
\centering
\includegraphics[width=0.75\linewidth]{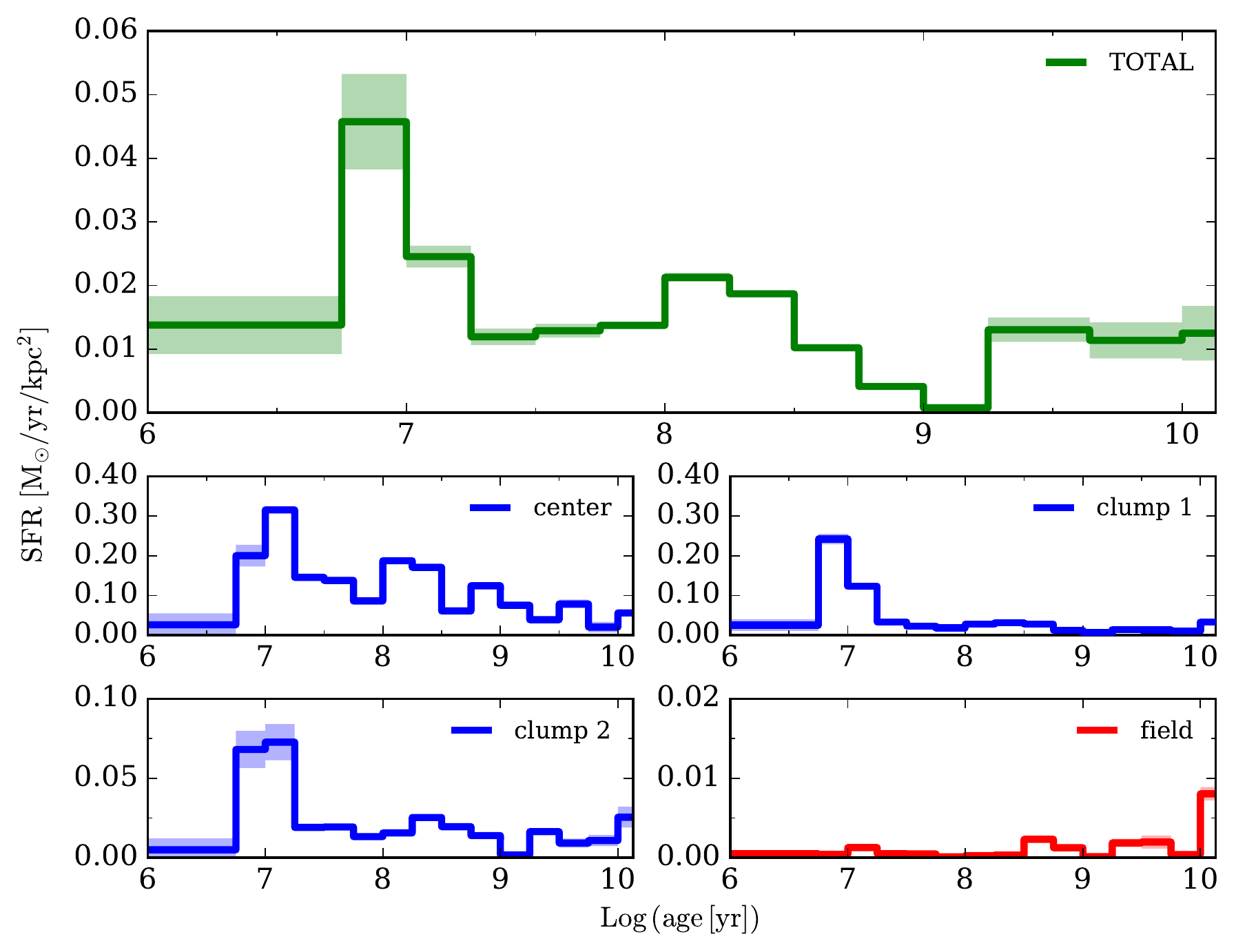}
\caption{SFR surface densities (SFR/area) of the whole galaxy and the separate fields of NGC~4449. The colors indicate the total region (green), and the regions where we find a high (blue) and low (red) SF activity. Notice the different scales of the plots.}
\label{sfh-all}
\end{figure*}

\subsection{External field}
The external region shows a prevalence of old stellar populations, its CMD consisting mainly of TP-AGB and RGB stars. 
This outer field was selected to avoid all the SF regions and the streams infalling in the galaxy, and, as expected, it shows no sign of recent SF, and the SFH is somewhat complementary to that of the three central regions. The peak is in the oldest bins, but a very low activity lasted until $\sim 300$ Myr ago.\\

A summary of the SFRs and stellar masses formed in different epochs for both the whole galaxy and the 4 sub-regions is given in Table \ref{table}.

\section{Results}
In this paper we derived and analyzed the detailed star formation history of the Magellanic irregular galaxy NGC~4449 on the basis of its optical color-magnitude diagram. Here we summarize the main results and compare them with the literature, in order to place them in a broader context of the formation and evolution of this class of galaxy.

NGC~4449 is often considered a starburst galaxy and is indeed one of the most actively star-forming systems of the local Universe \citep{Lee2009}. However, there is no unique definition of a starburst, which may depend on the properties considered in the analysis. One definition is based on a short measured duration of the starburst activity and a current SFR that exceeds the average past value by a factor of at least $2-3$ \citep{McQuinn2010}. In the case of NGC~4449 these values are less than $\sim 20$ Myr and $\sim 4$, respectively. Another related quantity is the integrated H$\alpha$ equivalent width, i.e., the H$\alpha$ flux divided by the continuum flux density under the line. \cite{Lee2009} fix a threshold at 100 \AA, which is not matched by NGC~4449 (which has an H$\alpha$ equivalent width of $\sim 72$~\AA). However, since the H$\alpha$ emission has a very short timescale, it might be tracing only starbursts with an enhanced activity in the last $\sim 5$ Myr, which is indeed not the case of NGC~4449.

Our study shows that NGC~4449 has experienced a significant enhancement in the SF activity about 10 Myr ago, but we do not consider it a real burst, since its SFR is only a factor of $\leq 4$ higher than in the quietest phases.  We are inclined to consider  a SF regime as ``bursting'' when the SFR of the burst is at least 10 times higher than average, its duration is short, and significantly long quiescent (or very quiet) phases are present. None of these conditions are met in NGC~4449.

Our results on the stellar populations and SFH clearly highlight the complexity of this galaxy, revealing a very peculiar boxy shape, a centrally concentrated S-shaped bar and H~\textsc{ii} regions in various clumps, also quite far from the most central region (at the edge of the optical galaxy). The SFH has a peak at $\sim 10$ Myr, lasting roughly until 20 Myr ago, in very good agreement within the uncertainties with the results from \cite{McQuinn2010}, who presented the only other SFH available for this galaxy determined from optical photometric data (which are the same \textit{HST} data we use here). 
In the very central region we compared the SFHs from optical and UV data: the COLIBRI solutions show a very good agreement (within $2\sigma$), while the MIST solution have more differences (they still agree within $3\sigma$). 
The fraction of stellar mass formed in the central region in the most recent 100 Myr is $\sim 89\%$, in very good agreement with the $84\pm4\%$ found by \cite{McQuinn2012}.


Figure \ref{sfh-all} shows the results for the SFHs of the whole galaxy and the sub-regions of NGC~4449, divided by the area of each region (chosen as the area including 90\% of the stars in each field) in order to obtain a SFR surface density and to properly compare the SFH even among fields of very different spatial dimensions. In fact, even though smaller, the central region has the highest SFR density, as expected from the great amount of ionized emission. Both the clumps have also a significant recent activity, and Clump~1 shows signs of ongoing SF. These three regions together provide almost the totality of the SFR($\leq 10$ Myr) of the galaxy. The external field is less active, showing a prevalence of older SF (see Table \ref{table} for the details). 
These results suggest that the star formation proceeded from outside in, so the SFR in the center is increasing with time and the SFR in the outer regions is decreasing with time. The middle regions, Clumps 1 and 2 are in between, with the exception of the 10 Myr peaks. Outside-in SF seems to be common in dwarf galaxies \citep{Gallart2008,Zhang2012,Meschin2014,Pan2015} whereas in spirals this process seems to proceed inside-out \citep{Williams2009}.\\

\begin{table*}
\caption{Summary of the star formation rates and stellar masses in the different fields.}
\begin{center}
\begin{tabular}{cccccc}
    \toprule
    \midrule
    \addlinespace[0.3em]
    \multirow{2}{*}{region} & $\mathrm{\langle SFR \rangle}$ & $\mathrm{SFR_{\, peak}}$ &  \multirow{2}{*}{$\mathrm{age_{\, peak}}$} & $\mathrm{M_{\ast}(age \leq 10\ Myr)}$ & $\mathrm{M_{\ast}(age > 1\ Gyr)}$ \\
    \addlinespace[0.3em]
    & $\mathrm{[M_{\odot}/yr/kpc^2]}$ & $\mathrm{[M_{\odot}/yr/kpc^2]}$ & & $\mathrm{[10^6\ M_{\odot}]}$ & $\mathrm{[10^9\ M_{\odot}]}$ \\
    \midrule
    Total   & $0.015 \pm 0.002$ & $0.046 \pm 0.008$ & $\,\ 7.8$ Myr & $5.06 \pm 0.72$ & $2.04 \pm 0.36$ \\
    \addlinespace[0.3em]
    Center  & $0.105 \pm 0.013$ & $0.315 \pm 0.008$ & $13.9$ Myr & $1.51 \pm 0.25$ & $0.87 \pm 0.11$ \\
    Clump 1 & $0.045 \pm 0.007$ & $0.242 \pm 0.013$ & $\,\ 7.8$ Myr & $2.39 \pm 0.16$ & $0.44 \pm 0.06$ \\
    Clump 2 & $0.021 \pm 0.005$ & $0.073 \pm 0.011$ & $13.9$ Myr & $0.44 \pm 0.08$ & $0.25 \pm 0.04$ \\
    \addlinespace[0.3em]
    Field & $0.005 \pm 0.001$ & $0.028 \pm 0.003$ & $11.7$ Gyr & $0.06 \pm 0.03$ & $0.50 \pm 0.05$ \\
    \midrule
    \bottomrule
\end{tabular}
\end{center}
\label{table}
\end{table*}

\section{Discussion and conclusions}
The results presented above suggest the impact of possible recent phenomena to be stronger in the Northern regions, on a East-West axis (crossing the Center and Clump~1). Interestingly, this trend resembles what was discussed by \cite{Annibali2011} in their analysis of the cluster population of this galaxy (previously shown in Fig. \ref{pop_distr}): they find that some old stellar clusters in NGC~4449, instead of following a uniform distribution across the galaxy as one would expect, seem to follow some linear structures possibly linked to a past accretion event (see their Figure 16). One of these structures crosses the regions where we find the highest activity and is roughly perpendicular to the structure in Clump~1. All these hints could again be evidence of a common event triggering significant activity in these regions, a scenario also suggested by \cite{Valdez2002} on the basis of their kinematic and dynamical study of the high perturbed velocity field of the ionized gas in NGC~4449.
Very young clusters (age $<10$ Myr) are instead found only in the very central regions (six in Center, two in Clump 1 and one in Clump 2) and, as expected, tightly follow the distribution of the H$\alpha$ emission \citep{Annibali2011,Gelatt2001}; here \cite{Reines2008} also detected 13 embedded massive star clusters with thermal radio emission.

The elongated structure shown in Figure \ref{pop_distr} and included in Clump~1 seems to be a coeval population probably formed as a result of the interaction, possibly a merging, with a smaller galaxy tidally disrupted by the main body of NGC~4449. Both the luminosity of these stars and the SFH we recover (see the results for Clump~1 in Fig. \ref{sfh-clumps}), suggest an age of a few tens of Myr, while the embedded massive WR cluster, also contained here, has a minimum age of $\sim 3$ Myr \citep{Sokal2015}. Indeed, \cite{Theis2001} performed several N-body simulations to reproduce the H~\textsc{i} morphology of NGC~4449, finding that the observed features could be created by an encounter with a smaller dwarf galaxy. Moreover, as discussed by \cite{Lelli2014a,Lelli2014c}, galaxies with young ($\lesssim 100$ Myr) bursts of star formation usually show a very asymmetric H~\textsc{i} morphology, which is also related to past and ongoing interaction/accretion events.

In many works it is outlined how starburst dwarf galaxies usually show very different morphological, dynamical and environmental characteristics. Among them, \cite{Lelli2014c} explore the H~\textsc{i} morphology of several active dwarf irregulars (NGC~4449 included), trying to understand the main process triggering the starburst activity. They find that this enhanced activity usually correlates well with a disturbed H~\textsc{i} morphology, suggesting that the starburst is likely triggered by external mechanisms (merging/gas inflow) rather than by internal ones (stellar feedback). The discovery of a stellar tidal stream from a disrupted dwarf galaxy in the halo of NGC~4449 \citep{Rich2012,Martinez-Delgado2012} and of a possible remnant of a gas-rich accreted satellite \citep{Annibali2012} also points at this scenario, and the dynamical timescale to see the tidal features of the encounters ($< 10^8$ yr from \citealt{Penarrubia2009}) is compatible with the beginning of the higher SF activity we find in the galaxy. The H~\textsc{i} tails are possibly led also by the companion galaxy DDO~125, an irregular galaxy at a projected distance of 41 kpc from NGC~4449.

On the other hand, \cite{ElBadry2016} investigated through cosmological hydrodynamic simulations the effect that stellar feedback has on the stellar component of isolated dwarf galaxies; they find that gas outflows and inflows can severely affect the stellar kinematics and radial gradients in low-mass galaxies ($\mathrm{M_{\ast} \sim 10^{7-9.6}}$). The two effects might indeed be in action at the same time: stellar feedback on small scales (less than 10 kpc) where we actually see the stellar component of the galaxy, and external mechanisms on scales where the H~\textsc{i} component becomes dominant (several 10 kpc).

As mentioned in the Introduction, LEGUS is also studying in detail the star clusters in NGC~4449 (Adamo et al., in preparation, Cook et al., in preparation), with a multi-band approach that should provide soon a clearer scenario not only on their formation and evolution, but also on that of the whole galaxy.

Even though our photometry can constrain populations as old as a few Gyr only, there are several spectroscopic studies revealing features from older stars. \cite{Strader2012} found globular clusters consistent with ages of $7-10$ Gyr. \cite{Karczewski2013} performed a fit to the multiwavelength spectral energy distribution of NGC~4449, from the far-ultraviolet to the submillimetre, and found models consistent with a first onset of star formation around 12 Gyr ago. They assumed a simplified SFH consisting of only three episodes, old (between 12 Gyr and 400 Myr ago) with a SFR of 0.09 M$_{\odot}$ yr$^{-1}$, intermediate (between 400 and 10 Myr ago) with a SFR of 0.14 M$_{\odot}$ yr$^{-1}$, young (last 10 Myr) with a SFR of 0.28 M$_{\odot}$ yr$^{-1}$. If we average our results to compare them with the listed ones, we find slightly higher values (old: 0.18 M$_{\odot}$ yr$^{-1}$, intermediate 0.26 M$_{\odot}$ yr$^{-1}$, young 0.51 M$_{\odot}$ yr$^{-1}$) but consistent within the uncertainties. Finally, \cite{Annibali2018} studied 14 globular clusters in the near IR, finding ages typically older than $\sim 9$~Gyr.

To statistically improve the comparison with other galaxies, we consider the work by \cite{Weisz2011}, who provide the SFHs of 60 dwarf galaxies within the ANGST program. They find a huge diversity in the evolution of the SF among different morphological types, and also within the dwarf irregular sub-sample (see their Figures 3 and 4). However, all the galaxies in the sample seem to have formed the bulk of their stars earlier than $\sim 1$~Gyr ago, in agreement with what we find for NGC~4449.

\acknowledgments{
These data are associated with the \textit{HST} GO Programs 10585 (PI A. Aloisi) and 13364 (PI D. Calzetti). Support for this programs was provided by NASA through grants from the Space Telescope Science Institute. E.S. is supported by INAF through a Ph.D. grant at the University of Bologna. D.A.G. kindly acknowledges financial support by the German Research Foundation (DFG) through program GO 1659/3-2. The authors profoundly thank Kristen McQuinn for the data provided and the useful discussions about this work.\\
}

\appendix
\section{SFERA}
\subsection{Description of the code and initial parameters}
In this Appendix we recall the main steps of the code SFERA, already implemented in \citet{Cignoni2015} and here upgraded with new stellar models, providing new benchmarks which are appropriate for distant galaxies like NGC~4449.

\begin{figure}
\centering
\includegraphics[width=\columnwidth]{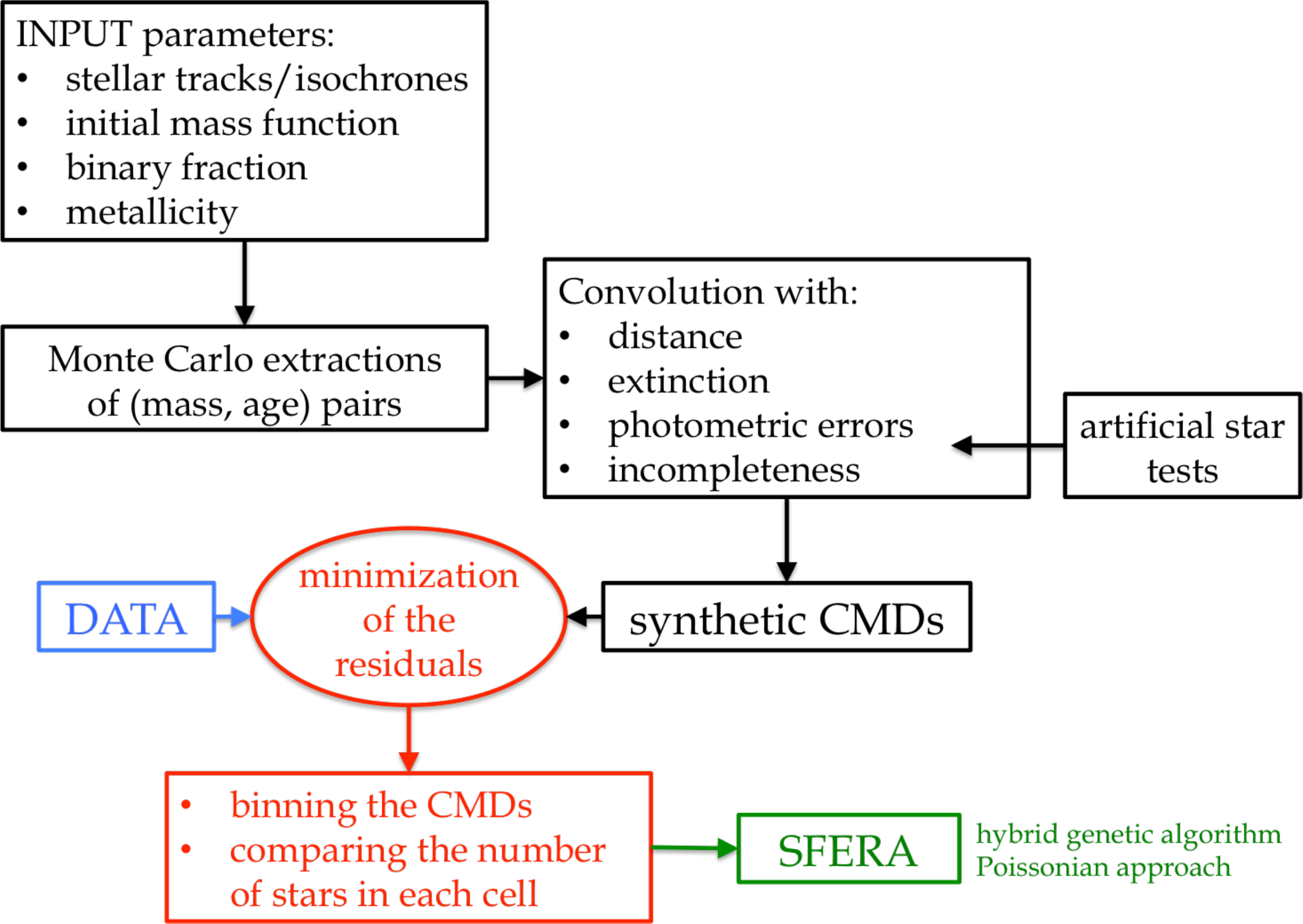}
\caption{Schematic representation of the steps implemented in SFERA.}
\label{sfera}
\end{figure}

SFERA performs a complete implementation of the synthetic CMD method, deriving the synthetic stars from the adopted stellar evolution models, estimating the observational effects through artificial star tests, and finally performing the minimization of the residuals between observed and synthetic CMDs, with an accurate estimate of the uncertainties of these procedures. Figure \ref{sfera} shows a schematic flow-chart of the algorithm that we describe in this appendix.

As already mentioned in Section \ref{sec_sfh}, we start from two sets of stellar evolution libraries, in particular we use isochrones already calibrated in the photometric filters from the observations. In general, we do not put constraints on the metallicity, but in the case of distant galaxies, like NGC~4449, we require that the metallicity in the most recent time bin varies in a small range around the value corresponding to the oxygen abundance derived from spectroscopy of the H~\textsc{ii} regions of the galaxy. At any other epoch, the metallicity cannot increase more than 25\% with respect to the adjacent younger bin. The isochrones are populated through a Monte Carlo extraction of stars, with a constant SFR and a Kroupa IMF ($\xi(m) \propto m^{-\alpha}$ with $\alpha=1.3$ for $0.1 \leq m < 0.5$ M$_{\odot}$ and $\alpha=2.3$ for $0.5 \leq m \leq 300$ M$_{\odot}$). Each star will be alive (thus, visible on the synthetic CMD) or dead (thus, contributing only to the astrated total mass) according to its lifetime from the stellar evolution models. In order to avoid statistical uncertainties from the models, we require at least $5 \times 10^6$ stars in each metallicity bin. 30\% of the stars are randomly chosen to have a companion, whose mass is assigned from the same IMF of the primary star.

These models are then convolved with the characteristics of the data: distance modulus in a range around the chosen literature value, varying in steps of 0.05 mag, photometric errors and incompleteness as resulting from the $m_{output}-m_{input}$ distribution of the artificial star tests (see Section \ref{sec_ast}). The reddening distribution is modeled as a Gaussian (with negative values excluded) with mean value and dispersion (representing total -foreground plus internal- and differential reddening) as free parameters \citep{Harris1997}.

To choose the time steps to adopt in the SFH, given the quality of the data, we decided to use logarithmic bins in the age range $\log(t)=6.0-10.13$ (corresponding to $t=1$ Myr $- 13.7$ Gyr) with a logarithmic step of 0.25, except for the first bin which is $\log(t)=6.0-6.75$. This is to follow the decreasing time resolution with increasing lookback time, which is also explored by changing the starting point and dimension of the time bins and checking for possible variations in the SFH.

\begin{figure}
\centering
\includegraphics[width=\columnwidth]{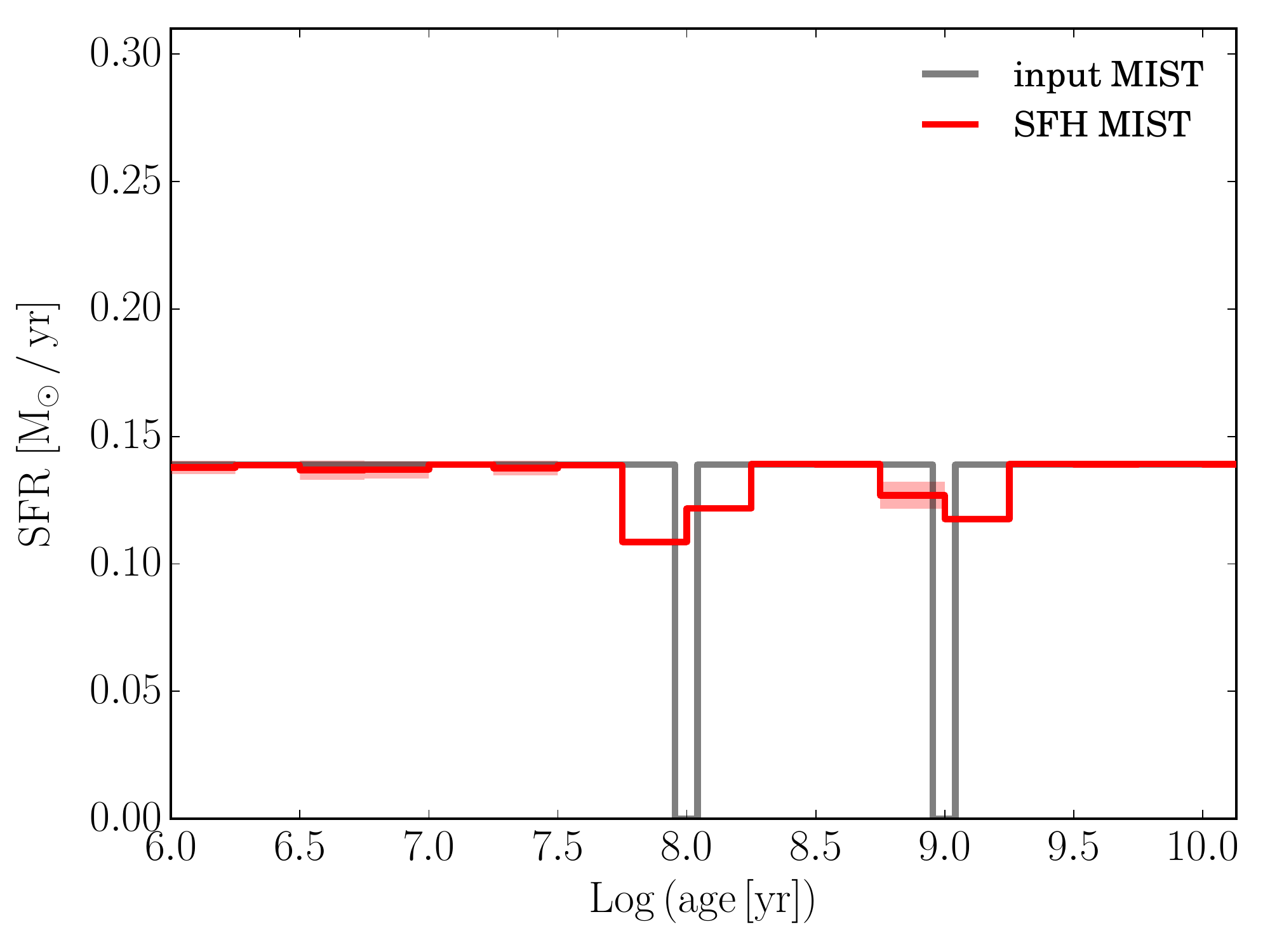}
\caption{Test of the time resolution in recovering quiescent phases in the SFH with SFERA. The input SFH was a constant rate with two gaps at 100 Myr and 1 Gyr lasting $\pm 10\%$ of the age (20 Myr and 200 Myr, respectively). The models used are the MIST ones both for creating the input CMD and for recovering the SFH.}
\label{test-mesa10}
\end{figure}
\begin{figure}
\centering
\includegraphics[width=\columnwidth]{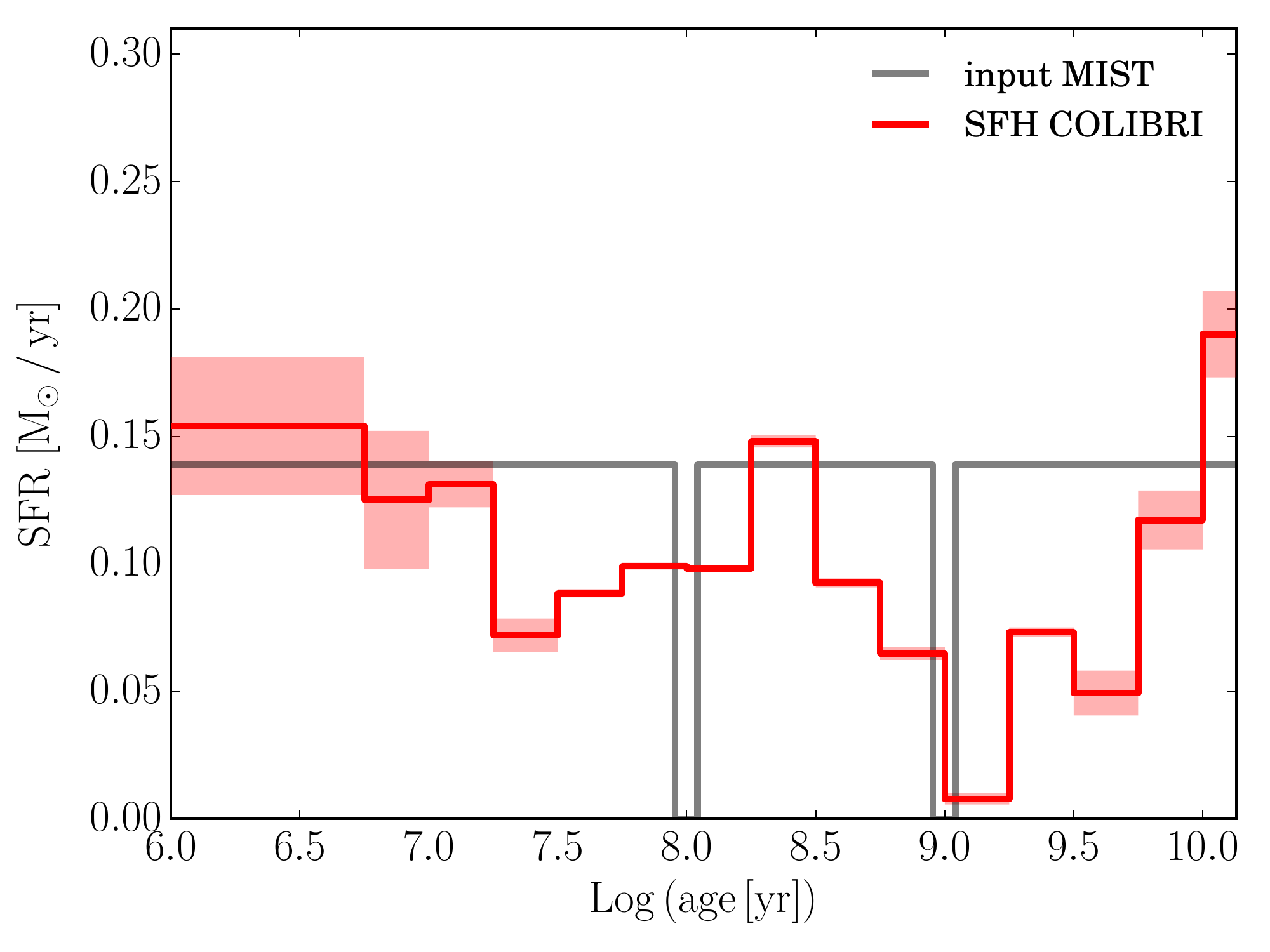}
\caption{Test of the time resolution in recovering quiescent phases in the SFH with SFERA. The input SFH was a constant rate with two gaps at 100 Myr and 1 Gyr lasting $\pm 10\%$ of the age (20 Myr and 200 Myr, respectively). The models used are the MIST ones for creating the input CMD and the PARSEC-COLIBRI ones for recovering the SFH, in order to test for systematic uncertainties due to the different stellar models.}
\label{test-parsec10}
\end{figure}
\begin{figure}
\centering
\includegraphics[width=\columnwidth]{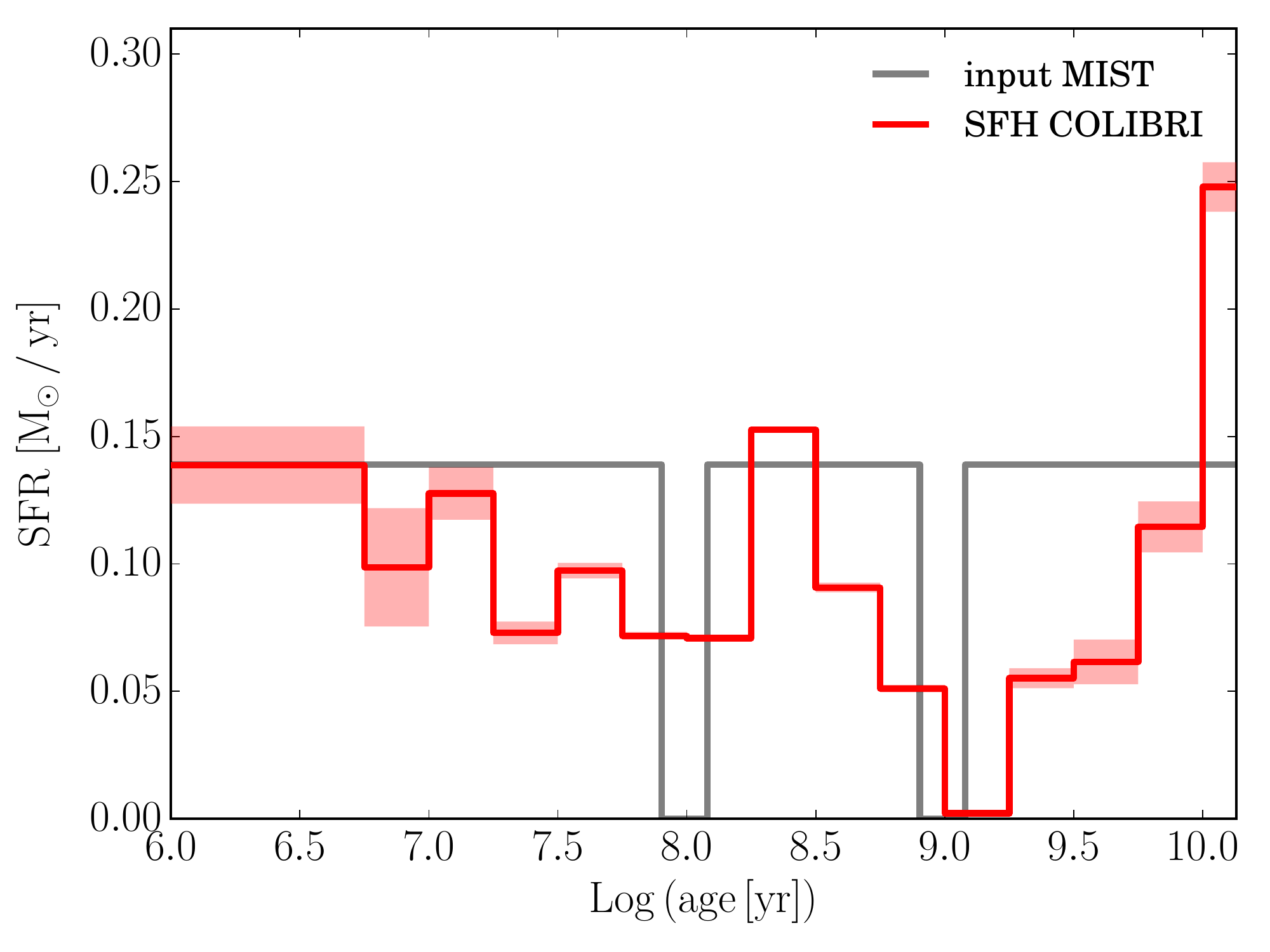}
\caption{Same as Figure \ref{test-parsec10} but with a longer duration of the quiescent phases ($\pm 20\%$ of the age).}
\label{test-parsec20}
\end{figure}
\begin{figure}
\centering
\includegraphics[width=\columnwidth]{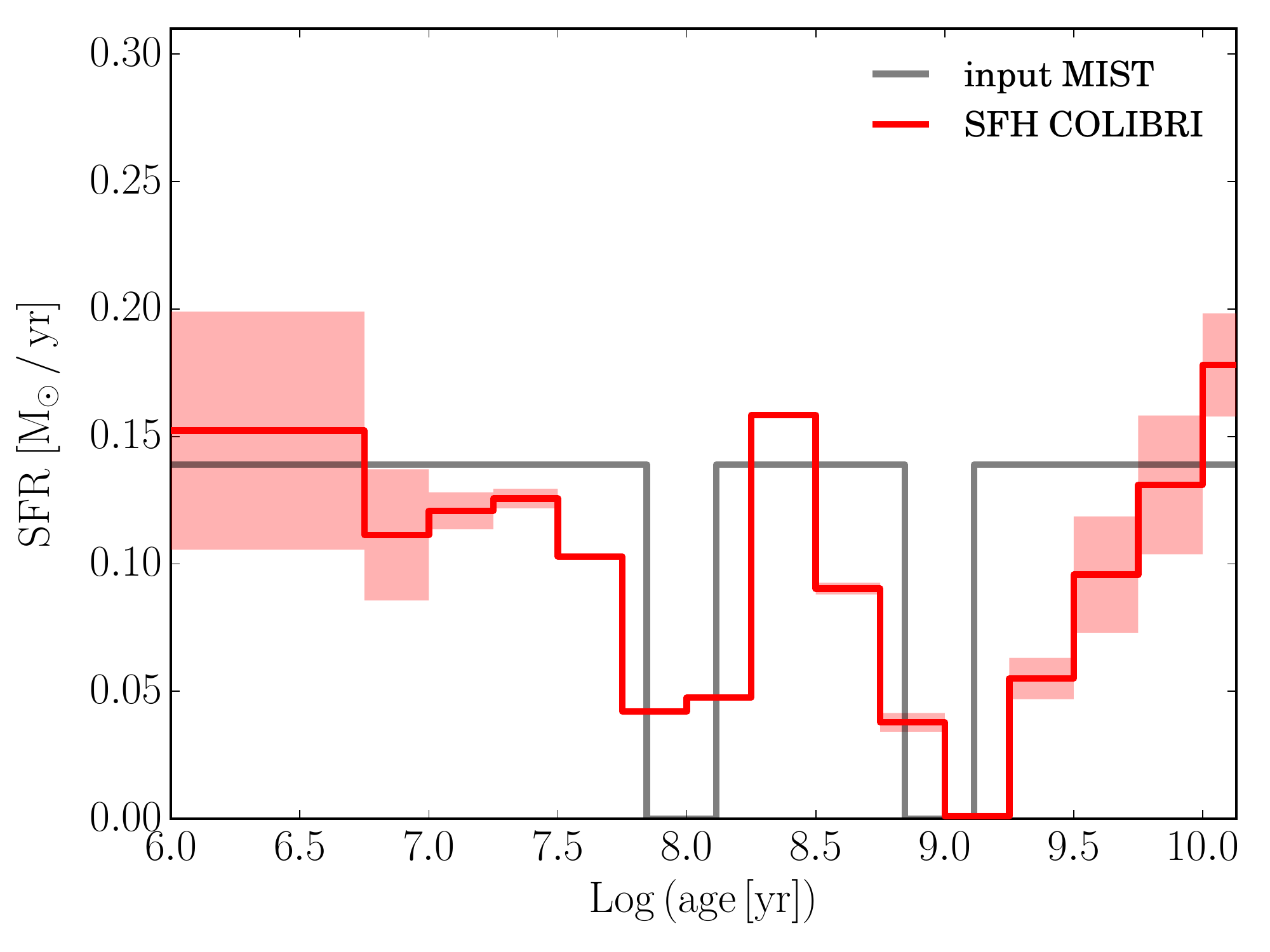}
\caption{Same as Figures \ref{test-parsec10} and \ref{test-parsec20} but with a longer duration of the quiescent phases ($\pm 30\%$ of the age).}
\label{test-parsec30}
\end{figure}

We need to bin both the synthetic and observed CMDs to perform the comparison of the star counts in each grid cell. To do so, the standard approach is to bin different regions trying to sample as much as possible the observed evolutionary features. In this process, there must be a combination of good statistics in each cell, to minimize the Poisson noise, and good sampling of the fine structure of the CMD, to maximize the time resolution of the results. One of the best procedures is to use an \textit{ad hoc} grid whose size varies depending on the density of stars on the CMD and on the reliability of the considered phase. This approach is more ``human dependent'' than a uniform grid, but it can represent a good balance of the different aspects that need to be taken into account. In principle, the impact of this choice on the SFH may be relevant, so we explored different combinations for a safer result. In the end we chose a quite large cell size (0.25 both in color and magnitude) for the brightest stars ($\mathrm{m_{F814W}} < 20$) to balance the low statistics caused by the low number of bright stars; we chose an intermediate cell size (0.1 both in color and magnitude) for the blue and red plumes and for the lower MS; for the RGB, we implemented a variable random binning from 10 to 900 cells in the box $0.75 < \mathrm{m_{F555W}} - \mathrm{m_{F814W}} < 4$ and $24 < \mathrm{m_{F814W}} < 26.5$, which we changed at every bootstrap (see later) in order to minimize the bin dependence of the results. Small random shifts both in color and magnitude are also applied to the whole grid.

The minimization is implemented taking into account the low number counts in some CMD cells, so we follow a Poissonian statistics, looking for the combination of synthetic CMDs that minimizes a likelihood distance between model and data, which corresponds to the most likely SFH for these data. The Poisson-based likelihood function we use is:
\begin{eqnarray}  
\chi_P= \sum_{i=1}^{N
    bin} obs_{i}\ln\frac{obs_{i}}{mod_{i}}-obs_{i}+mod_{i}
\label{pois1} 
\end{eqnarray}
where $mod_{i}$ and $obs_{i}$ are the model and the data histograms in the $i-$bin \citep{Cash1979,Dolphin2002}. This likelihood is minimized with the hybrid-genetic algorithm, a combination of PIKAIA (the free code available at \url{http://www.hao.ucar.edu/modeling/pikaia/pikaia.php} and already implemented in IAC-pop, see \citealt{Aparicio2009}), and a simulated annealing procedure (as implemented in, e.g., \citealt{Cole2007}).
 
 \begin{figure}
\centering
\includegraphics[width=\columnwidth]{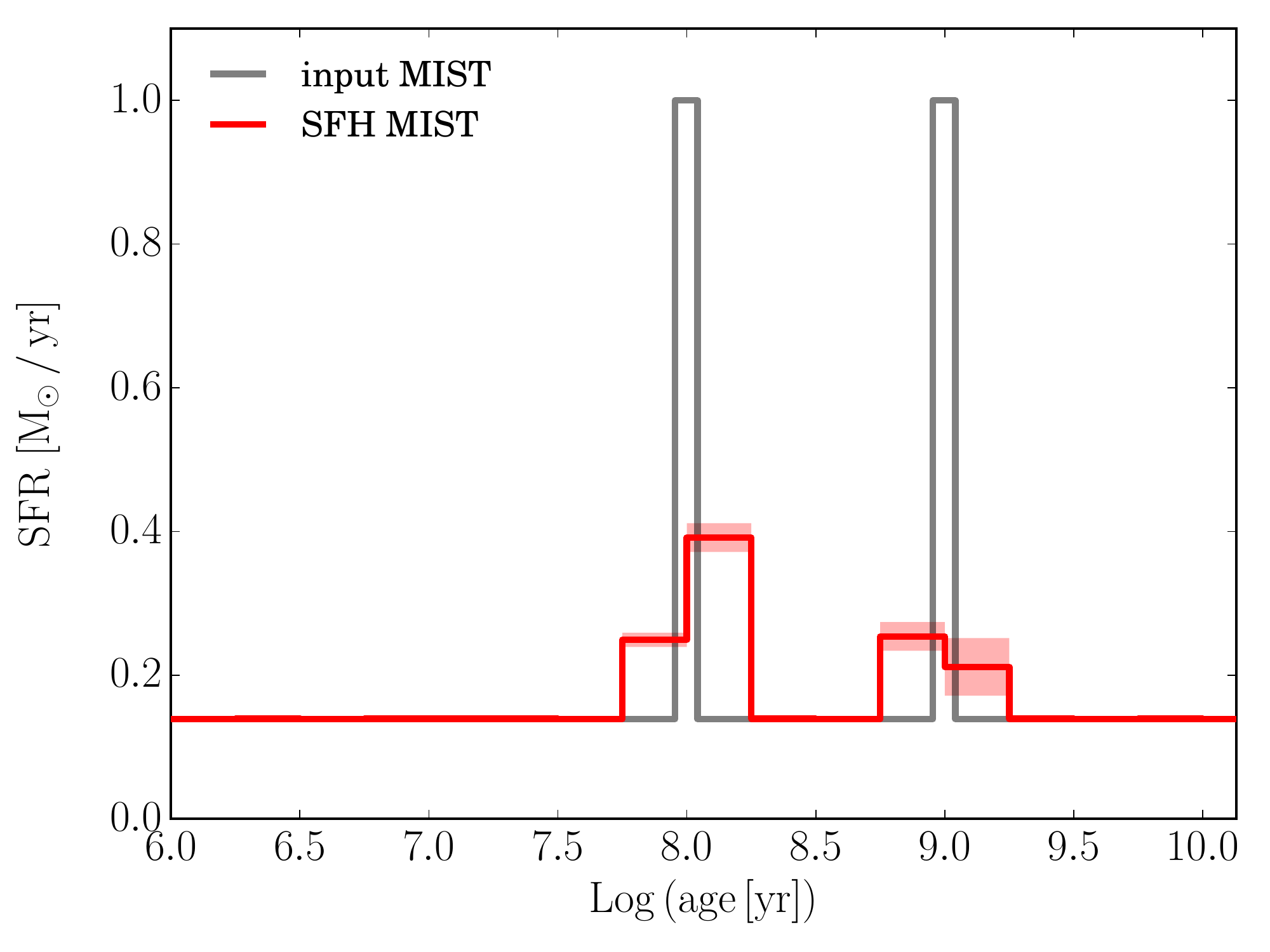}
\caption{Test of the time resolution in recovering bursts in the SFH with SFERA. The input SFH was a constant rate with two bursts ($\sim 10$ times higher than the average) at 100 Myr and 1 Gyr lasting $\pm 10\%$ of the age (20 Myr and 200 Myr, respectively). The models used are the MIST ones both for creating the input CMD and for recovering the SFH.}
\label{burst-mesa10}
\end{figure}
\begin{figure}
\centering
\includegraphics[width=\columnwidth]{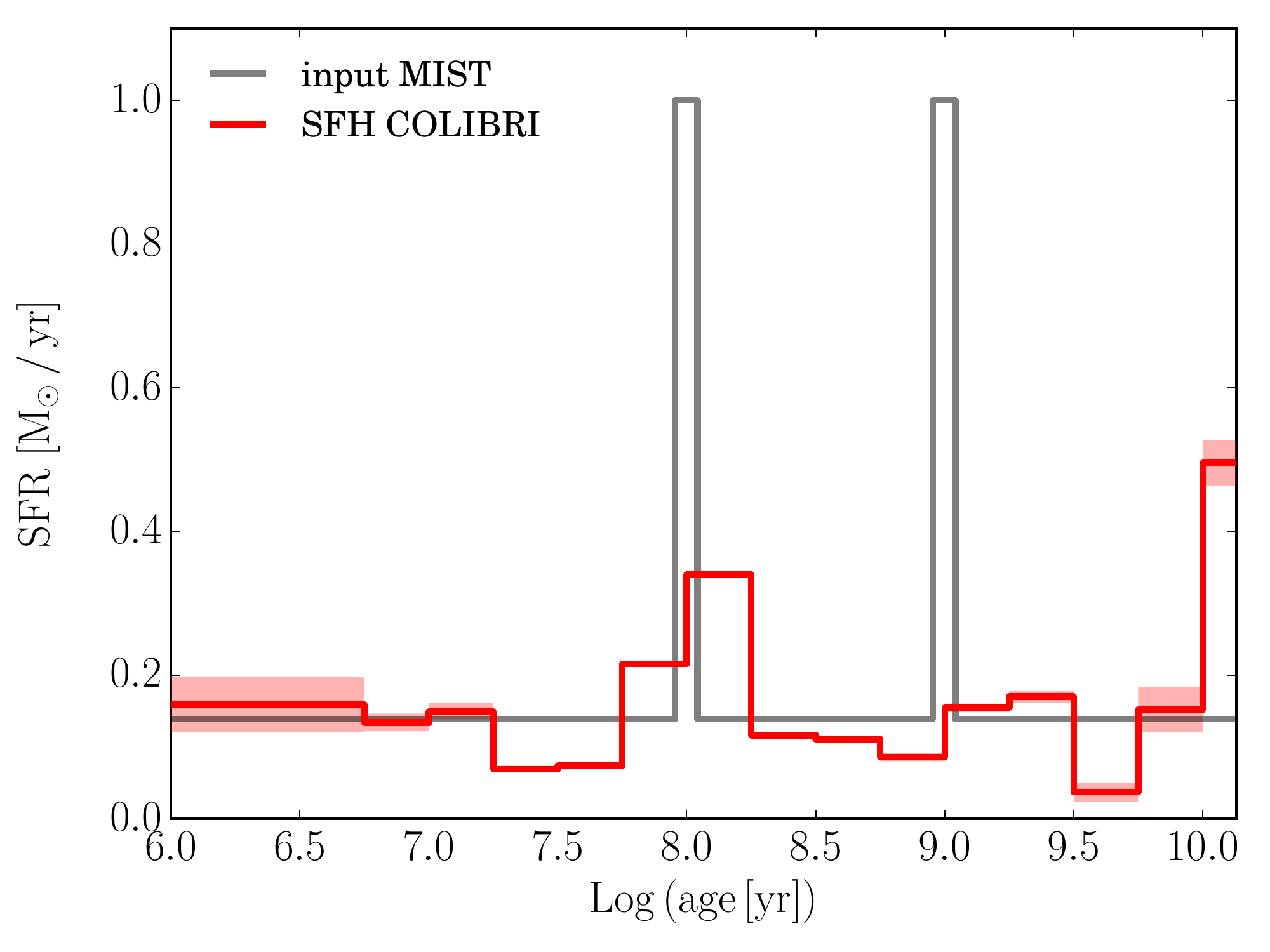}
\caption{Test of the time resolution in recovering bursts in the SFH with SFERA. The input SFH was a constant rate with two bursts ($\sim 10$ times higher than the average) at 100 Myr and 1 Gyr lasting $\pm 10\%$ of the age (20 Myr and 200 Myr, respectively). The models used are the MIST ones for creating the input CMD and the PARSEC-COLIBRI ones for recovering the SFH, in order to test for systematic uncertainties due to the different stellar models.}
\label{burst-parsec10}
\end{figure}
\begin{figure}
\centering
\includegraphics[width=\columnwidth]{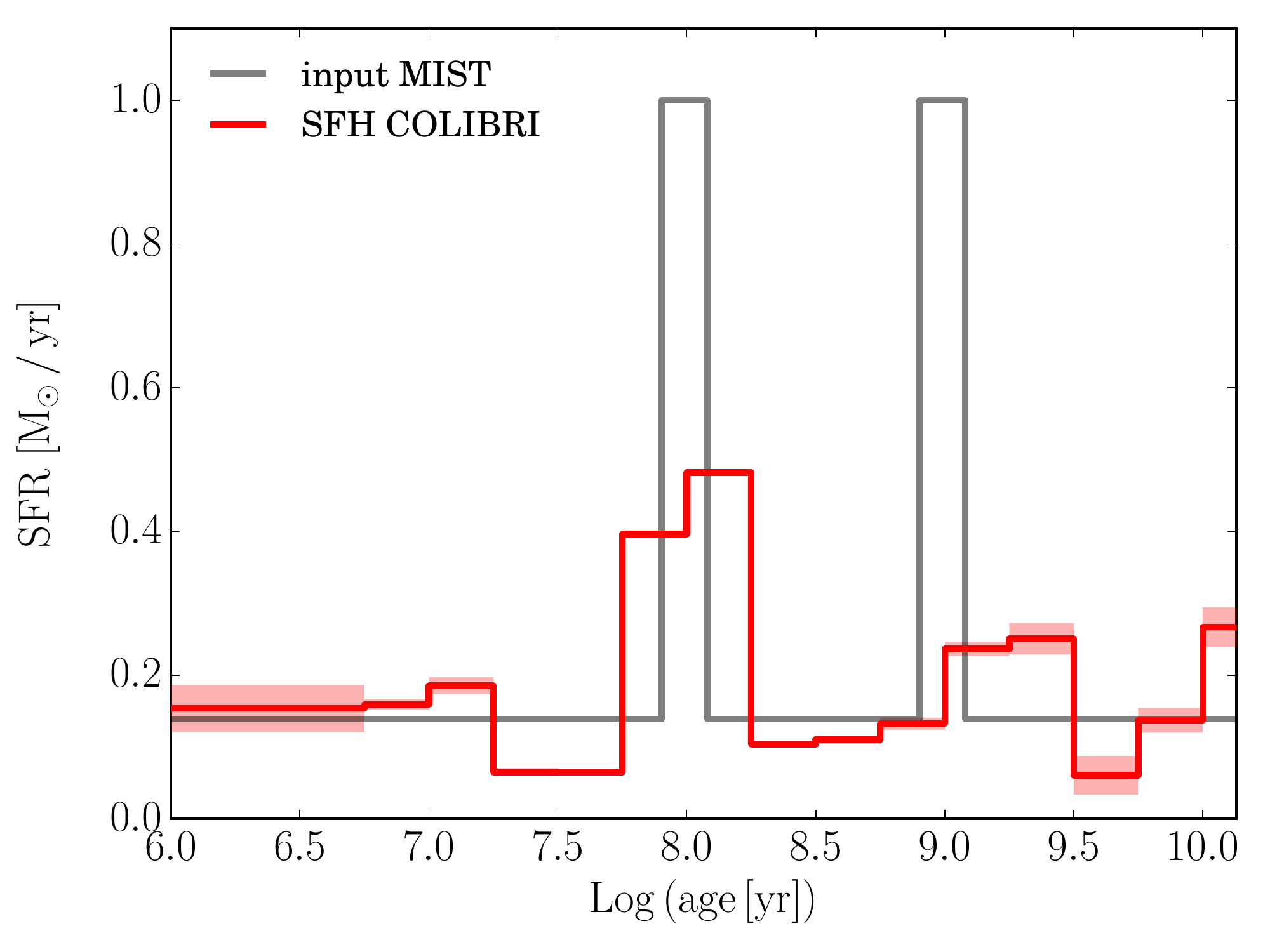}
\caption{Same as Figure \ref{burst-parsec10} but with a longer duration of the bursts ($\pm 20\%$ of the age).}
\label{burst-parsec20}
\end{figure}
\begin{figure}
\centering
\includegraphics[width=\columnwidth]{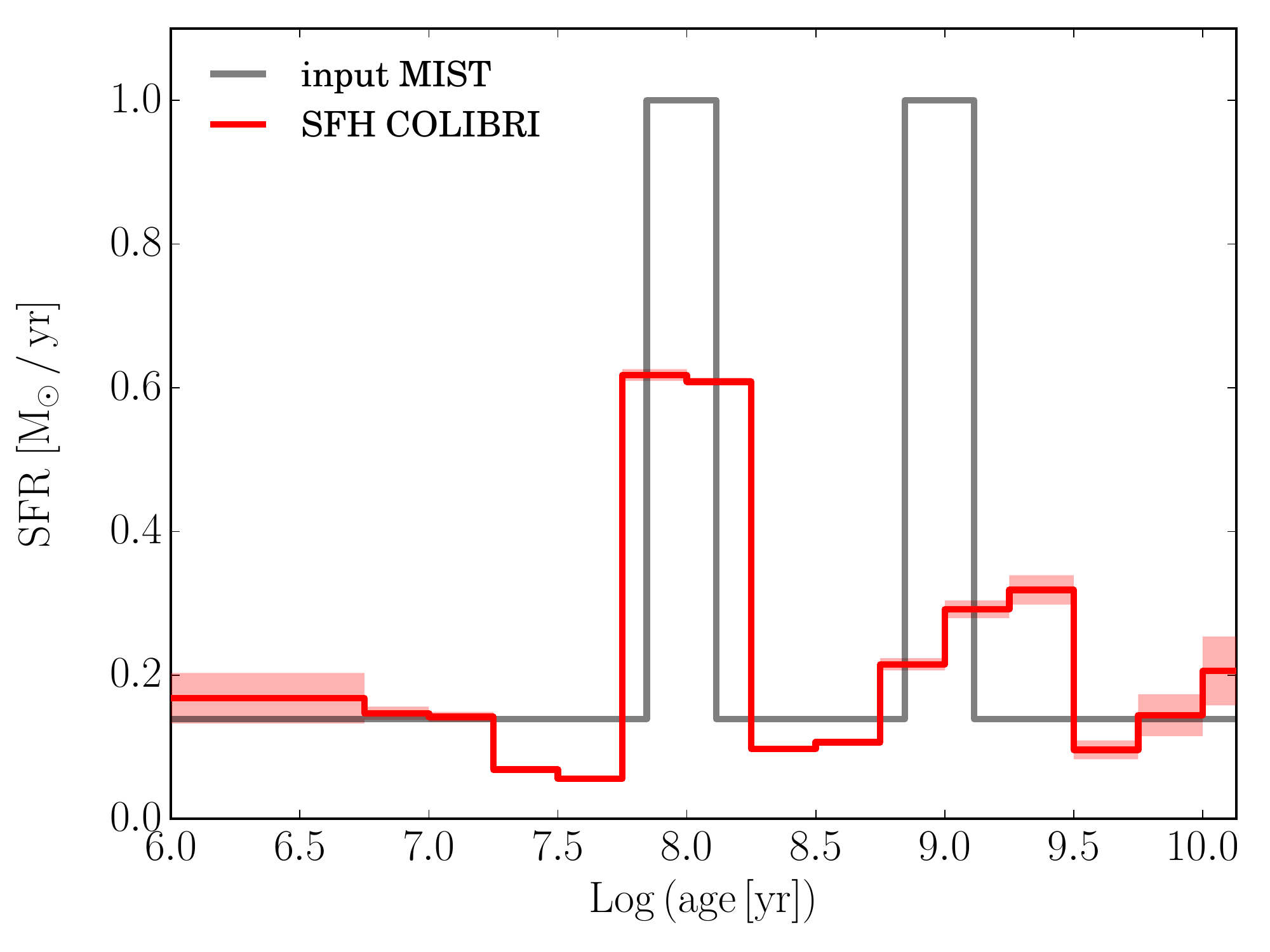}
\caption{Same as Figures \ref{burst-parsec10} and \ref{burst-parsec20} but with a longer duration of the bursts ($\pm 30\%$ of the age).}
\label{burst-parsec30}
\end{figure}

\subsection{Uncertainty treatment}
The statistical uncertainty is computed using a bootstrap technique on the data, i.e., applying small shifts in color and magnitude to the observed CMD and re-deriving the SFH for each version of it. We average the different solutions and take the rms deviation as statistical error on the mean value.

Systematic uncertainties, probably the most impacting ones (see the tests below), are accounted for by re-deriving the SFH with different age and CMD binnings, and using different sets of isochrones. These are added in quadrature to the statistical error.

The distance search is not included in the minimization. With all the other parameters fixed, we vary the distance and look for the value that minimizes the likelihood. This value is then fixed, and for the case of NGC~4449 is within the uncertainty range from the literature. The uncertainty due to the distance variations is not included in the SFH uncertainties, since it is negligible compared to the random and systematic differences included in the given error bars (varying the distance modulus up to $\pm 0.20$ affects the SFRs less than 5\%).

Other effects (e.g., variations of the binary fraction) have been shown to be negligible when compared to other kinds of uncertainties \citep{Cignoni2010,Monelli2010,Dolphin2013,Lewis2015}.

\subsection{Time resolution tests}
In order to quantify the impact of time resolution on our conclusions on the continuity (no strong burst and no long quiescent phases) of the SFH derived for NGC~4449, we built some tests for our code by using synthetic CMDs generated with a known SFH.

To test if we are able to identify and resolve quiescent phases in the life of a galaxy  at the distance of NGC~4449, we generated a CMD with a constant SFR with a gap at 100 Myr with a duration of 20 Myr, and another gap at 1 Gyr with a duration of 200 Myr (age$\pm$10\%). We simulated the same distance, photometry and completeness conditions as for NGC~4449. We then run SFERA with either the same stellar models used for constructing the fake data (MIST) or with a different set of isochrones (PARSEC-COLIBRI). The results are shown in Figures \ref{test-mesa10} and \ref{test-parsec10}, and it is evident that, when the same models are used, the lower SFR is perfectly recovered by the code; when we use the other set, although there are some hints, the gaps are not well recovered. This improves by increasing the duration of the quiescent phase to $\pm 20\%$ and $\pm 30\%$ of the age (see Figures \ref{test-parsec20} and \ref{test-parsec30}).

An analogous test was performed by adding two bursts to the constant SFR with the same criteria used before. Again, when the same isochrones are used for both constructing and recovering the SFH, the age resolution is within $\pm 10\%$ of the age (Figure \ref{burst-mesa10}), while if we use different sets, a duration of at least $\pm 30\%$ of the age is required to well constrain the bursts, in particular for the older one (see Figures \ref{burst-parsec10}, \ref{burst-parsec20} and \ref{burst-parsec30}).

These results strongly depend on the fact that our CMDs do not reach the ancient main sequence turnoff (MSTO), so the resolution at the oldest epochs is a few Gyr only, and the SFH relies on poorer age chronometers (RGB, AGB and TP-AGB phases). As thoroughly illustrated by \cite{Weisz2011}, the ancient MSTO constrains so tightly the SFH that the systematic uncertainties are strongly reduced independently of the stellar models used for the SFH recovery. They indeed find uncertainties larger than 50\% when the CMD is too shallow to reach the ancient MSTO. However, limiting the SFH studies to galaxies within $1-2$ Mpc, where we see the ancient MSTO, would prevent us to cover all the morphological types (e.g., blue compact dwarf galaxies), metallicity ranges and environments that populate the Local Universe.\\

\bibliography{bib}

\begin{thebibliography}{}
\expandafter\ifx\csname natexlab\endcsname\relax\def\natexlab#1{#1}\fi

\bibitem[{{Annibali} {et~al.}(2008){Annibali}, {Aloisi}, {Mack}, {Tosi}, {van
  der Marel}, {Angeretti}, {Leitherer}, \& {Sirianni}}]{Annibali2008}
{Annibali}, F., {Aloisi}, A., {Mack}, J., {et~al.} 2008, \aj, 135, 1900

\bibitem[{{Annibali} {et~al.}(2011){Annibali}, {Tosi}, {Aloisi}, \& {van der
  Marel}}]{Annibali2011}
{Annibali}, F., {Tosi}, M., {Aloisi}, A., \& {van der Marel}, R.~P. 2011, \aj,
  142, 129

\bibitem[{{Annibali} {et~al.}(2012){Annibali}, {Tosi}, {Aloisi}, {van der
  Marel}, \& {Martinez-Delgado}}]{Annibali2012}
{Annibali}, F., {Tosi}, M., {Aloisi}, A., {van der Marel}, R.~P., \&
  {Martinez-Delgado}, D. 2012, \apjl, 745, L1

\bibitem[{{Annibali} {et~al.}(2017){Annibali}, {Tosi}, {Romano}, {Buzzoni},
  {Cusano}, {Fumana}, {Marchetti}, {Mignoli}, {Pasquali}, \&
  {Aloisi}}]{Annibali2017}
{Annibali}, F., {Tosi}, M., {Romano}, D., {et~al.} 2017, \apj, 843, 20

\bibitem[{{Annibali} {et~al.}(2018){Annibali}, {Morandi}, {Watkins}, {Tosi},
  {Aloisi}, {Buzzoni}, {Cusano}, {Fumana}, {Marchetti}, {Mignoli},
  {Mucciarelli}, {Romano}, \& {van der Marel}}]{Annibali2018}
{Annibali}, F., {Morandi}, E., {Watkins}, L.~L., {et~al.} 2018, \mnras,
  arXiv:1802.02338

\bibitem[{{Aparicio} \& {Hidalgo}(2009)}]{Aparicio2009}
{Aparicio}, A., \& {Hidalgo}, S.~L. 2009, \aj, 138, 558

\bibitem[{{Berg} {et~al.}(2012){Berg}, {Skillman}, {Marble}, {van Zee},
  {Engelbracht}, {Lee}, {Kennicutt}, {Calzetti}, {Dale}, \&
  {Johnson}}]{Berg2012}
{Berg}, D.~A., {Skillman}, E.~D., {Marble}, A.~R., {et~al.} 2012, \apj, 754, 98

\bibitem[{{Boyer} {et~al.}(2009){Boyer}, {Skillman}, {van Loon}, {Gehrz}, \&
  {Woodward}}]{Boyer2009}
{Boyer}, M.~L., {Skillman}, E.~D., {van Loon}, J.~T., {Gehrz}, R.~D., \&
  {Woodward}, C.~E. 2009, \apj, 697, 1993

\bibitem[{{Calzetti} {et~al.}(2015){Calzetti}, {Lee}, {Sabbi}, {Adamo},
  {Smith}, {Andrews}, {Ubeda}, {Bright}, {Thilker}, {Aloisi}, {Brown},
  {Chandar}, {Christian}, {Cignoni}, {Clayton}, {da Silva}, {de Mink}, {Dobbs},
  {Elmegreen}, {Elmegreen}, {Evans}, {Fumagalli}, {Gallagher}, {Gouliermis},
  {Grebel}, {Herrero}, {Hunter}, {Johnson}, {Kennicutt}, {Kim}, {Krumholz},
  {Lennon}, {Levay}, {Martin}, {Nair}, {Nota}, {{\"O}stlin}, {Pellerin},
  {Prieto}, {Regan}, {Ryon}, {Schaerer}, {Schiminovich}, {Tosi}, {Van Dyk},
  {Walterbos}, {Whitmore}, \& {Wofford}}]{Calzetti2015}
{Calzetti}, D., {Lee}, J.~C., {Sabbi}, E., {et~al.} 2015, \aj, 149, 51

\bibitem[{{Cash}(1979)}]{Cash1979}
{Cash}, W. 1979, \apj, 228, 939

\bibitem[{{Choi} {et~al.}(2016){Choi}, {Dotter}, {Conroy}, {Cantiello},
  {Paxton}, \& {Johnson}}]{Choi2016}
{Choi}, J., {Dotter}, A., {Conroy}, C., {et~al.} 2016, \apj, 823, 102

\bibitem[{{Cignoni} {et~al.}(2012){Cignoni}, {Cole}, {Tosi}, {Gallagher},
  {Sabbi}, {Anderson}, {Grebel}, \& {Nota}}]{Cignoni2012}
{Cignoni}, M., {Cole}, A.~A., {Tosi}, M., {et~al.} 2012, \apj, 754, 130

\bibitem[{{Cignoni} {et~al.}(2013){Cignoni}, {Cole}, {Tosi}, {Gallagher},
  {Sabbi}, {Anderson}, {Grebel}, \& {Nota}}]{Cignoni2013}
{Cignoni}, M., {Cole}, A.~A., {Tosi}, M., {et~al.} 2013, \apj, 775, 83

\bibitem[{{Cignoni} \& {Tosi}(2010)}]{Cignoni2010}
{Cignoni}, M., \& {Tosi}, M. 2010, Advances in Astronomy, 2010, 158568

\bibitem[{{Cignoni} {et~al.}(2015){Cignoni}, {Sabbi}, {van der Marel}, {Tosi},
  {Zaritsky}, {Anderson}, {Lennon}, {Aloisi}, {de Marchi}, {Gouliermis},
  {Grebel}, {Smith}, \& {Zeidler}}]{Cignoni2015}
{Cignoni}, M., {Sabbi}, E., {van der Marel}, R.~P., {et~al.} 2015, \apj, 811,
  76

\bibitem[{{Cignoni} {et~al.}(2016){Cignoni}, {Sabbi}, {van der Marel},
  {Lennon}, {Tosi}, {Grebel}, {Gallagher}, {Aloisi}, {de Marchi}, {Gouliermis},
  {Larsen}, {Panagia}, \& {Smith}}]{Cignoni2016}
{Cignoni}, M., {Sabbi}, E., {van der Marel}, R.~P., {et~al.} 2016, \apj, 833,
  154

\bibitem[{{Cignoni} {et~al.}(2018){Cignoni}, {Sacchi}, {Aloisi}, {Tosi},
  {Calzetti}, {Lee}, {Sabbi}, {Adamo}, {Cook}, {Dale}, {Elmegreen},
  {Gallagher}, {Gouliermis}, {Grasha}, {Grebel}, {Hunter}, {Johnson}, {Messa},
  {Smith}, {Thilker}, {Ubeda}, \& {Whitmore}}]{Cignoni2018}
{Cignoni}, M., {Sacchi}, E., {Aloisi}, A., {et~al.} 2018, ArXiv e-prints,
  arXiv:1802.06792

\bibitem[{{Cole} {et~al.}(2007){Cole}, {Skillman}, {Tolstoy}, {Gallagher},
  {Aparicio}, {Dolphin}, {Gallart}, {Hidalgo}, {Saha}, {Stetson}, \&
  {Weisz}}]{Cole2007}
{Cole}, A.~A., {Skillman}, E.~D., {Tolstoy}, E., {et~al.} 2007, \apjl, 659, L17

\bibitem[{{Dolphin}(2016)}]{Dolphin2016}
{Dolphin}, A. 2016, {DOLPHOT: Stellar photometry}, Astrophysics Source Code
  Library, ascl:1608.013

\bibitem[{{Dolphin}(2002)}]{Dolphin2002}
{Dolphin}, A.~E. 2002, \mnras, 332, 91

\bibitem[{{Dolphin}(2013)}]{Dolphin2013}
{Dolphin}, A.~E. 2013, \apj, 775, 76

\bibitem[{{Dolphin} {et~al.}(2003){Dolphin}, {Saha}, {Skillman}, {Dohm-Palmer},
  {Tolstoy}, {Cole}, {Gallagher}, {Hoessel}, \& {Mateo}}]{Dolphin2003}
{Dolphin}, A.~E., {Saha}, A., {Skillman}, E.~D., {et~al.} 2003, \aj, 126, 187

\bibitem[{{El-Badry} {et~al.}(2016){El-Badry}, {Wetzel}, {Geha}, {Hopkins},
  {Kere{\v s}}, {Chan}, \& {Faucher-Gigu{\`e}re}}]{ElBadry2016}
{El-Badry}, K., {Wetzel}, A., {Geha}, M., {et~al.} 2016, \apj, 820, 131

\bibitem[{{Gallart} {et~al.}(2008){Gallart}, {Stetson}, {Meschin}, {Pont}, \&
  {Hardy}}]{Gallart2008}
{Gallart}, C., {Stetson}, P.~B., {Meschin}, I.~P., {Pont}, F., \& {Hardy}, E.
  2008, \apjl, 682, L89

\bibitem[{{Gallart} {et~al.}(2015){Gallart}, {Monelli}, {Mayer}, {Aparicio},
  {Battaglia}, {Bernard}, {Cassisi}, {Cole}, {Dolphin}, {Drozdovsky},
  {Hidalgo}, {Navarro}, {Salvadori}, {Skillman}, {Stetson}, \&
  {Weisz}}]{Gallart2015}
{Gallart}, C., {Monelli}, M., {Mayer}, L., {et~al.} 2015, \apjl, 811, L18

\bibitem[{{Gelatt} {et~al.}(2001){Gelatt}, {Hunter}, \&
  {Gallagher}}]{Gelatt2001}
{Gelatt}, A.~E., {Hunter}, D.~A., \& {Gallagher}, J.~S. 2001, \pasp, 113, 142

\bibitem[{{Gouliermis} {et~al.}(2017){Gouliermis}, {Elmegreen}, {Elmegreen},
  {Calzetti}, {Cignoni}, {Gallagher}, {Kennicutt}, {Klessen}, {Sabbi},
  {Thilker}, {Ubeda}, {Aloisi}, {Adamo}, {Cook}, {Dale}, {Grasha}, {Grebel},
  {Johnson}, {Sacchi}, {Shabani}, {Smith}, \& {Wofford}}]{Gouliermis2017}
{Gouliermis}, D.~A., {Elmegreen}, B.~G., {Elmegreen}, D.~M., {et~al.} 2017,
  \mnras, 468, 509

\bibitem[{{Grasha} {et~al.}(2017{\natexlab{a}}){Grasha}, {Elmegreen},
  {Calzetti}, {Adamo}, {Aloisi}, {Bright}, {Cook}, {Dale}, {Fumagalli},
  {Gallagher}, {Gouliermis}, {Grebel}, {Kahre}, {Kim}, {Krumholz}, {Lee},
  {Messa}, {Ryon}, \& {Ubeda}}]{Grasha2017a}
{Grasha}, K., {Elmegreen}, B.~G., {Calzetti}, D., {et~al.} 2017{\natexlab{a}},
  \apj, 842, 25

\bibitem[{{Grasha} {et~al.}(2017{\natexlab{b}}){Grasha}, {Calzetti}, {Adamo},
  {Kim}, {Elmegreen}, {Gouliermis}, {Dale}, {Fumagalli}, {Grebel}, {Johnson},
  {Kahre}, {Kennicutt}, {Messa}, {Pellerin}, {Ryon}, {Smith}, {Shabani},
  {Thilker}, \& {Ubeda}}]{Grasha2017b}
{Grasha}, K., {Calzetti}, D., {Adamo}, A., {et~al.} 2017{\natexlab{b}}, \apj,
  840, 113

\bibitem[{{Harris} {et~al.}(1997){Harris}, {Zaritsky}, \&
  {Thompson}}]{Harris1997}
{Harris}, J., {Zaritsky}, D., \& {Thompson}, I. 1997, \aj, 114, 1933

\bibitem[{{Hunter} {et~al.}(1999){Hunter}, {van Woerden}, \&
  {Gallagher}}]{Hunter1999}
{Hunter}, D.~A., {van Woerden}, H., \& {Gallagher}, J.~S. 1999, \aj, 118, 2184

\bibitem[{{Johnson} {et~al.}(2016){Johnson}, {Seth}, {Dalcanton}, {Beerman},
  {Fouesneau}, {Lewis}, {Weisz}, {Williams}, {Bell}, {Dolphin}, {Larsen},
  {Sandstrom}, \& {Skillman}}]{Johnson2016}
{Johnson}, L.~C., {Seth}, A.~C., {Dalcanton}, J.~J., {et~al.} 2016, \apj, 827,
  33

\bibitem[{{Karczewski} {et~al.}(2013){Karczewski}, {Barlow}, {Page}, {Kuin},
  {Ferreras}, {Baes}, {Bendo}, {Boselli}, {Cooray}, {Cormier}, {De Looze},
  {Galametz}, {Galliano}, {Lebouteiller}, {Madden}, {Pohlen}, {R{\'e}my-Ruyer},
  {Smith}, \& {Spinoglio}}]{Karczewski2013}
{Karczewski}, O.~{\L}., {Barlow}, M.~J., {Page}, M.~J., {et~al.} 2013, \mnras,
  431, 2493

\bibitem[{{Kroupa}(2001)}]{Kroupa2001}
{Kroupa}, P. 2001, \mnras, 322, 231

\bibitem[{{Lee} {et~al.}(2009){Lee}, {Kennicutt}, {Funes}, {Sakai}, \&
  {Akiyama}}]{Lee2009}
{Lee}, J.~C., {Kennicutt}, Jr., R.~C., {Funes}, S.~J.~J.~G., {Sakai}, S., \&
  {Akiyama}, S. 2009, \apj, 692, 1305

\bibitem[{{Lelli} {et~al.}(2014{\natexlab{a}}){Lelli}, {Fraternali}, \&
  {Verheijen}}]{Lelli2014a}
{Lelli}, F., {Fraternali}, F., \& {Verheijen}, M. 2014{\natexlab{a}}, \aap,
  563, A27

\bibitem[{{Lelli} {et~al.}(2014{\natexlab{b}}){Lelli}, {Verheijen}, \&
  {Fraternali}}]{Lelli2014b}
{Lelli}, F., {Verheijen}, M., \& {Fraternali}, F. 2014{\natexlab{b}}, \aap,
  566, A71

\bibitem[{{Lelli} {et~al.}(2014{\natexlab{c}}){Lelli}, {Verheijen}, \&
  {Fraternali}}]{Lelli2014c}
{Lelli}, F., {Verheijen}, M., \& {Fraternali}, F. 2014{\natexlab{c}}, \mnras,
  445, 1694

\bibitem[{{Lewis} {et~al.}(2015){Lewis}, {Dolphin}, {Dalcanton}, {Weisz},
  {Williams}, {Bell}, {Seth}, {Simones}, {Skillman}, {Choi}, {Fouesneau},
  {Guhathakurta}, {Johnson}, {Kalirai}, {Leroy}, {Monachesi}, {Rix}, \&
  {Schruba}}]{Lewis2015}
{Lewis}, A.~R., {Dolphin}, A.~E., {Dalcanton}, J.~J., {et~al.} 2015, \apj, 805,
  183

\bibitem[{{Marconi} {et~al.}(1995){Marconi}, {Tosi}, {Greggio}, \&
  {Focardi}}]{Marconi1995}
{Marconi}, G., {Tosi}, M., {Greggio}, L., \& {Focardi}, P. 1995, \aj, 109, 173

\bibitem[{{Marigo} {et~al.}(2013){Marigo}, {Bressan}, {Nanni}, {Girardi}, \&
  {Pumo}}]{Marigo2013}
{Marigo}, P., {Bressan}, A., {Nanni}, A., {Girardi}, L., \& {Pumo}, M.~L. 2013,
  \mnras, 434, 488

\bibitem[{{Mart{\'{\i}}nez-Delgado} {et~al.}(2012){Mart{\'{\i}}nez-Delgado},
  {Romanowsky}, {Gabany}, {Annibali}, {Arnold}, {Fliri}, {Zibetti}, {van der
  Marel}, {Rix}, {Chonis}, {Carballo-Bello}, {Aloisi}, {Macci{\`o}},
  {Gallego-Laborda}, {Brodie}, \& {Merrifield}}]{Martinez-Delgado2012}
{Mart{\'{\i}}nez-Delgado}, D., {Romanowsky}, A.~J., {Gabany}, R.~J., {et~al.}
  2012, \apjl, 748, L24

\bibitem[{{McQuinn} {et~al.}(2012){McQuinn}, {Skillman}, {Dalcanton}, {Cannon},
  {Dolphin}, {Holtzman}, {Weisz}, \& {Williams}}]{McQuinn2012}
{McQuinn}, K.~B.~W., {Skillman}, E.~D., {Dalcanton}, J.~J., {et~al.} 2012,
  \apj, 759, 77

\bibitem[{{McQuinn} {et~al.}(2010){McQuinn}, {Skillman}, {Cannon}, {Dalcanton},
  {Dolphin}, {Hidalgo-Rodr{\'{\i}}guez}, {Holtzman}, {Stark}, {Weisz}, \&
  {Williams}}]{McQuinn2010}
{McQuinn}, K.~B.~W., {Skillman}, E.~D., {Cannon}, J.~M., {et~al.} 2010, \apj,
  721, 297

\bibitem[{{Meschin} {et~al.}(2014){Meschin}, {Gallart}, {Aparicio}, {Hidalgo},
  {Monelli}, {Stetson}, \& {Carrera}}]{Meschin2014}
{Meschin}, I., {Gallart}, C., {Aparicio}, A., {et~al.} 2014, \mnras, 438, 1067

\bibitem[{{Monelli} {et~al.}(2010){Monelli}, {Hidalgo}, {Stetson}, {Aparicio},
  {Gallart}, {Dolphin}, {Cole}, {Weisz}, {Skillman}, {Bernard}, {Mayer},
  {Navarro}, {Cassisi}, {Drozdovsky}, \& {Tolstoy}}]{Monelli2010}
{Monelli}, M., {Hidalgo}, S.~L., {Stetson}, P.~B., {et~al.} 2010, \apj, 720,
  1225

\bibitem[{{Pan} {et~al.}(2015){Pan}, {Li}, {Lin}, {Wang}, {Fan}, \&
  {Kong}}]{Pan2015}
{Pan}, Z., {Li}, J., {Lin}, W., {et~al.} 2015, \apjl, 804, L42

\bibitem[{{Pe{\~n}arrubia} {et~al.}(2009){Pe{\~n}arrubia}, {Navarro},
  {McConnachie}, \& {Martin}}]{Penarrubia2009}
{Pe{\~n}arrubia}, J., {Navarro}, J.~F., {McConnachie}, A.~W., \& {Martin},
  N.~F. 2009, \apj, 698, 222

\bibitem[{{Reines} {et~al.}(2008){Reines}, {Johnson}, \& {Goss}}]{Reines2008}
{Reines}, A.~E., {Johnson}, K.~E., \& {Goss}, W.~M. 2008, \aj, 135, 2222

\bibitem[{{Rich} {et~al.}(2012){Rich}, {Collins}, {Black}, {Longstaff}, {Koch},
  {Benson}, \& {Reitzel}}]{Rich2012}
{Rich}, R.~M., {Collins}, M.~L.~M., {Black}, C.~M., {et~al.} 2012, \nat, 482,
  192

\bibitem[{{Rosenfield} {et~al.}(2016){Rosenfield}, {Marigo}, {Girardi},
  {Dalcanton}, {Bressan}, {Williams}, \& {Dolphin}}]{Rosenfield2016}
{Rosenfield}, P., {Marigo}, P., {Girardi}, L., {et~al.} 2016, \apj, 822, 73

\bibitem[{{Sabbi} {et~al.}(2018){Sabbi}, {Calzetti}, {Ubeda}, {Adamo},
  {Cignoni}, {Thilker}, {Aloisi}, {Elmegreen}, {Elmegreen}, {Gouliermis},
  {Grebel}, {Messa}, {Smith}, {Tosi}, {Dolphin}, {Andrews}, {Ashworth},
  {Bright}, {Brown}, {Chandar}, {Christian}, {Clayton}, {Cook}, {Dale}, {de
  Mink}, {Dobbs}, {Evans}, {Fumagalli}, {Gallagher}, {Grasha}, {Herrero},
  {Hunter}, {Johnson}, {Kahre}, {Kennicutt}, {Kim}, {Krumholz}, {Lee},
  {Lennon}, {Martin}, {Nair}, {Nota}, {Ostlin}, {Pellerin}, {Prieto}, {Regan},
  {Ryon}, {Sacchi}, {Schaerer}, {Schiminovich}, {Shabani}, {Van Dyk},
  {Walterbos}, {Whitmore}, \& {Wofford}}]{Sabbi2018}
{Sabbi}, E., {Calzetti}, D., {Ubeda}, L., {et~al.} 2018, ArXiv e-prints,
  arXiv:1801.05467

\bibitem[{{Sacchi} {et~al.}(2016){Sacchi}, {Annibali}, {Cignoni}, {Aloisi},
  {Sohn}, {Tosi}, {van der Marel}, {Grocholski}, \& {James}}]{Sacchi2016}
{Sacchi}, E., {Annibali}, F., {Cignoni}, M., {et~al.} 2016, \apj, 830, 3

\bibitem[{{Schlafly} \& {Finkbeiner}(2011)}]{Schlafly2011}
{Schlafly}, E.~F., \& {Finkbeiner}, D.~P. 2011, \apj, 737, 103

\bibitem[{{Skillman} {et~al.}(2017){Skillman}, {Monelli}, {Weisz}, {Hidalgo},
  {Aparicio}, {Bernard}, {Boylan-Kolchin}, {Cassisi}, {Cole}, {Dolphin},
  {Ferguson}, {Gallart}, {Irwin}, {Martin}, {Mart{\'{\i}}nez-V{\'a}zquez},
  {Mayer}, {McConnachie}, {McQuinn}, {Navarro}, \& {Stetson}}]{Skillman2017}
{Skillman}, E.~D., {Monelli}, M., {Weisz}, D.~R., {et~al.} 2017, \apj, 837, 102

\bibitem[{{Sokal} {et~al.}(2015){Sokal}, {Johnson}, {Indebetouw}, \&
  {Reines}}]{Sokal2015}
{Sokal}, K.~R., {Johnson}, K.~E., {Indebetouw}, R., \& {Reines}, A.~E. 2015,
  \aj, 149, 115

\bibitem[{{Strader} {et~al.}(2012){Strader}, {Seth}, \&
  {Caldwell}}]{Strader2012}
{Strader}, J., {Seth}, A.~C., \& {Caldwell}, N. 2012, \aj, 143, 52

\bibitem[{{Theis} \& {Kohle}(2001)}]{Theis2001}
{Theis}, C., \& {Kohle}, S. 2001, \aap, 370, 365

\bibitem[{{Tolstoy} {et~al.}(2009){Tolstoy}, {Hill}, \& {Tosi}}]{Tolstoy2009}
{Tolstoy}, E., {Hill}, V., \& {Tosi}, M. 2009, \araa, 47, 371

\bibitem[{{Tosi} {et~al.}(1991){Tosi}, {Greggio}, {Marconi}, \&
  {Focardi}}]{Tosi1991}
{Tosi}, M., {Greggio}, L., {Marconi}, G., \& {Focardi}, P. 1991, \aj, 102, 951

\bibitem[{{Valdez-Guti{\'e}rrez} {et~al.}(2002){Valdez-Guti{\'e}rrez},
  {Rosado}, {Puerari}, {Georgiev}, {Borissova}, \&
  {Ambrocio-Cruz}}]{Valdez2002}
{Valdez-Guti{\'e}rrez}, M., {Rosado}, M., {Puerari}, I., {et~al.} 2002, \aj,
  124, 3157

\bibitem[{{Weisz} {et~al.}(2008){Weisz}, {Skillman}, {Cannon}, {Dolphin},
  {Kennicutt}, {Lee}, \& {Walter}}]{Weisz2008}
{Weisz}, D.~R., {Skillman}, E.~D., {Cannon}, J.~M., {et~al.} 2008, \apj, 689,
  160

\bibitem[{{Weisz} {et~al.}(2011){Weisz}, {Dalcanton}, {Williams}, {Gilbert},
  {Skillman}, {Seth}, {Dolphin}, {McQuinn}, {Gogarten}, {Holtzman}, {Rosema},
  {Cole}, {Karachentsev}, \& {Zaritsky}}]{Weisz2011}
{Weisz}, D.~R., {Dalcanton}, J.~J., {Williams}, B.~F., {et~al.} 2011, \apj,
  739, 5

\bibitem[{{Williams} {et~al.}(2009){Williams}, {Dalcanton}, {Dolphin},
  {Holtzman}, \& {Sarajedini}}]{Williams2009}
{Williams}, B.~F., {Dalcanton}, J.~J., {Dolphin}, A.~E., {Holtzman}, J., \&
  {Sarajedini}, A. 2009, \apjl, 695, L15

\bibitem[{{Zhang} {et~al.}(2012){Zhang}, {Hunter}, {Elmegreen}, {Gao}, \&
  {Schruba}}]{Zhang2012}
{Zhang}, H.-X., {Hunter}, D.~A., {Elmegreen}, B.~G., {Gao}, Y., \& {Schruba},
  A. 2012, \aj, 143, 47

\end{thebibliography}

\end{document}